\documentclass[11pt]{article}
\usepackage{amsmath}   
\usepackage{amsfonts}    
\usepackage{xfrac} 
\usepackage{tabu}
\usepackage{color}
\definecolor{myred}{RGB}{255, 0, 0}
\definecolor{myblue}{RGB}{0, 0, 255}
\newcommand {\bu} {\mbox{\boldmath $u$}}
\newcommand {\btildeu} {\mbox{\boldmath $\tilde{u}$}}

\newcommand {\bx} {\mbox{\boldmath $x$}}
\newcommand {\btildex} {\mbox{\boldmath $\tilde{x}$}}
\newcommand {\by} {\mbox{\boldmath $y$}}
\newcommand {\bz} {\mbox{\boldmath $z$}}

\newcommand {\bU} {\mbox{\boldmath $U$}}

\newcommand {\bX} {\mbox{\boldmath $X$}}
\newcommand {\bY} {\mbox{\boldmath $Y$}}
\newcommand {\bZ} {\mbox{\boldmath $Z$}}
\newcommand{\calA}{{\cal A}}
\newcommand{\calB}{{\cal B}}
\newcommand{\calC}{{\cal C}}

\newcommand{\calG}{{\cal G}}

\newcommand{\calI}{{\cal I}}
\newcommand{\calJ}{{\cal J}}
\newcommand{\calK}{{\cal K}}

\newcommand{\calP}{{\cal P}}
\newcommand{\calQ}{{\cal Q}}

\newcommand{\calS}{{\cal S}}
\newcommand{\calT}{{\cal T}}
\newcommand{\calU}{{\cal U}}

\newcommand{\calX}{{\cal X}}
\newcommand{\calY}{{\cal Y}}
\newcommand{\calZ}{{\cal Z}}

\begin{document}
\thispagestyle{empty}
\title{Expurgated Bounds for the Asymmetric Broadcast Channel\\}
\author{\\ Ran Averbuch, Nir Weinberger$^{*}$, and Neri Merhav\\}
\maketitle
\begin{center}
The Andrew \& Erna Viterbi Faculty of Electrical Engineering \\
Technion - Israel Institute of Technology \\
Technion City, Haifa 3200003, ISRAEL \\
nir.wein@gmail.com, \{rans@campus, merhav@ee\}.technion.ac.il
\end{center}
\vspace{1.5\baselineskip}
\setlength{\baselineskip}{1.5\baselineskip}

\begin{abstract}
\let\thefootnote\relax\footnote{* This work was done while N. Weinberger was at the Technion. Currently, he is with the School of Electrical Engineering at Tel--Aviv University.}
This work contains two main contributions concerning the 
expurgation of hierarchical ensembles for the asymmetric broadcast channel. 
The first is an analysis of 
the optimal maximum likelihood (ML) decoders for the weak and strong user. 
Two different methods of code expurgation will be used, that will provide two competing error exponents.
The second is the derivation of expurgated exponents under the generalized 
stochastic likelihood decoder (GLD).
We prove that the GLD exponents are at least as tight as the maximum between the random coding error exponents derived in an earlier work by Averbuch and Merhav (2017) and one of our ML--based expurgated exponents. 
By that, we actually prove the existence of hierarchical codebooks 
that achieve the best of the random coding exponent and the expurgated exponent simultaneously for both users.   \\

\end{abstract}

\newpage
\section{Introduction}
One of the most elementary system configuation models in multi-user information theory 
is the broadcast channel (BC). It has been introduced more than four decades ago by Cover \cite{COVER72}, and since then, a vast amount of papers and books, analyzing different aspects of 
the broadcast model, have been published. 
Although the characterization of the capacity region of the general BC is still 
an open problem, some special cases have been solved. 
Most notably, the broadcast channel with degraded message sets, also known as the asymmetric broadcast channel (ABC), was introduced and solved by K\"{o}rner and Marton \cite{KM77}.  

The direct part of their coding theorem relies on  
Bergmans' scheme \cite{BERGMANS1}, which suggested the use of an hierarchical random code: 
first generate ``cloud centers", 
which designate messages 
intended to both the receiver with
the relatively high channel quality, 
henceforth referred to as the {\it strong user}, and the receiver with
the relatively low channel quality, henceforth referred to as the {\it weak user}.
Then, in the second step, ``around" each cloud center,
generate a ``sattelite" codeword for each message 
that is intended to the strong user
only. The transmitter sends a codeword pertaining to one of the clouds. 
The strong decoder fully decodes both the common message (cloud center) and his private message (sattelite),
whereas the weak decoder decodes the common message only. 

While the capacity region of the ABC has been known for many years, 
only little is known about its reliability functions.
The earliest work on error exponents for the general ABC 
is of K\"{o}rner and Sgarro \cite{KS80}. Later, Kaspi and Merhav \cite{KM11} have derived tighter lower bounds to the reliability functions of both users
by analyzing random coding error exponents of their optimal decoders.
Most recently \cite{AM17}, the exact random coding error exponents have been determined for both the strong user and the weak user, for the ensemble of fixed composition codes.

Even in the single--user case, it is known for many years  
that the random coding error exponent is not tight (with respect to the reliability function) for relatively low coding rates, and may be improved by expurgation \cite{CKM77}, \cite{GAL68}.  
Specifically, improved bounds are obtained by eliminating codewords that contribute relatively highly to the error probability, 
and asserting that some upper bound holds for all remaining codewords.
More recent papers, where the method of expurgation is studied are \cite{MERHAVetal2014}, \cite{MERHAVlist2014} (list decoding), \cite{PRAD2008} (Gaussian BC) and \cite{PRAD2014} (discrete memoryless multiple--access channels), among many others.  

The main objective of this paper is to study expurgation techniques for the hierarchical ensemble used over the ABC. 
Expurgating a code for the ABC is not a trivial extension of expurgation in the single-user case, because there might be conflicting goals from the viewpoints of the two users. 
Nonetheless, we were able to define expurgation procedures that guarantee no harm to the performance of either user. 
This has paved the way to derive tighter lower bounds on 
the reliability functions of the ABC.

We start by analyzing the optimal maximum likelihood (ML) decoder, and derive some expurgated bounds, that are natural generalizations of the single--user expurgated bound due to Csisz\'ar, K\"orner and Marton (CKM) \cite{CKM77}. Although our first process of code expurgation is fairly intuitive, there is at least one specific step in our first derivation where exponential tightness might be compromised. 
This point gives rise to a possible room for improvement upon the results of our first theorem, and indeed, such an improvement is achieved by a second method of expurgation. 
Here, one starts by expurgating cloud centers, and only afterwards, single codewords. The intuition behind this technique is the following. When the exponential rate of the codewords within a cloud is too high, the weak user can still make a good estimation, merely by relying on the set of cloud centers. 
The expurgated bounds of our second method, however, are not always tighter than those of the first method, because of other differences in their derivations.

We then expand the scope and consider the generalized likelihood decoder (GLD), 
which is a more general family of stochastic likelihood decoders. 
For such decoders, the probability of deciding on a given message is proportional to a general exponential function of the joint empirical distribution of the cloud--center, the codeword and the received channel output vector.
The random coding error exponent of the ordinary and the mismatched likelihood decoders for single--user have been derived by Scarlett {\it et al.} \cite{Scar2015}. In a more recent paper by Merhav \cite{MERHAV2017}, the expurgated exponent of the GLD has been derived and compared to the classical expurgated bound of \cite{CKM77}, showing an explicit improvement at relatively high coding rates. 
In this paper, we consider GLD's for both the strong and the weak users of an ABC, and derive expurgated exponents under these decoders. 
These bounds generalize the bound of \cite{MERHAV2017}, and prove that they are at least as tight as the maximum between the random coding error exponents of \cite{AM17} and the expurgated bounds of our first theorem, which are based on the ML decoder. By that, we actually prove the existence of hierarchical codebooks 
that attain the best of the random coding exponent and the expurgated exponent simultaneously for both users.
The main drawback of those error exponents is that they are not easy to calculate since they involve minimizations over relatively cumbersome auxiliary channels, 
and hence, efficient computation algorithms for the GLD 
bound are under current research. 
From this viewpoint, the exponents of our first theorems are much more attractive.  

The remaining part of the paper is organized as follows. 
In Section 2, we establish notation conventions, and review some preliminaries.
In Section 3, we formalize the model, the decoders and reliability functions. 
In Section 4, we summarize the main theoretical results of this paper. 
Sections 5, 6 and 7 provide the proofs of our main theorems.

\section{Notation Conventions and Preliminaries}

\subsection{Notation Conventions}
Throughout the paper, random variables will be denoted by capital letters, specific values they 
may take will be denoted by the corresponding lower case letters, and their alphabets will be 
denoted by calligraphic letters. Random vectors and their realizations will be denoted, 
respectively, by capital letters and the corresponding lower case letters, both in the bold face 
font. Their alphabets will be superscripted by their dimensions. 
For example, the random vector $\bX = (X_{1}, \dotsc , X_{n})$, ($n$ - positive integer) 
may take a specific vector value $\bx = (x_{1}, \dotsc , x_{n})$ in $\mathcal{X}^{n}$, 
the $n$-th order Cartesian power of $\mathcal{X}$, which is the alphabet of each component of this vector. Sources and channels will be subscripted by the names of the relevant random 
variables/vectors and their conditionings, whenever applicable, 
following the standard notation conventions, e.g., $Q_{X}$, $Q_{Y|X}$, and so on. 
When there is no room for ambiguity, these subscripts will be omitted. For a generic joint 
distribution $Q_{XY} = \{Q_{XY}(x,y), x \in \mathcal{X}, y \in \mathcal{Y} \}$, which will sometimes 
be abbreviated by $Q$, information measures will be denoted in the conventional manner, but
with a subscript $Q$, that is, $H_{Q}(X)$ 
is the marginal entropy of $X$, $H_{Q}(X|Y)$ is the conditional entropy of $X$ given $Y$, 
$I_{Q}(X;Y) = H_{Q}(X) - H_{Q}(X|Y)$ is the mutual information between $X$ and $Y$, 
and so on. The weighted divergence between 
two conditional distributions (channels), say, $Q_{Z|X}$ and $W = \{W(z|x), 
x \in \mathcal{X}, z \in \mathcal{Z} \}$, with weighting $Q_{X}$ is defined as
\begin{equation}
  D(Q_{Z|X} || W | Q_{X}) = \sum_{x \in \mathcal{X}} Q_{X}(x) 
\sum_{z \in \mathcal{Z}} Q_{Z|X}(z|x) \log \frac{Q_{Z|X}(z|x)}{W(z|x)},
\end{equation}
where logarithms, here and throughout the sequel, are taken to the natural base.
The probability of an event $\mathcal{E}$ will be denoted by 
$\mathrm{Pr} \{ \mathcal{E} \}$, and the expectation operator with respect to (w.r.t.) a 
probability distribution $P$ will be denoted by $\mathbb{E}_{P} \{\cdot \}$, where the subscript will often be omitted. 
For two positive sequences $a_{n}$ and $b_{n}$, the notation $a_{n} \doteq b_{n}$ will stand 
for equality in the exponential scale, that is, $\lim_{n \to \infty} \frac{1}{n} 
\log \frac{a_{n}}{b_{n}} = 0$. The indicator function of an event $\mathcal{E}$ 
will be denoted by $\mathcal{I} \{ \mathcal{E} \}$. 
The notation $[x]_{+}$ will stand for $\max \{0, x\}$. 

The empirical distribution of a sequence $\bx \in \mathcal{X}^{n}$, which will be denoted by $\hat{P}_{\bx}$, is the vector of relative frequencies, $\hat{P}_{\bx}(x)$, 
of each symbol $x \in \mathcal{X}$ in $\bx$. 
The type class of $\bx \in \mathcal{X}^{n}$, denoted $\mathcal{T}(\bx)$, is the set of all vectors $\bx'$ with $\hat{P}_{\bx'} = \hat{P}_{\bx}$. 
When we wish to emphasize the dependence of the type class on the empirical distribution $\hat{P}$, we will denote it by $\mathcal{T}(\hat{P})$. Information measures associated with empirical distributions will be denoted with 'hats' and will be subscripted by the sequences from which they are induced. 
For example, the entropy associated 
with $\hat{P}_{\bx}$, which is the empirical entropy of $\bx$, will be denoted by $\hat{H}_{\bx}(X)$. Similar conventions will apply to the joint empirical distribution, 
the joint type class, the conditional empirical distributions and the conditional type classes 
associated with pairs (and multiples) of sequences of length $n$. 
Accordingly, $\hat{P}_{\bx\by}$ would be the joint empirical distribution of $(\bx, \by) = \{(x_{i}, y_{i})\}_{i=1}^{n}$, 
$\mathcal{T}(\hat{P}_{\bx\by})$ will denote the joint type class of $(\bx, \by)$, $\mathcal{T}(\bx| \by)$ will stand for the conditional type class of $\bx$ given $\by$, 
$\hat{I}_{\bx\by}(X;Y)$ will denote the empirical mutual information, and so on. 
When we wish to emphasize the dependence of $\mathcal{T}(\bx| \by)$ upon 
$\by$ and the relevant empirical conditional distribution, $Q_{X|Y} = 
\hat{P}_{\bx|\by}$, we denote it by $\mathcal{T}( Q_{X|Y} | \by)$. 
Similar conventions will apply to triples of sequences, say, 
$\{(\bx, \by, \bz)\}$, etc. Likewise, when we wish to emphasize the 
dependence of empirical information measures upon a given empirical distribution, $Q$, we denote them using the subscript $Q$, as described above.

\subsection{Preliminaries}

As already mentioned in the Introduction, the exact random coding error exponents of the ABC have been derived and analyzed in \cite{AM17}. 
For the weak user, it is given by
\begin{align}
E_{\mbox{\tiny w}}(R_{y},R_{z}) =  \min_{Q_{Z|UX} }  \left\{ D(Q_{Z|UX} \| W_{Z|X} | Q_{UX})    
+  \left[ I_{Q}(U;Z) + \left[ I_{Q}(X;Z|U) - R_{y} \right]_{+} - R_{z} \right]_{+}    \right\},
\end{align}
while the exact error exponent of the strong user is given by
\begin{align}
E_{\mbox{\tiny s}}(R_{y},R_{z}) &=  \min_{Q_{Y|UX} }  \bigg\{  D(Q_{Y|UX} \| W_{Y|X} | Q_{UX})  \nonumber \\
&  ~~~~~~~~~~~~~~~~~~~~ + \min \left\{ \left[ I_{Q}(UX;Y)  - R_{y}  - R_{z} \right]_{+} , \left[ I_{Q}(X;Y|U) - R_{y} \right]_{+} \right\}  \bigg\}.
\end{align}  
Along the proofs in the current paper, some mathematical results are used extensively. Instead of explaining them each time repeatedly, let us summerize them: \\
$\bullet$ We abbreviate the union bounds and Markov's inequality by UB and MI, respectively. \\
$\bullet$ We refer to the following inequality as the {\it power distribution inequality} (PD),
\begin{align}
\left(\sum_{j \in \calJ} a_{j}  \right)^{s} \leq \sum_{j \in \calJ} a_{j}^{s}
\end{align}  
which holds whenever $s \in [0,1]$ and $a_{j} \geq 0$ for all $j \in \calJ$ \cite[Exercise 4.15(f)]{GAL68}. A special case occurs when the cardinality of $\calJ$ is subexponential in the blocklength $n$ (e.g., when we sum over type--classes) and then
\begin{align}
\left(\sum_{j \in \calJ} e^{n a_{j}}  \right)^{s}
\doteq \left(\max_{j \in \calJ} e^{n a_{j}}  \right)^{s}
= \max_{j \in \calJ} e^{n s a_{j}} 
\doteq  \sum_{j \in \calJ} e^{n s a_{j}}.
\end{align}
$\bullet$ Let $N$ be a binomial random variable with $e^{nR}$ ($R \geq 0$) trials and success rate of the exponential order of $e^{-n I}$ ($I \geq 0$). It is shown in \cite[Chap. 6.3]{MERHAV09} that for $s > 0$
\begin{align}
\label{TEMdef}
\mathbb{E} \left\{  N^{s} \right\} 
\doteq         
          \left\{ 
          \begin{array}{l l}
\exp \left\{ n [ R  - I ] s \right\}        & \quad \text{$R  \geq  I$  } \\
\exp \left\{ n [ R -  I  ] \right\}            & \quad \text{$R  <  I$  } 
           \end{array} \right.  .
\end{align}

\section{Definitions and Problem Formulation}

We consider a memoryless ABC 
with a finite input alphabet $\mathcal{X}$ and finite output alphabets $\mathcal{Y}$ and $\mathcal{Z}$. 
Let $W_{Y|X} \equiv W_{1} = \{W_{1}(y|x),~x \in \mathcal{X},~y \in \mathcal{Y} \}$ 
and $W_{Z|X} \equiv W_{2} = \{W_{2}(z|x),~x \in \mathcal{X},~z \in \mathcal{Z} \}$
denote the single--letter input--output transition probability matrices,
associated with the strong user and the weak user, respectively.
When these channels are fed by an input vector $\bx \in \mathcal{X}^{n}$, 
they produce
the corresponding output vectors $\by \in \mathcal{Y}^{n}$ and $\bz \in \mathcal{Z}^{n}$, according to
\begin{align}
   W_{1}(\by|\bx)  &=  \prod_{t=1}^{n}  W_{1}(y_{t}|x_{t}), \\
   W_{2}(\bz|\bx)  &=  \prod_{t=1}^{n}  W_{2}(z_{t}|x_{t}).
\end{align}
We are interested in sending 
one out of $M_{y}M_{z}$ messages to the strong user, 
that observes $\by$, and 
one out of $M_{z}$ messages to the weak user, 
that observes $\bz$. 
The two messages are chosen with uniform probability.
Although our results prove the existence of a single sequence of {\it deterministic} hierarchical constant composition (HCC) codebooks, whose error probabilities are provably bounded, our proof techniques use extensively the following mechanism of random selection of an HCC code for the ABC.
Let $\mathcal{U}$ be a finite alphabet, 
let $P_{U}$ be a given probability distribution on $\calU$, 
and let $P_{X|U}$ be a given matrix of conditional probabilities of $X$ given $U$, 
such that the type--class $\calT(P_{U})$ and the conditional type--class $\calT(P_{X|U}|\bu)$ are non--empty.
We first select, independently at random, $M_{z} = \lceil e^{n R_{z}} \rceil$ $n$-vectors 
(``cloud centers''), 
$\bu_{0}, \bu_{1}, \dotsc, \bu_{M_{z}-1}$, 
all under the uniform distribution over the type--class $\calT(P_{U})$. 
Next, for each $m = 0,1, \dotsc, M_{z} - 1$, we select conditionally 
independently (given $\bu_{m}$), $M_{y} = \lceil e^{n R_{y}} \rceil$ codewords, 
$\bx_{m,0}, \bx_{m,1}, \dotsc, \bx_{m,(M_{y}-1)}$,
under the uniform distribution across the conditional type--class $\calT(P_{X|U}|\bu_{m})$. 
We denote the sub--code for each cloud by $\mathcal{C}_{m}(n) = \{ \bx_{m,0}, \bx_{m,1}, \dotsc, \bx_{m,(M_{y}-1)} \}$. 
Thus, the communication rate to the weak user is $R_{z}$, while the total communication rate to the strong user is $R_{z} + R_{y}$.
Once selected, the entire codebook $\calC(n)=\cup_{m=0}^{M_z-1}\calC_{m}(n)$, 
and the collection of cloud centers, $\{ \bu_{0},
\bu_{1}, \dotsc, \bu_{M_{z}-1} \}$, are
revealed to the encoder and to both decoders. 
We usually omit the dependence on $n$ from the notation of the code, 
and use $\calC$ and $\calC_{m}$, for short.

For any of the following described decoding rules, denote by $[\hat{m}(\by), \hat{i}(\by)]$ the decoded pair of the strong user, and by $\tilde{m}(\bz)$ the decoded cloud of the weak user. The ML decoder for the strong user is given by
\begin{equation}
\label{MLstrong}
[\hat{m}(\by), \hat{i}(\by)] = \operatorname*{arg\,max}_{0 \leq m \leq M_{z}-1, 0 \leq i \leq M_{y}-1} W_{1}(\by|\bx_{mi}) ,
\end{equation}
and the optimal ML decoder for the weak user (the bin index decoder) is given by
\begin{equation}
\label{MLweak}
\tilde{m}(\bz) = \operatorname*{arg\,max}_{0 \leq m \leq M_{z}-1} W_{2}(\bz|\mathcal{C}_{m})  ,
\end{equation}
where
\begin{equation}
W_{2}(\bz|\mathcal{C}_{m})  \overset{\Delta}{=} 
\frac {1}{M_{y}} \sum_{\bx \in \mathcal{C}_{m}} 
W_{2}(\bz|\bx) = \frac {1}{M_{y}} \sum_{i=0}^{M_{y}-1} W_{2}(\bz|\bx_{mi}).
\end{equation}
The likelihood decoder is a stochastic decoder, that chooses the decoded message according to the posterior probability mass function, induced by the channel output (either $\by$ or $\bz$). For the strong user, the ordinary likelihood decoder randomly selects the estimated message $(\hat{m},\hat{i})$ according to the following posterior distribution
\begin{align}
\mathrm{Pr} \left\{ \hat{m}=m, \hat{i}=i \middle| \by \right\}  =
\frac{W_{1}(\by|\bx_{mi}) } {\sum_{m'=0}^{M_{z}-1} \sum_{i'=0}^{M_{y}-1} W_{1}(\by|\bx_{m'i'}) } .
\end{align}
The generalized likelihood decoder (GLD) for the strong user is defined by
\begin{align}
\label{StrongGLD}
\mathrm{Pr} \left\{ \hat{m}=m, \hat{i}=i \middle| \by \right\}  =
\frac{\exp \{n g_{\mbox{\tiny s}}( \hat{P}_{\bu_{m} \bx_{mi} \by } ) \}} {\sum_{m'=0}^{M_{z}-1} \sum_{i'=0}^{M_{y}-1} 
\exp \{n g_{\mbox{\tiny s}}( \hat{P}_{\bu_{m'} \bx_{m'i'} \by } ) \} } ,
\end{align}
where $\hat{P}_{\bu_{m} \bx_{mi} \by }$ is the empirical distribution of $(\bu_{m}, \bx_{mi}, \by)$, and $g_{\mbox{\tiny s}}(\cdot)$ is a given continuous, real valued functional of this empirical distribution.
In the same manner, the ordinary likelihood decoder for the weak user randomly selects the estimated cloud $\tilde{m}$ according to
\begin{align}
\mathrm{Pr} \left\{ \tilde{m}=m \middle| \bz \right\}  =
\frac{\sum_{i=0}^{M_{y}-1} W_{2}(\bz|\bx_{mi}) }  {\sum_{m'=0}^{M_{z}-1} \sum_{i'=0}^{M_{y}-1} W_{2}(\bz|\bx_{m'i'}) } ,
\end{align}
while the GLD for the weak user is defined by
\begin{align}
\mathrm{Pr} \left\{ \tilde{m}=m \middle| \bz \right\}  =
\frac{\sum_{i=0}^{M_{y}-1} \exp \{n g_{\mbox{\tiny w}}( \hat{P}_{\bu_{m} \bx_{mi} \bz } ) \}} {\sum_{m'=0}^{M_{z}-1} \sum_{i'=0}^{M_{y}-1} \exp \{n g_{\mbox{\tiny w}}( \hat{P}_{\bu_{m'} \bx_{m'i'} \bz } ) \} }.
\end{align}
Exactly as the universal decoders derived in \cite{AM17}, generalized decoders may also depend on the cloud--centers,  
which may be helpful, since all of the codewords in each sub--code are highly correlated via their cloud--center.
One of the most important properties of the GLD is the following. The union bound, which is used in the first steps of the derivations for both users, actually provides an exact expression for the probability of error, unlike in the analyses of the ML decoders, where the union bound harms the exponential tightness, at least for relatively high rates. 

The generalized likelihood decoders cover several important special cases for the strong user. The choice
\begin{align}
g_{\mbox{\tiny s}}( \hat{P}_{\bu_{m} \bx_{mi} \by } ) = \sum_{u,x,y} \hat{P}_{\bu_{m} \bx_{mi} \by }(u,x,y) \log W_{1}(y|x) = \sum_{x,y} \hat{P}_{\bx_{mi} \by }(x,y) \log W_{1}(y|x)
\end{align}
corresponds to the ordinary likelihood decoder. More generally, one may introduce an ``inverse temperature" parameter $\beta \geq 0$ and define
\begin{align}
g_{\mbox{\tiny s}}( \hat{P}_{\bu_{m} \bx_{mi} \by } ) = \beta \sum_{x,y} \hat{P}_{\bx_{mi} \by }(x,y) \log W_{1}(y|x) .
\end{align}
Here, $\beta$ controls the degree of skewedness of the distribution (\ref{StrongGLD}): while $\beta=1$ corresponds to the ordinary likelihood decoder, $\beta \to \infty$ leads to the deterministic ML decoder. In the same manner,
\begin{align}
g_{\mbox{\tiny s}}( \hat{P}_{\bu_{m} \bx_{mi} \by } ) = \beta \sum_{x,y} \hat{P}_{\bx_{mi} \by }(x,y) \log W'(y|x) ,
\end{align}
for $W'$ being different from $W_{1}$, defines a family of mismatched likelihood decoders.
Yet another interesting choice is
\begin{align}
g_{\mbox{\tiny s}}( \hat{P}_{\bu_{m} \bx_{mi} \by } ) = \beta \hat{I}_{\bu_{m} \bx_{mi} \by } (UX;Y),
\end{align}
which is a parametric family of stochastic mutual information decoders, where the limit of $\beta \to \infty$ yields the ordinary maximum mutual information (MMI) universal decoder \cite{CK11}. 
Similarly for the weak user, the choice
\begin{align}
g_{\mbox{\tiny w}}( \hat{P}_{\bu_{m} \bx_{mi} \bz } ) 
 = \sum_{x,z} \hat{P}_{\bx_{mi} \bz }(x,z) \log W_{2}(z|x)
\end{align}
corresponds to the ordinary likelihood decoder for the weak user.
From now on, we will assume that $g_{\mbox{\tiny s}}(\cdot)$ and $g_{\mbox{\tiny w}}(\cdot)$ are the same functional, and denote both of them by $g(\cdot)$.

Let $\bY \in \mathcal{Y}^{n}$ and $\bZ \in \mathcal{Z}^{n}$ be the random channel outputs resulting from the transmission of $\bx_{mi}$. For a given code $\calC(n)$, define the error probabilities as
\begin{align}
P_{\mbox{\tiny e}|mi} (\calC(n)) &=  \mathrm{Pr} 
\left\{ [\hat{m}(\bY), \hat{i}(\bY)] \neq (m,i) \middle| \bx_{mi}~\mbox{sent} \right\} ,
\end{align}
and
\begin{align}
P_{\mbox{\tiny e}|m} (\calC(n)) &=  
\frac {1}{M_{y}} \sum_{i=0}^{M_{y}-1} \mathrm{Pr} \left\{ \tilde{m}(\bZ) \neq m  \middle| \bx_{mi}~\mbox{sent}  \right\} ,
\end{align}
where in both definitions, $\mathrm{Pr} \{ \cdot \}$ 
designates the probability measure associated with the 
randomness of the channel outputs given its input, and the (possibly) stochastic decoder.
Moreover, the error probabilities are defined to be zero whenever the blocklength is such that no code can be generated.
Our main objective is to prove the existence of sequences of HCC codes and obtain the tightest possible single--letter expressions that lower bound the following limits
\begin{align}
E_{\mbox{\tiny su}}(R_{y}, R_{z}) = \liminf_{n \to \infty} \left[  - \frac{\log \max_{m,i} P_{\mbox{\tiny e}|mi}(\calC(n))}{n} \right] , 
\end{align}
and
\begin{align}
E_{\mbox{\tiny wu}}(R_{y}, R_{z}) = \liminf_{n \to \infty} \left[  - \frac{\log \max_{m} P_{\mbox{\tiny e}|m}(\calC(n))}{n}  \right], 
\end{align}
both for the ML decoder and the GLD.

In a recent paper \cite{AM17}, exact random coding error exponents have been derived for both users of the ABC. We may expect to improve these error exponents, at least when one of the coding rates is low, by code expurgation. 
In this paper, we derive expurgated exponents for the ABC under ML decoding in two different methods. In addition, we discuss the GLD, that enables us to achieve the best between the random coding bound and one of the ML--based expurgated bounds.

\section {Main Results}
For maximum likelihood decoding, we distinguish between two different methods of expurgation for the HCC ensemble. The first methodology is based on the following technique of expurgation: we randomly draw a HCC codebook, and then simultaneously expurgate both bad clouds and bad codewords within the remaining clouds. The resulting expurgated bounds are given in Theorem 1, which is proved in Section 5. A full discussion of the first two theorems follows the second theorem. 
In order to state our first theorem, we start with the following definitions. We define the following sets of distributions
\begin{align}
\calS &\overset{\Delta}{=} \big\{Q_{UXX'}:~ Q_{UX'} = Q_{UX} = P_{UX}  \big\} , \\
\calP &\overset{\Delta}{=} \big\{Q_{UU'XX'}:~ Q_{U'X'} = Q_{UX} = P_{UX}  \big\} . 
\end{align}
For a general channel $W$, we define the averaged Chernoff distance function by
\begin{align}
\label{Ddefinition}
D_{s}(Q_{XX'}) \overset{\Delta}{=} 
 - \sum_{(x,x') \in \calX^{2}} Q_{XX'}(x,x') \log \left[ \sum_{y \in \calY}  W^{1-s}(y| x)  \cdot  W^{s}(y| x')  \right] ,
\end{align}
where in the following, we choose $W=W_{1}$ or $W=W_{2}$, depending on the user we are relating to.
For the weak user, define an error exponent function as
\begin{align}
\label{expur_BC_weak1}
E_{\mbox{\tiny wu}}^{\mbox{\tiny ML1}}  (R_{y},R_{z})
&\overset{\Delta}{=} \max_{0 \leq t \leq 1}
\min_{\substack{Q_{UU'XX'} \in \calP \\ I_{Q}(UX;U'X') \leq  2R_{y} + R_{z} }}
\big[ I_{Q}(UX;U'X') + D_{t}(Q_{XX'}) \big] - R_{y}- R_{z} .
\end{align}
Next, for the strong user we define the following error exponent functions
\begin{align}
\label{expur_BC_strong1}
E_{\mbox{\tiny su-1}}^{\mbox{\tiny ML1}}  (R_{y},s)
&\overset{\Delta}{=} \min_{\substack{Q_{UXX'} \in \calS \\ I_{Q}(X;X'|U) \leq  R_{y} }}
\big[ I_{Q}(X;X'|U) + D_{s}(Q_{XX'}) \big] - R_{y}, \\
\label{expur_BC_strong111}
E_{\mbox{\tiny su-2}}^{\mbox{\tiny ML1}}  (R_{y},R_{z},s)
&\overset{\Delta}{=} \min_{\substack{Q_{UU'XX'} \in \calP \\ I_{Q}(UX;U'X') \leq  R_{y} + R_{z} }}
\big[  I_{Q}(UX;U'X') + D_{s}(Q_{XX'}) \big] - R_{y}- R_{z}, \\
\label{expur_BC_strong2}
E_{\mbox{\tiny su}}^{\mbox{\tiny ML1}}  (R_{y},R_{z})
&\overset{\Delta}{=} \max_{0 \leq s \leq 1} \min
\left\{ E_{\mbox{\tiny su-1}}^{\mbox{\tiny ML1}}  (R_{y},s), 
        E_{\mbox{\tiny su-2}}^{\mbox{\tiny ML1}}  (R_{y},R_{z},s) \right\} .
\end{align}

\textbf{Theorem 1.} There exists a sequence of HCC codes, $\{\calC(n),~ n=1,2, \dotsc\}$, with a rate pair $(R_{y},R_{z})$ for which both
\begin{align}
\label{THM1STRONG}
\liminf_{n \to \infty} \bigg[  - \frac{\log \max_{m,i} P_{\mbox{\tiny e}|mi}(\calC(n))}{n}    \bigg]  
&\geq E_{\mbox{\tiny su}}^{\mbox{\tiny ML1}}  (R_{y},R_{z}), 
\end{align}
and
\begin{align}
\label{THM1WEAK}
\liminf_{n \to \infty} \bigg[  - \frac{\log \max_{m} P_{\mbox{\tiny e}|m}(\calC(n))}{n}    \bigg]  
&\geq E_{\mbox{\tiny wu}}^{\mbox{\tiny ML1}}  (R_{y},R_{z}).
\end{align}

The second method is somewhat different, and the idea behind it is the following. At the first step, we expurgate sub--codes, merely according to their cloud--centers. 
Then, at the second step, we fix the set of cloud--centers of the remaining clouds from the first step, and then expurgate specific codewords, as well as clouds, according to some collective behavior of their codewords.  
The resulting expurgated bounds are given in Theorem 2, and as can be seen below, the expressions are more complicated than those of Theorem 1, at least for the weak user. The proof can be found in Section 6. 
In order to state our second theorem, we need a few definitions. 

For a given marginal $Q_{UZ}$, let $\mathcal{S}(Q_{UZ})$ denote the set of conditional distributions $\{Q_{X|UZ}\}$ such that $\sum_{z} Q_{UZ}(u,z)Q_{X|UZ}(x|u,z) = P_{UX}(u,x)$ for every $(u,x) \in \mathcal{U} \times \mathcal{X}$, where $P_{UX} = P_{U} \times P_{X|U}$. We denote $\bar{t} = 1-t$.
For the weak user, define 
\begin{align}
\label{DhatDEF}
&\hat{D}_{t}(R_{y}, Q_{UU'}) \overset{\Delta}{=} \nonumber \\
&\min_{Q_{Z|UU'}} \min_{Q_{X|UZ} \in \calS(Q_{UZ})} \min_{Q_{X'|U'Z} \in \calS(Q_{U'Z})} 
\Big\{ \bar{t} \cdot D(Q_{Z|UX} \| W_{Z|X} | Q_{UX}) + t \cdot D(Q_{Z|U'X'} \| W_{Z|X'} | Q_{U'X'}) \nonumber \\
&+\bar{t} \cdot I_{Q}(Z;U'|U) + t \cdot I_{Q}(Z;U|U') + t \cdot [I_{Q}(X;Z|U) - R_{y}]_{+}  + \bar{t} \cdot [I_{Q}(X';Z|U') - R_{y}]_{+}   \Big\}.
\end{align}
We define the following set of distributions
\begin{align}
\calQ  &\overset{\Delta}{=}  \big\{Q_{UU'}:~ Q_{U} = Q_{U'} = P_{U}  \big\} ,
\end{align}
and an error exponent function
\begin{align}
\label{expur_BC_weak21}
E_{\mbox{\tiny wu}}^{\mbox{\tiny ML2}}  (R_{y},R_{z})
&\overset{\Delta}{=} \max_{0 \leq t \leq 1}  \min_{\substack{Q_{UU'} \in \calQ \\ I_{Q}(U;U') \leq  R_{z} }}
 \big[ I_{Q}(U;U') + \hat{D}_{t}(R_{y}, Q_{UU'})  \big] - R_{z} .
\end{align}
Next, for the strong user we define the following error exponent functions
\begin{align}
\label{expur_BC_strong21}
E_{\mbox{\tiny su-1}}^{\mbox{\tiny ML2}}  (R_{y},s)
&\overset{\Delta}{=} \min_{\substack{Q_{UXX'} \in \calS \\ I_{Q}(X;X'|U) \leq  R_{y} }}
\big[ I_{Q}(X;X'|U) + D_{s}(Q_{XX'})  \big] - R_{y}, \\
\label{expur_BC_strong222}
E_{\mbox{\tiny su-2}}^{\mbox{\tiny ML2}}  (R_{y},R_{z},s)
&\overset{\Delta}{=} \min_{\substack{Q_{UU'XX'} \in \calP \\ I_{Q}(U;U') \leq  R_{z} }}
\big[  I_{Q}(UX;U'X') + D_{s}(Q_{XX'}) \big] - R_{y} - R_{z}, \\
\label{expur_BC_strong22}
E_{\mbox{\tiny su}}^{\mbox{\tiny ML2}}  (R_{y},R_{z})
&\overset{\Delta}{=} \max_{0 \leq s \leq 1} \min
\left\{ E_{\mbox{\tiny su-1}}^{\mbox{\tiny ML2}}  (R_{y},s), 
        E_{\mbox{\tiny su-2}}^{\mbox{\tiny ML2}}  (R_{y},R_{z},s) \right\} .
\end{align}
\textbf{Theorem 2.} There exists a sequence of HCC codes, $\{\calC(n),~ n=1,2, \dotsc\}$, with a rate pair $(R_{y},R_{z})$ for which both
\begin{align}
\label{THM2STRONG}
\liminf_{n \to \infty} \bigg[  - \frac{\log \max_{m,i} P_{\mbox{\tiny e}|mi}(\calC(n))}{n}    \bigg]  
&\geq E_{\mbox{\tiny su}}^{\mbox{\tiny ML2}}  (R_{y},R_{z}), 
\end{align}
and
\begin{align}
\label{THM2WEAK}
\liminf_{n \to \infty} \bigg[  - \frac{\log \max_{m} P_{\mbox{\tiny e}|m}(\calC(n))}{n}    \bigg]  
&\geq E_{\mbox{\tiny wu}}^{\mbox{\tiny ML2}}  (R_{y},R_{z}).
\end{align}
\textbf{Discussion} 

First, all of the expressions in Theorems 1 and 2 
generalize
the well--known CKM expurgated bound \cite{CKM77}. 
For example, it can be easily recovered from the expression $E_{\mbox{\tiny wu}}^{\mbox{\tiny ML1}}  (R_{y},R_{z})$ of Theorem 1, when degenerating the hierarchical codebook by choosing $R_{y}=0$, as well as $P_{X|U}(x|u) = \delta(x-u)$ (assuming that $\calX=\calU$), in order to get back to the CKM expurgated bound. 

Concerning the strong user, each bound is given by the minimum between two different expressions. The first expression is related to error events within the cloud of the true codeword.
In fact, we have that $E_{\mbox{\tiny su-1}}^{\mbox{\tiny ML1}}  (R_{y},s) =E_{\mbox{\tiny su-1}}^{\mbox{\tiny ML2}}  (R_{y},s)$, where the difference is given by the second components, $E_{\mbox{\tiny su-2}}^{\mbox{\tiny ML1}}(R_{y}, R_{z}, s)$ and $E_{\mbox{\tiny su-2}}^{\mbox{\tiny ML2}}(R_{y}, R_{z}, s)$, for which the method of expurgation is relevant and cause a change in the final expressions. Although the objectives in (\ref{expur_BC_strong111}) and (\ref{expur_BC_strong222}) are exactly the same, the constraints are different, and are not subsets of one another.  

Concerning the weak user, the situation is much more complicated, because of the structure of the optimal decoder. The derivation in the proof of Theorem 1 contains a passage [(\ref{PDI11}) to (\ref{PDI12})] that may harm the exponential tightness of the result. Specifically, we use the PD inequality over the 
sums that stems from the definition of the optimal decoder, i.e.,
\begin{align}
& \left\{ \frac{1}{M_{y}} \sum_{i=0}^{M_{y}-1} W_{2}(\bz| \bx_{mi}) \right\}^{1-t}  \left\{\frac{1}{M_{y}}  \sum_{j=0}^{M_{y}-1} W_{2}(\bz| \bx_{m'j})  \right\}^{t}  \\
&\leq   \frac{1}{M_{y}} \sum_{i=0}^{M_{y}-1} \sum_{j=0}^{M_{y}-1}   W_{2}^{1-t}(\bz| \bx_{mi})  \cdot  W_{2}^{t}(\bz| \bx_{m'j}) ,
\end{align}
and therefore, the resulting bound of Theorem 1 is, in fact, a natural generalization of the classical single--user expurgated bound, which only depends on the Chernoff distance between pairs of codewords. Because of this passage, the bound of Theorem 1 is inferior to the bound of Theorem 2, at relatively high values of $R_{y}$. However, the resulting exponent of Theorem 1 still outperforms the result of Theorem 2, at least for relatively low $R_{y}$ values (see Fig.\ 1). 
The reason for the bound of Theorem 2 to be inferior at relatively low rates is because of the remaining parts of the two proofs.
While both proofs use Markov's inequality in order to show the existence of good codebooks, some logical arguments that can only be claimed in the proof of Theorem 1 provide tighter upper bounds. 
More specifically, the proof of Theorem 1 relies merely on type--class enumerators, which takes only integer values. It is shown that there exists codebooks, for which those enumerators must equal to zero in some range of relatively low rates. 
On the other hand, the proof of Theorem 2 relies on some more complicated quantities, that are not necessarily integer--valued, and hence, they cannot be assured to be equal to zero at any range of rates.     
Since the derivation in the proof of Theorem 2 is exponentially tight after the first two steps, and does not compromise on the optimal decoders (as Theorem 1 does in the passage we mentioned above), it provides a better result at relatively high $R_{y}$. Specifically, the expression given in Theorem 2 reaches a plateau at high $R_{y}$, while the expression of Theorem 1 reaches zero. 
One should note that the improvement at high rates is obtained by expressions which are more complicated to compute.  

\begin{figure}[ht!]
\centering
\includegraphics[width=130mm]{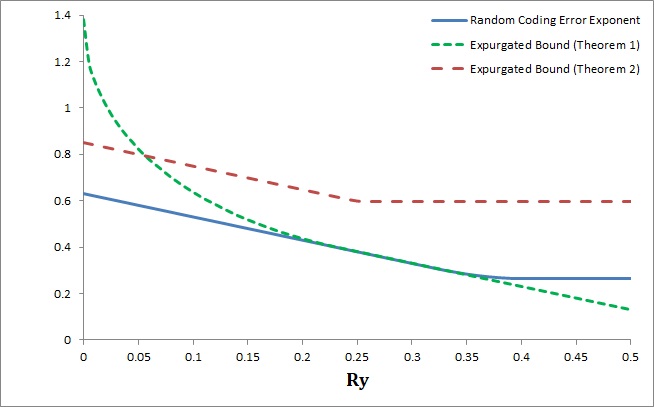}
\caption{Expurgated bounds for the weak user ($R_{z}=0$). \label{overflow}}
\end{figure} 
We next provide some numerical results, comparing our expurgated bounds for the weak user (Fig.\ 1) and for the strong user (Fig.\ 2), as given by Theorems 1 and 2. 
Let $W_1$ and $W_2$ be two binary symmetric channels (BSC´s) with
crossover parameters $p_y$ and $p_z$, respectively ($p_z > p_y$).
Let $\mathcal{U}$ be binary as well and let $P_{U}$ be 
uniformly distributed over $\{0,1\}$. 
Also, let $P_{X|U}$ be a BSC with crossover parameter $p_{x|u} =0.15$. Let us choose the channel probabilities to be $p_{z} = 0.001$ and $p_{y} = 0.0005$. 

In Fig.\ 2, the orange (dot-dashed) curve describes the expression of $E_{\mbox{\tiny su-1}}^{\mbox{\tiny ML1}}  (R_{y},s)$, which is common in both exponents. As can be seen, each of the two exponents is dominated by this expression at relatively high sattelite rates. Note that at least for this specific example, $E_{\mbox{\tiny su}}^{\mbox{\tiny ML1}}(R_{y}, R_{z})$ is higher than $E_{\mbox{\tiny su}}^{\mbox{\tiny ML2}}(R_{y}, R_{z})$ at any pair of coding rates. 
\begin{figure}[ht!]
\centering
\includegraphics[width=130mm]{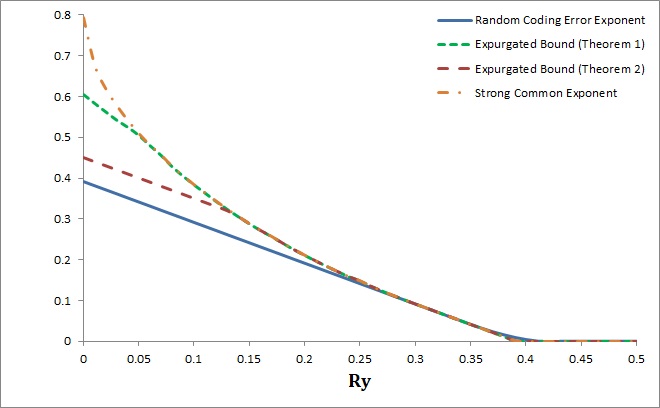}
\caption{Expurgated bounds for the strong user ($R_{z}=0.2$). \label{overflow}}
\end{figure}


We now move on to the GLD. As was already mentioned earlier, the GLD enables us 
to make a tighter derivation for the probability of error, 
and therefore, the resulting expurgated bounds are strictly tighter, at least at relatively high rates. The drawback of the expressions of Theorem 3 is that they are quite cumbersome, at least when compared to those of Theorems 1 or 2.
In order to characterize the expurgated bounds of the GLD, we define first a few quantities. Let
\begin{align}
\label{phiDEF}
\phi(R_{y}, Q_{UY})
 = \max_{\{Q_{X|UY}:~ I_{Q}(X;Y|U) \leq R_{y} \}} [g(Q)  - I_{Q}(X;Y|U)] + R_{y},
\end{align}
and
\begin{align}
\label{psiDEF}
\psi(R_{y}, R_{z}, Q_{Y})
 = \max_{\{Q_{UX|Y}:~ I_{Q}(U;Y) \leq R_{z} , ~ I_{Q}(UX;Y)  \leq R_{z} + R_{y} \}} [ g(Q)  - I_{Q}(UX;Y) ] + R_{z} + R_{y} .
\end{align}
Also, define
\begin{align}
&\Upsilon (Q_{UXX'}, R_{y}, R_{z})
= \min_{Q_{Y|UXX'}} \bigg( D(Q_{Y|UX} \| W_{Y|X} | Q_{UX}) 
   + I_{Q}(X';Y|UX)   
\nonumber \\ &~~~~~~~~~~~~~~~~~~~~~~~~~~~ + \Big[ \max \big\{ g( Q_{UXY} ), \phi(R_{y}, Q_{UY}), \psi(R_{y}, R_{z}, Q_{Y})  \big\} - g( Q_{UX'Y} ) \Big]_{+}  \bigg),   
\end{align}
and
\begin{align}
\label{OMEGAdefinition}
&\Omega (Q_{UU'XX'}, R_{y}, R_{z})
= \min_{Q_{Y|UU'XX'}} \bigg( D(Q_{Y|UX} \| W_{Y|X} | Q_{UX}) + I_{Q}(U'X';Y|UX) 
\nonumber \\ &~~~~~~~~~~~~~~~~~~~~~~~~~~~ + \Big[ \max \big\{ g( Q_{UXY} ), \phi(R_{y}, Q_{UY}), \psi(R_{y}, R_{z}, Q_{Y})  \big\} - g( Q_{U'X'Y} ) \Big]_{+}  \bigg).   
\end{align}
We define the following error exponent functions. For the weak user,
\begin{align}
\label{THM3WEAK}
E_{\mbox{\tiny wu}}^{\mbox{\tiny GLD}}  (R_{y},R_{z}) 
&\overset{\Delta}{=} \min_{\substack{Q_{UU'XX'} \in \calP \\ I_{Q}(UX;U'X')  < 2R_{y} + R_{z}, \\ I_{Q}(U;U')<R_{z} }} [I_{Q}(UX;U'X') + \Omega (Q, R_{y}, R_{z}) ] - R_{y} - R_{z} ,
\end{align}
and for the strong user 
\begin{align}
\label{THM3STRONG1}
E_{\mbox{\tiny su-1}}^{\mbox{\tiny GLD}}  (R_{y},R_{z})
&\overset{\Delta}{=} \min_{\substack{Q_{UXX'} \in \calS \\ I_{Q}(X;X'|U) < R_{y}}}   
 [ I_{Q}(X;X'|U) + \Upsilon (Q, R_{y}, R_{z}) ] - R_{y}  , \\
E_{\mbox{\tiny su-2}}^{\mbox{\tiny GLD}}  (R_{y},R_{z})
&\overset{\Delta}{=} \min_{\substack{Q_{UU'XX'} \in \calP \\ I_{Q}(UX;U'X') < R_{y}+R_{z}, \\ I_{Q}(UX;U') < R_{z}}}   [ I_{Q}(UX;U'X') + \Omega (Q, R_{y}, R_{z}) ]  - R_{y} - R_{z} , \\
\label{THM3STRONG2}
E_{\mbox{\tiny su}}^{\mbox{\tiny GLD}}  (R_{y},R_{z})
&\overset{\Delta}{=}  \min
\left\{ E_{\mbox{\tiny su-1}}^{\mbox{\tiny GLD}}  (R_{y},R_{z}), 
        E_{\mbox{\tiny su-2}}^{\mbox{\tiny GLD}}  (R_{y},R_{z}) \right\} .
\end{align}
\textbf{Theorem 3.} There exists a sequence of HCC codes, $\{\calC(n),~ n=1,2, \dotsc\}$, with a rate pair $(R_{y},R_{z})$ for which both
\begin{align}
\liminf_{n \to \infty} \bigg[  - \frac{\log \max_{m,i} P_{\mbox{\tiny e}|mi}(\calC(n))}{n}    \bigg]  
&\geq E_{\mbox{\tiny su}}^{\mbox{\tiny GLD}}  (R_{y},R_{z}), 
\end{align}
and
\begin{align}
\liminf_{n \to \infty} \bigg[  - \frac{\log \max_{m} P_{\mbox{\tiny e}|m}(\calC(n))}{n}    \bigg]  
&\geq E_{\mbox{\tiny wu}}^{\mbox{\tiny GLD}}  (R_{y},R_{z}) .
\end{align}
\textbf{Discussion} 

$\bullet$ An expurgated bound for the GLD in the single user regime has been derived by Merhav \cite{MERHAV2017}. It should be noticed that the resulting expressions of Theorem 3, as well as some parts of its proof (in Section 7) are direct generalizations of the single--user case. In those cases, we will omit some parts of the proof that highly resemble those in \cite{MERHAV2017}.

$\bullet$ The expression of eq.\ (\ref{THM3WEAK}) has the same structure as the bound given in Theorem 1, except that here the functional $\Omega (Q, R_{y}, R_{z})$ replaces the expected Chernoff distance, and an additional constraint ($I_{Q}(U;U')<R_{z}$) has been added. We prove in Appendix A that at least for the choice $g(Q) = \mathbb{E}_{Q} \log W_{2}(Z|X)$, the expurgated bound of Theorem 3, $E_{\mbox{\tiny wu}}^{\mbox{\tiny GLD}}(R_{y},R_{z})$, is at least as tight as the bound of Theorem 1, $E_{\mbox{\tiny wu}}^{\mbox{\tiny ML1}}(R_{y},R_{z})$.  

$\bullet$ One of the main advantages of the GLD, is the fact that the derivation of its probability of error may be exponentially tighter than the derivations in the proofs of Theorems 1 or 2, 
because its circumvents the use of the union bound in the proofs of Theorems 1 and 2 [eqs.\ (\ref{INITinequality1})--(\ref{INITinequality2})].
As a consequence, we show in Appendix B that $E_{\mbox{\tiny wu}}^{\mbox{\tiny GLD}}(R_{y},R_{z})$ cannot be smaller than the random coding error exponent of the weak user at any pair of rates, by examining the former for the suboptimal GLD based on the universal metric $g(Q) = I_{Q}(UX;Z)$. 
We conclude that $E_{\mbox{\tiny wu}}^{\mbox{\tiny GLD}}(R_{y},R_{z})$ is at least as tight as the maximum between $E_{\mbox{\tiny wu}}^{\mbox{\tiny ML1}}(R_{y},R_{z})$ and the random coding error exponent, $E_{\mbox{\tiny w}}(R_{y},R_{z})$. 

$\bullet$ The same can be proved for the strong user, i.e., that $E_{\mbox{\tiny su}}^{\mbox{\tiny GLD}}(R_{y},R_{z})$ is at least as tight as the maximum between $E_{\mbox{\tiny su}}^{\mbox{\tiny ML1}}(R_{y},R_{z})$ and the random coding error exponent, $E_{\mbox{\tiny s}}(R_{y},R_{z})$. We conclude, that there exist a HCC codebook, for which one user works in the ``expurgated region" (slope greater than $1$), while the other user works in the ``random coding region" (slope smaller than $1$). For example, it may be the case when the channel to the strong user is quite clean, while the channel to the weak user is very noisy, compared to the required rates. 

$\bullet$ Keeping in mind the discussion after Theorem 2, it is now clear that the expurgated bound for the weak user of Theorem 3, $E_{\mbox{\tiny wu}}^{\mbox{\tiny GLD}}(R_{y},R_{z})$, is strictly tighter than the bound $E_{\mbox{\tiny wu}}^{\mbox{\tiny ML2}}(R_{y},R_{z})$ of Theorem 2, at relatively low rates, and is strictly tighter than the bound $E_{\mbox{\tiny wu}}^{\mbox{\tiny ML1}}(R_{y},R_{z})$ of Theorem 1, at relatively high rates. However, we were not able to determine whether $E_{\mbox{\tiny wu}}^{\mbox{\tiny GLD}}(R_{y},R_{z})$ is at least as tight as $E_{\mbox{\tiny wu}}^{\mbox{\tiny ML2}}(R_{y},R_{z})$ of Theorem 2, at relatively high rates. In other words, it is not clear whether the bound of Theorem 3 is at least as tight as the maximum between the bounds of the first two theorems, although we conjecture that it is indeed the case when choosing one of the decoding metrics $g(Q) = \beta \mathbb{E}_{Q} \log W_{2}(Z|X)$ or $g(Q) = \beta I_{Q}(UX;Z)$, and letting $\beta \to \infty$.


\section{Proof of Theorem 1}
The proof has three main parts: In part 1, we upper bound the error probabilities and express each one of them using suitable type-class enumerators. In part 2, which is the main step of the proof, we show the existence of an hierarchical codebook for which these enumerators are upper bounded by specific deterministic functions of information measures. In part 3, we put back these deterministic bounds in order to get the desired results.

\subsection{Part 1} 
We define the Chernoff distance between the codewords $\bx$ and $\bx'$ by 
($W=W_{1}$ or $W=W_{2}$)
\begin{align}
d_{s}(\bx, \bx') \overset{\Delta}{=}  -\log \left[ \sum_{\by}  W^{1-s}(\by| \bx)  \cdot  W^{s}(\by| \bx')  \right].
\end{align}
For the strong user, we have the following upper bound for any $s \in [0,1]$
\begin{align}
   P_{\mbox{\tiny e}|mi} (\calC)  
\label{INITinequality1} 
&= \sum_{\by} W_{1}(\by| \bx_{mi}) \cdot \calI \left\{ \bigcup_{(m',j) \neq (m,i)} \Big\{  W_{1}(\by| \bx_{m'j}) \geq  W_{1}(\by| \bx_{mi})  \Big\} \right\}  \\
\label{INITinequality2}
&\overset{\mbox{\tiny UB}}{\leq}  \sum_{\by} W_{1}(\by| \bx_{mi}) \cdot \left[ \sum_{(m',j) \neq (m,i)} \frac{W_{1}(\by| \bx_{m'j}) }{W_{1}(\by| \bx_{mi})} \right] ^{s} \\
&\overset{\mbox{\tiny PD}}{\leq}  \sum_{\by} W_{1}(\by| \bx_{mi}) \cdot \frac{\sum_{(m',j) \neq (m,i)} W_{1}^{s}(\by| \bx_{m'j}) }{W_{1}^{s}(\by| \bx_{mi})} \\
&= \sum_{(m',j) \neq (m,i)} \sum_{\by}   W_{1}^{1-s}(\by| \bx_{mi})  \cdot  W_{1}^{s}(\by| \bx_{m'j}) \\
&= \sum_{(m',j) \neq (m,i)} \exp \left\{- d_{s}(\bx_{mi}, \bx_{m'j})\right\}   \\
\label{TheSameUpperBoundStrong}
&= \sum_{j \neq i} \exp \left\{- d_{s}(\bx_{mi}, \bx_{mj})\right\} 
+  \sum_{m' \neq m} \sum_{j=0}^{M_{y}-1}  \exp \left\{- d_{s}(\bx_{mi}, \bx_{m'j})\right\}  \\
&=  \sum_{Q_{UXX'} \in \calS} N_{mi}^{\mbox{\tiny IN}}(Q_{UXX'}, \calC) \cdot  e^{- nD_{s}(Q_{XX'})} \nonumber \\
&~~~~~+ \sum_{Q_{UU'XX'} \in \calP} N_{mi}^{\mbox{\tiny OUT}}(Q_{UU'XX'}, \calC) \cdot  e^{- nD_{s}(Q_{XX'})},
\end{align}
where $N_{mi}^{\mbox{\tiny IN}}(Q_{UXX'}, \calC)$ denotes the number of codewords $\bx_{mj} \in \mathcal{C}_{m}$, for any $j \neq i$, such that the joint empirical distribution of $\bx_{mj}$ with $(\bu_{m}, \bx_{mi})$ is $Q_{UXX'}$, that is
\begin{align}
\label{ENUMstrong1}
N_{mi}^{\mbox{\tiny IN}}(Q_{UXX'}, \calC) = \sum_{j \neq i}   
  \mathcal{I} \Big\{ (\bu_{m}, \bx_{mi}, \bx_{mj} ) \in \mathcal{T}(Q_{UXX'}) \Big\},
\end{align}
and $N_{mi}^{\mbox{\tiny OUT}}(Q_{UU'XX'}, \calC)$ denotes the number of pairs $(\bu_{m'}, \bx_{m'j})$, $\bx_{m'j} \in \mathcal{C}_{m'}$, for any $m' \neq m$, such that the joint empirical distribution of them with $(\bu_{m}, \bx_{mi})$ is $Q_{UU'XX'}$, that is
\begin{align}
\label{ENUMstrong2}
N_{mi}^{\mbox{\tiny OUT}}(Q_{UU'XX'}, \calC) = \sum_{m' \neq m}   
   \sum_{j=0}^{M_{y}-1} 
 \mathcal{I} \Big\{ (\bu_{m}, \bx_{mi}, \bu_{m'}, \bx_{m'j} ) \in \mathcal{T}(Q_{UU'XX'}) \Big\}.
\end{align}
For the weak user, we have for any $t \in [0,1]$ 
\begin{align}  
   P_{\mbox{\tiny e}|m} (\calC)  &= \frac {1}{M_{y}} \sum_{i=0}^{M_{y}-1} \sum_{\bz} W_{2}(\bz| \bx_{mi}) \cdot \calI \left\{ \bigcup_{m' \neq m} \Big\{  W_{2}(\bz| \mathcal{C}_{m'}) \geq  W_{2}(\bz| \mathcal{C}_{m})  \Big\} \right\}  \\
&\overset{\mbox{\tiny UB}}{\leq} \frac {1}{M_{y}} \sum_{i=0}^{M_{y}-1} \sum_{\bz} W_{2}(\bz| \bx_{mi}) \cdot \left[ \sum_{m' \neq m} \frac{ W_{2}(\bz| \mathcal{C}_{m'}) }{W_{2}(\bz| \mathcal{C}_{m})} \right]^{t} \\
&\overset{\mbox{\tiny PD}}{\leq} \frac {1}{M_{y}} \sum_{i=0}^{M_{y}-1} \sum_{\bz} W_{2}(\bz| \bx_{mi}) \cdot \frac{\sum_{m' \neq m} W_{2}^{t}(\bz| \mathcal{C}_{m'}) }{W_{2}^{t}(\bz| \mathcal{C}_{m})} \\
&= \sum_{m' \neq m} \sum_{\bz}   W_{2}^{1-t}(\bz| \mathcal{C}_{m})  \cdot  W_{2}^{t}(\bz| \mathcal{C}_{m'}) \\  
\label{PDI11}
&=   \sum_{m' \neq m} \sum_{\bz}  
\left[ \left\{ \frac{1}{M_{y}} \sum_{i=0}^{M_{y}-1} W_{2}(\bz| \bx_{mi}) \right\}^{1-t}  \left\{\frac{1}{M_{y}}  \sum_{j=0}^{M_{y}-1} W_{2}(\bz| \bx_{m'j})  \right\}^{t} \right]  \\
\label{PDI12}
&\overset{\mbox{\tiny PD}}{\leq}   \sum_{m' \neq m} \sum_{\bz}  
 \frac{1}{M_{y}} \sum_{i=0}^{M_{y}-1} \sum_{j=0}^{M_{y}-1}   W_{2}^{1-t}(\bz| \bx_{mi})  \cdot  W_{2}^{t}(\bz| \bx_{m'j})   \\
\label{StartingPointML}
&=  \frac{1}{M_{y}}  \sum_{m' \neq m}   
  \sum_{i=0}^{M_{y}-1} \sum_{j=0}^{M_{y}-1} \sum_{\bz}  W_{2}^{1-t}(\bz| \bx_{mi})  \cdot  W_{2}^{t}(\bz| \bx_{m'j})   \\
&=  \frac{1}{M_{y}}  \sum_{m' \neq m}   
  \sum_{i=0}^{M_{y}-1} \sum_{j=0}^{M_{y}-1} \exp \left\{- d_{t}(\bx_{mi}, \bx_{m'j})  \right\}   \\
&=  \frac{1}{M_{y}}  \sum_{Q_{UU'XX'} \in \calP} \hat{N}_{m}(Q_{UU'XX'}, \calC)   e^ {  - nD_{t}(Q_{XX'})  },
\end{align}
where $\hat{N}_{m}(Q_{UU'XX'}, \calC)$ denotes the number of triplets $(\bu_{m'}, \bx_{mi},  \bx_{m'j})$, $\bx_{mi} \in \mathcal{C}_{m}$, $\bx_{m'j} \in \mathcal{C}_{m'}$, for any $m' \neq m$, such that the joint empirical distribution of them with $\bu_{m}$ is $Q_{UU'XX'}$, that is
\begin{align}
\hat{N}_{m}(Q_{UU'XX'}, \calC) = \sum_{m' \neq m}   
  \sum_{i=0}^{M_{y}-1} \sum_{j=0}^{M_{y}-1} 
 \mathcal{I} \Big\{ (\bu_{m}, \bu_{m'}, \bx_{mi}, \bx_{m'j} ) \in \mathcal{T}(Q_{UU'XX'}) \Big\}.
\end{align}
\textbf{Remark:}
It is important to notice that the main weakness of this derivation is the passage between (\ref{PDI11}) and (\ref{PDI12}), where we are using the PD inequality over the two exponential sums of sizes $M_{y}$, but in spite of that, the resulting exponent is still higher at relatively low coding rates, as can be seen in Fig.\ 1.

\subsection{Part 2}
At this part we prove the existence of codebooks with prescribed upper bounds on the enumerators. The main idea here is that for every $\epsilon > 0$ and all sufficiently large $n$, there exists a code $\calC$ with a rate pair $(R_{y},R_{z})$, that satisfies, for every codeword $(m,i)$ and every $Q_{UU'XX'}$,
\begin{align}
&N_{mi}^{\mbox{\tiny IN}}(Q_{UXX'}, \calC)   \leq N_{\mbox{\tiny IN}}^{*}(Q_{UXX'}) \nonumber \\
&\overset{\Delta}{=}
   \left\{ 
               \begin{array}{l l}
                \exp \Big\{n [R_{y}  - I_{Q}(X;X'|U) + \epsilon]  \Big\}                 &             \quad \text{  $R_{y}  \geq  I_{Q}(X;X'|U) - \epsilon$  }\\
                 0        & \quad \text{ $R_{y}  <  I_{Q}(X;X'|U) - \epsilon,$  } 
               \end{array} \right.
\end{align}
and
\begin{align}
&N_{mi}^{\mbox{\tiny OUT}}(Q_{UU'XX'}, \calC)  \leq N_{\mbox{\tiny OUT}}^{*}(Q_{UU'XX'}) \nonumber \\
&\overset{\Delta}{=}
   \left\{ 
               \begin{array}{l l}
                \exp \Big\{n [R_{y} + R_{z} - I_{Q}(UX;U'X') + \epsilon]  \Big\}                 &             \quad \text{  $R_{y} + R_{z} \geq  I_{Q}(UX;U'X') - \epsilon$  }\\
                 0        & \quad \text{ $R_{y} + R_{z} <  I_{Q}(UX;U'X') - \epsilon,$  } 
               \end{array} \right.
\end{align}
and furthermore, for every cloud $m$ and every $Q_{UU'XX'}$,
\begin{align}
&\hat{N}_{m}(Q_{UU'XX'}, \calC)  \leq \hat{N}^{*}(Q_{UU'XX'})  \nonumber \\
&\overset{\Delta}{=}
   \left\{ 
               \begin{array}{l l}
                \exp \Big\{n [2R_{y} + R_{z} - I_{Q}(UX;U'X') + \epsilon]  \Big\}                 &             \quad \text{  $2R_{y} + R_{z} \geq  I_{Q}(UX;U'X') - \epsilon$  }\\
                 0        & \quad \text{ $2R_{y} + R_{z} <  I_{Q}(UX;U'X') - \epsilon$  } 
               \end{array} \right. .
\end{align}
To show this, we first compute ensemble averages.
Define
\begin{align}
\hat{N}(Q_{UU'XX'}, \calC) &\overset{\Delta}{=} \frac{1}{M_{z}}  \sum_{m=0}^{M_{z}-1}  \hat{N}_{m}(Q_{UU'XX'}, \calC),  \\
N^{\mbox{\tiny IN}}(Q_{UXX'}, \calC) &\overset{\Delta}{=} \frac{1}{M_{y}M_{z}}  \sum_{m=0}^{M_{z}-1} \sum_{i=0}^{M_{y}-1}  N_{mi}^{\mbox{\tiny IN}}(Q_{UXX'}, \calC) , \\
N^{\mbox{\tiny OUT}}(Q_{UU'XX'}, \calC) &\overset{\Delta}{=} \frac{1}{M_{y}M_{z}}  \sum_{m=0}^{M_{z}-1} \sum_{i=0}^{M_{y}-1}  N_{mi}^{\mbox{\tiny OUT}}(Q_{UU'XX'}, \calC)  .
\end{align}
For the weak user, we have that 
\begin{align}
\mathbb{E}\{\hat{N}(Q_{UU'XX'}, \calC)\} &= \frac{1}{M_{z}}  \sum_{m=0}^{M_{z}-1} \mathbb{E} \{\hat{N}_{m}(Q_{UU'XX'}, \calC) \} \\
&= \mathbb{E} \{\hat{N}_{0}(Q_{UU'XX'}, \calC) \} \\
&= (M_{z}-1) \cdot M_{y}^{2} \cdot \mathrm{Pr} \big\{ (\bU, \bU', \bX, \bX' ) \in \mathcal{T}(Q_{UU'XX'}) \big\} \\
&= (M_{z}-1) \cdot M_{y}^{2} \cdot \frac {|\mathcal{T}(Q_{UU'XX'})|}{|\mathcal{T}(P_{U})|^{2} \cdot |\mathcal{T}(P_{X|U})|^{2}}  \\
&\doteq (M_{z}-1) \cdot M_{y}^{2} \cdot \frac { \exp\{nH_{Q}(U,U',X,X')\} }{ e^{2nH(U)} \cdot e^{2nH(X|U)} }  \\
&= \exp \Big\{n [2R_{y} + R_{z} - I_{Q}(UX;U'X')]  \Big\}.
\end{align}
Similarly, for the strong user, we have that
\begin{align}
\mathbb{E}\{N^{\mbox{\tiny IN}}(Q_{UXX'}, \calC)\} &= \frac{1}{M_{y}M_{z}}  \sum_{m=0}^{M_{z}-1} \sum_{i=0}^{M_{y}-1} \mathbb{E} \big\{ N_{mi}^{\mbox{\tiny IN}}(Q_{UXX'}, \calC) \big\} \\
&\doteq \exp \Big\{n [R_{y}  - I_{Q}(X;X'|U)]  \Big\},
\end{align}
and
\begin{align}
\mathbb{E}\{N^{\mbox{\tiny OUT}}(Q_{UU'XX'}, \calC)\} &= \frac{1}{M_{y}M_{z}}  \sum_{m=0}^{M_{z}-1} \sum_{i=0}^{M_{y}-1} \mathbb{E} \big\{ N_{mi}^{\mbox{\tiny OUT}}(Q_{UU'XX'}, \calC) \big\} \\
&\doteq \exp \Big\{n [R_{y} + R_{z} -  I_{Q}(UX;U'X')]  \Big\}.
\end{align}
It then follows that
\begin{align}
&\mathrm{Pr}  \left\{\calC:~ \bigcup_{Q_{UU'XX'}} \left\{  \hat{N}(Q_{UU'XX'}, \calC) > \mathbb{E}\{\hat{N}(Q_{UU'XX'}, \calC)\} \cdot e^ {n \epsilon/2  }  \right\}  \right\} \nonumber \\
&~~~~~ \bigcup \left\{ \bigcup_{Q_{UXX'}} \left\{ N^{\mbox{\tiny IN}}(Q_{UXX'}, \calC)  > \mathbb{E}\left\{N^{\mbox{\tiny IN}}(Q_{UXX'}, \calC)\right\} \cdot e^ { n \epsilon/2 }  \right\} \right\} \nonumber \\
&~~~~~ \bigcup \left\{ \bigcup_{Q_{UU'XX'}} \left\{  N^{\mbox{\tiny OUT}}(Q_{UU'XX'}, \calC)  > \mathbb{E}\left\{ N^{\mbox{\tiny OUT}}(Q_{UU'XX'}, \calC)\right\} \cdot e^ { n \epsilon/2 }  \right\} \right\} \\
&\overset{\mbox{\tiny UB}}{\leq}   \sum_{Q_{UU'XX'}} \mathrm{Pr} \left\{\calC:~ \hat{N}(Q_{UU'XX'}, \calC) > \mathbb{E}\{\hat{N}(Q_{UU'XX'}, \calC)\} \cdot e^ {n \epsilon/2}  \right\}  \nonumber \\
&~~~~~ +  \sum_{Q_{UXX'}} \mathrm{Pr} \left\{\calC:~ N^{\mbox{\tiny IN}}(Q_{UXX'}, \calC)  > \mathbb{E}\left\{N^{\mbox{\tiny IN}}(Q_{UXX'}, \calC)\right\} \cdot e^ { n \epsilon/2 }  \right\} \nonumber \\
&~~~~~ +\sum_{Q_{UU'XX'}} \mathrm{Pr}  \left\{\calC:~ N^{\mbox{\tiny OUT}}(Q_{UU'XX'}, \calC)  > \mathbb{E}\left\{ N^{\mbox{\tiny OUT}}(Q_{UU'XX'}, \calC)\right\} \cdot e^ { n \epsilon/2 }  \right\}   \\
&\overset{\mbox{\tiny MI}}{\leq}  \sum_{Q_{UU'XX'}} e^{-n \epsilon/2}  + \sum_{Q_{UXX'}} e^{-n \epsilon/2} + \sum_{Q_{UU'XX'}} e^{-n \epsilon/2} \\
& \leq  3 \cdot  (n+1)^{|\calU|^{2} \cdot |\calX|^{2}} \cdot  e^{-n \epsilon/2} \to 0,
\end{align}
which means that with high probability
\begin{align}
\frac{1}{M_{z}}  \sum_{m=0}^{M_{z}-1} \hat{N}_{m}(Q_{UU'XX'}, \calC)  &\leq e^ {n [2R_{y} + R_{z} - I_{Q}(UX;U'X') + \epsilon/2]  }~~  \forall{Q_{UU'XX'}},   \\
\frac{1}{M_{y}M_{z}}  \sum_{m=0}^{M_{z}-1} \sum_{i=0}^{M_{y}-1}  N_{mi}^{\mbox{\tiny IN}}(Q_{UXX'}, \calC)  &\leq  e^ { n [R_{y} - I_{Q}(X;X'|U) + \epsilon/2] }~~  \forall{Q_{UXX'}},  \\ 
\frac{1}{M_{y}M_{z}}  \sum_{m=0}^{M_{z}-1} \sum_{i=0}^{M_{y}-1}  N_{mi}^{\mbox{\tiny OUT}}(Q_{UU'XX'}, \calC)  &\leq  e^ { n [R_{y} + R_{z} - I_{Q}(UX;U'X') + \epsilon/2] }~~  \forall{Q_{UU'XX'}}.
\end{align}
For a given such code and every given $Q_{UU'XX'}$, there must exist at least $(1 - 3e^{-n \epsilon/2}) \cdot M_{z}$ values of $m$ and at least $(1 - 3e^{-n \epsilon/2}) \cdot M_{y}M_{z}$ pairs $(m,i)$ such that 
\begin{align}
\hat{N}_{m}(Q_{UU'XX'}, \calC) &\leq  e^ {n [2R_{y} + R_{z} - I_{Q}(UX;U'X') + \epsilon]  }, \\
 N_{mi}^{\mbox{\tiny IN}}(Q_{UXX'}, \calC)  &\leq  e^ { n [R_{y} - I_{Q}(X;X'|U) + \epsilon] },  \\ 
 N_{mi}^{\mbox{\tiny OUT}}(Q_{UU'XX'}, \calC)  &\leq  e^ { n [R_{y} + R_{z} - I_{Q}(UX;U'X') + \epsilon] }.
\end{align}
Upon eliminating the exceptional clouds and exceptional codewords from the code, for all $Q_{UU'XX'}$, we end up with at least $[1- 3(n+1)^{|\calU|^{2} \cdot |\calX|^{2}} \cdot  e^{-n \epsilon/2}] \cdot M_{z}$ clouds and at least $[1- 3(n+1)^{|\calU|^{2} \cdot |\calX|^{2}} \cdot  e^{-n \epsilon/2}] \cdot M_{y}M_{z}$ codewords for which
\begin{align}
\hat{N}_{m}(Q_{UU'XX'}, \calC) &\leq e^ {n [2R_{y} + R_{z} - I_{Q}(UX;U'X') + \epsilon]  }~~  \forall{Q_{UU'XX'}},  \\
 N_{mi}^{\mbox{\tiny IN}}(Q_{UXX'}, \calC)  &\leq  e^ { n [R_{y} - I_{Q}(X;X'|U) + \epsilon] }~~  \forall{Q_{UXX'}},  \\ 
 N_{mi}^{\mbox{\tiny OUT}}(Q_{UU'XX'}, \calC)  &\leq  e^ { n [R_{y} + R_{z} - I_{Q}(UX;U'X') + \epsilon] }~~  \forall{Q_{UU'XX'}}.
\end{align}
Let $\calC'$ denote the sub-code formed by these $[1- 3(n+1)^{|\calU|^{2} \cdot |\calX|^{2}} \cdot  e^{-n \epsilon/2}] \cdot M_{z}$ remaining clouds and $[1- 3(n+1)^{|\calU|^{2} \cdot |\calX|^{2}} \cdot  e^{-n \epsilon/2}] \cdot M_{y}M_{z}$ remaining codewords. Since $\hat{N}_{m}(Q_{UU'XX'}, \calC') \leq \hat{N}_{m}(Q_{UU'XX'}, \calC)$,  $N_{mi}^{\mbox{\tiny IN}}(Q_{UXX'}, \calC') \leq N_{mi}^{\mbox{\tiny IN}}(Q_{UXX'}, \calC)$ and $N_{mi}^{\mbox{\tiny OUT}}(Q_{UU'XX'}, \calC') \leq N_{mi}^{\mbox{\tiny OUT}}(Q_{UU'XX'}, \calC)$, then the sub-code certainly satisfies
\begin{align}
\label{upup311}
\hat{N}_{m}(Q_{UU'XX'}, \calC') &\leq e^ {n [2R_{y} + R_{z} - I_{Q}(UX;U'X') + \epsilon]  }~~  \forall{m,Q_{UU'XX'}}, \\
\label{upup411} 
 N_{mi}^{\mbox{\tiny IN}}(Q_{UXX'}, \calC')  &\leq  e^ { n [R_{y} - I_{Q}(X;X'|U) + \epsilon] }~~  \forall{(m,i),Q_{UXX'}},  \\
\label{upup511} 
 N_{mi}^{\mbox{\tiny OUT}}(Q_{UU'XX'}, \calC')  &\leq  e^ { n [R_{y} + R_{z} - I_{Q}(UX;U'X') + \epsilon] }~~  \forall{(m,i),Q_{UU'XX'}}.
\end{align}
Finally, observe that since $\hat{N}_{m}(Q_{UU'XX'}, \calC')$ is a non-negative integer, then for $Q_{UU'XX'}$ with $2R_{y} + R_{z} - I_{Q}(UX;U'X') + \epsilon < 0$, the inequality of (\ref{upup311}) means $\hat{N}_{m}(Q_{UU'XX'}, \calC') = 0$, in which case the right hand side of (\ref{upup311}) becomes $\hat{N}^{*}(Q_{UU'XX'})$, and similarly for  (\ref{upup411}) with $N_{\mbox{\tiny IN}}^{*}(Q_{UXX'})$ and for (\ref{upup511}) with $N_{\mbox{\tiny OUT}}^{*}(Q_{UU'XX'})$. Thus, we have shown that there exists a code $\calC'$ consisting of $ [1- 3(n+1)^{|\calU|^{2} \cdot |\calX|^{2}} \cdot  e^{-n \epsilon/2}] \cdot e^{nR_{z}}$ clouds and $[1- 3(n+1)^{|\calU|^{2} \cdot |\calX|^{2}} \cdot  e^{-n \epsilon/2}] \cdot e^{n(R_{y}+R_{z})}$ codewords, for which all clouds satisfy $\hat{N}_{m}(Q_{UU'XX'}, \calC') \leq \hat{N}^{*}(Q_{UU'XX'})$, and all codewords satisfy $N_{mi}^{\mbox{\tiny IN}}(Q_{UXX'}, \calC') \leq N_{\mbox{\tiny IN}}^{*}(Q_{UXX'})$ and $N_{mi}^{\mbox{\tiny OUT}}(Q_{UU'XX'}, \calC') \leq N_{\mbox{\tiny OUT}}^{*}(Q_{UU'XX'})$ for all joint types $Q_{UU'XX'}$. 

Since there are at least $\frac{1}{2} \cdot e^{nR_{z}}$ clouds in $\calC'$ (we take the factor of one half just for simplicity), and at least $\frac{1}{2} \cdot e^{n(R_{z}+R_{y})}$ codewords, then we conclude that at least $\frac{1}{4} \cdot e^{nR_{z}}$ clouds contain at least $\frac{1}{4} \cdot e^{nR_{y}}$ codewords each. Thus, we can eliminate some more clouds and codewords in order to obtain a codebook, for which each cloud contains exactly the same number of codewords.

\subsection{Part 3}
As a consequence of the above observation, we have seen the existence of a code $\calC'$ for which
\begin{align}
&\max_{m}  P_{\mbox{\tiny e}|m} (\calC') \nonumber \\  
&\leq  e^{-nR_{y}}  \sum_{Q_{UU'XX'} \in \calP} \hat{N}^{*}(Q_{UU'XX'})   \exp \big\{  - nD_{t}(Q_{XX'})  \big\} \\
&=  e^{-nR_{y}}  \sum_{\substack{Q_{UU'XX'} \in \calP \\ I_{Q}(UX;U'X') \leq  2R_{y} + R_{z}+ \epsilon }}  
  \exp \big\{ n[2R_{y} + R_{z} - I_{Q}(UX;U'X')  - D_{t}(Q_{XX'}) + \epsilon ]  \big\} \\
&=  \sum_{\substack{Q_{UU'XX'} \in \calP \\ I_{Q}(UX;U'X') \leq  2R_{y} + R_{z}+ \epsilon }}  
  \exp \big\{ n[R_{y} + R_{z} - I_{Q}(UX;U'X')  - D_{t}(Q_{XX'}) + \epsilon ]  \big\},
\end{align}
as well as
\begin{align}
 &\max_{m,i}  P_{\mbox{\tiny e}|mi} (\calC') \nonumber \\  
&\leq  \sum_{Q_{UXX'} \in \calS} N_{\mbox{\tiny IN}}^{*}(Q_{UXX'}) \cdot  e^{- nD_{s}(Q_{XX'})} + \sum_{Q_{UU'XX'} \in \calP} N_{\mbox{\tiny OUT}}^{*}(Q_{UU'XX'}) \cdot  e^{- nD_{s}(Q_{XX'})} \\
&=  \sum_{\substack{Q_{UXX'} \in \calS \\ I_{Q}(X;X'|U') \leq  R_{y}+ \epsilon }} \exp \Big\{ n[R_{y}  - I_{Q}(X;X'|U)  - D_{s}(Q_{XX'}) + \epsilon ]  \Big\} \nonumber \\
&~~~~~+ \sum_{\substack{Q_{UU'XX'} \in \calP \\ I_{Q}(UX;U'X') \leq  R_{y} + R_{z}+ \epsilon }} \exp \Big\{ n[R_{y} + R_{z} - I_{Q}(UX;U'X')  - D_{s}(Q_{XX'}) + \epsilon ]  \Big\},
\end{align}
and due to the arbitrariness of $\epsilon > 0$, this means that there exists a sequence of codes with a rate pair $(R_{y}, R_{z})$ for which (\ref{THM1STRONG}) and (\ref{THM1WEAK}) hold. 
This completes the proof of Theorem 1.

\section{Proof of Theorem 2} 
As we explained in Section 4, the expurgation of an hierarchical codebook can be done in two stages. In part 1, we concentrate only on the cloud-centers, and discard a negligible amount of them in order to have a good set, where by ``good" we mean that some specific structural properties are ensured to be valid. Then, we fix the set of cloud-centers, and move to the second stage, where we continue to expurgate clouds, as well as codewords, in order to obtain a complete codebook, for which some additional properties are assured. More specifically, in part 2, we upper bound the error probabilities and express each one of them using suitable quantities. Part 3 is rather technical and contains the derivations of the ensemble averages.
In part 4, which is the main step of the proof, we show the existence of an hierarchical codebook, for which these quantities are upper bounded by specific deterministic functions of information measures of the fixed cloud centers. In the final part, we put back these deterministic bounds in order to get the desired results.

\subsection{Part 1}
Let $N_{m}(Q_{UU'}, \calC)$ denote the number of cloud centers $\bu_{m'}$, for any $m' \neq m$, such that the joint empirical distribution of $\bu_{m'}$ with $\bu_{m}$ is $Q_{UU'}$, that is
\begin{align}
N_{m}(Q_{UU'}, \calC) =    
  \sum_{m' \neq m}  
 \mathcal{I} \Big\{ (\bu_{m}, \bu_{m'} ) \in \mathcal{T}(Q_{UU'}) \Big\}.
\end{align}
As in proof of Theorem 1, one can easily prove that for every $\epsilon > 0$ and all sufficiently large $n$, there exists a code $\calC$ (cloud-centers) with a rate $R_{z}$ (essentially), that satisfies, for every cloud $m$ and every $Q_{UU'}$,
\begin{align}
\label{weakResult}
&N_{m}(Q_{UU'}, \calC)   \leq N^{*}(Q_{UU'}) \nonumber \\
&\overset{\Delta}{=}
   \left\{ 
               \begin{array}{l l}
                \exp \Big\{n [R_{z}  - I_{Q}(U;U') + \epsilon]  \Big\}                 &             \quad \text{  $R_{z}  \geq  I_{Q}(U;U') - \epsilon$  }\\
                 0        & \quad \text{ $R_{z}  <  I_{Q}(U;U') - \epsilon.$  } 
               \end{array} \right.
\end{align}
Although the proof is omitted here, one important point should be noticed. It is shown that there exists a code $\calC'$ consisting of $M'_{z} \overset{\Delta}{=}  \big[1 - (n+1)^{|\calU|^{2}}  \cdot  e^{-n \epsilon/2} \big] \cdot M_{z}$ clouds, for which all of them satisfy $N_{m}(Q_{UU'}, \calC') \leq N^{*}(Q_{UU'})$ for all joint types $Q_{UU'}$, and hence, we stick with $M'_{z}$, instead of the original size of $M_{z}$. 
From now on, 
we assume without loss of generality that $\{ \bu_{l} \}_{l=0}^{M'_{z}-1}$ is the set of cloud centers for which this property holds. 
Now, let us continue to the second stage of expurgation.

\subsection{Part 2}
In order to continue, we need a few definitions. First, define the average log--likelihood 
\begin{align}
f(Q_{XZ}) = \frac{1}{n} \log W_{2}(\bz|\bx) 
= \sum_{(x,z) \in \calX \times \calZ} Q_{XZ}(x,z) \log W_{2}(z|x) ,
\end{align}
where $Q_{XZ}$ is understood to be the joint empirical distribution of $(\bx,\bz) \in \calX^n \times \calZ^n$, and let
\begin{align} 
\label{E0DEF}    
E_{0}(R_{y},Q_{UXZ}, t) &\overset{\Delta}{=} 
 \left\{ 
               \begin{array}{l l}
  \big[ R_{y} -  I_{Q}(X;Z|U)  \big] \cdot  t     & \quad \text{ $R_{y}   \geq  I_{Q}(X;Z|U)$  } \\
  \big[ R_{y} -  I_{Q}(X;Z|U)  \big]      & \quad \text{ $R_{y}  <  I_{Q}(X;Z|U)$  } 
               \end{array} \right. .
\end{align}
Next, define
\begin{align}
E_{1}(R_{y},Q_{UZ}, t)  \overset{\Delta}{=} 
\max_{Q_{X|UZ} \in \calS(Q_{UZ})}  \left[  E_{0}(R_{y},Q_{UXZ}, t) + tf(Q_{XZ})  \right],  
\end{align}
and furthermore, let
\begin{align}
\label{BDEF}
B(R_{y},t,Q_{UU'}) \overset{\Delta}{=}
\max_{Q_{Z|UU'}}   \left[   H_{Q}(Z|U,U') + E_{1}(R_{y},Q_{UZ}, 1-t) + E_{1}(R_{y},Q_{U'Z}, t)   \right].
\end{align}
For the strong user, let us denote
\begin{align}
\label{K_definition}
K(R_{y}, s, Q_{UU'}) \overset{\Delta}{=}  
\max_{Q_{XX'|UU'}} \big[ R_{y}  -I_{Q}(X;U'|U) -  I_{Q}(UX;X'|U') - D_{s}(Q_{XX'}) \big].
\end{align}
For the weak user, we use the same upper bound that we have derived in (\ref{PDI11}): 
\begin{align}
P_{\mbox{\tiny e}|m} (\calC)  
&\leq  \frac{1}{M_{y}} \sum_{m' \neq m} \sum_{\bz}  
\left[ \left\{ \sum_{i=0}^{M_{y}-1} W_{2}(\bz| \bx_{mi}) \right\}^{1-t}  \left\{\sum_{j=0}^{M_{y}-1} W_{2}(\bz| \bx_{m'j})  \right\}^{t} \right] .
\end{align}
Let us further define
\begin{align}
G(\calC_{m}, \calC_{m'}) &\overset{\Delta}{=} 
\sum_{\bz}  
\left[ \left\{ \sum_{i=0}^{M_{y}-1} W_{2}(\bz| \bx_{mi}) \right\}^{1-t}  \left\{\sum_{j=0}^{M_{y}-1} W_{2}(\bz| \bx_{m'j})  \right\}^{t} \right]  ,
\end{align}
such that we can write in short
\begin{align}
P_{\mbox{\tiny e}|m} (\calC)  
\leq  \frac{1}{M_{y}}  \sum_{m' \neq m}  G(\calC_{m}, \calC_{m'}) ,
\end{align}
to emphasize here the difference from Theorem 1. The ``price" for not using the upper bound $(\sum_{i} a_{i})^{t} \leq \sum_{i} a_{i}^{t}$ is that we need to keep the channel output $\bz$.
Next, for the strong user, we take the upper bound of (\ref{TheSameUpperBoundStrong})
\begin{align}
P_{\mbox{\tiny e}|mi} (\calC)  
&\leq \sum_{j \neq i} \exp \left\{- d_{s}(\bx_{mi}, \bx_{mj}  )  \right\} 
+  \sum_{m' \neq m} \sum_{j=0}^{M_{y}-1}  \exp \left\{ - d_{s}(\bx_{mi}, \bx_{m'j} ) \right\}  \\
&=  \sum_{Q_{UXX'} \in \calS} N_{mi}^{\mbox{\tiny IN}}(Q_{UXX'}, \calC) \cdot  e^{- nD_{s}(Q_{XX'})} +  \sum_{m' \neq m} \sum_{j=0}^{M_{y}-1}  \exp \left\{ - d_{s}(\bx_{mi}, \bx_{m'j} ) \right\}, 
\end{align}
where $N_{mi}^{\mbox{\tiny IN}}(Q_{UXX'}, \calC)$ has already been defined in (\ref{ENUMstrong1}). 

We will prove in Part 4 that for every $\epsilon > 0$ and all sufficiently large $n$, there exists a code $\calC'' \subseteq \calC'$ with a rate pair $(R_{y},R_{z})$, that satisfies, for every codeword $(m,i)$ and every $Q_{UXX'}$,
\begin{align}
\label{LHD1}
&N_{mi}^{\mbox{\tiny IN}}(Q_{UXX'}, \calC'')   \leq N_{\mbox{\tiny IN}}^{*}(Q_{UXX'}) \nonumber \\
&\overset{\Delta}{=}
   \left\{ 
               \begin{array}{l l}
                \exp \Big\{n [R_{y}  - I_{Q}(X;X'|U) + \epsilon]  \Big\}                 &             \quad \text{  $R_{y}  \geq  I_{Q}(X;X'|U) - \epsilon$  }\\
                 0        & \quad \text{ $R_{y}  <  I_{Q}(X;X'|U) - \epsilon,$  } 
               \end{array} \right.
\end{align}
and
\begin{align}
\label{LHD2}
&\sum_{m' \neq m} \sum_{j=0}^{M_{y}-1}  \exp \left\{- d_{s}(\bx_{mi}, \bx_{m'j} )  \right\}  \leq \frac{1}{M'_{z}}  \sum_{m=0}^{M'_{z}-1} \sum_{m' \neq m} \exp \Big\{ n [ K(R_{y}, s, \hat{P}_{\bu_{m}\bu_{m'}})   + \epsilon] \Big\},
\end{align}
and furthermore, for every cloud $m$,
\begin{align}
\label{LHD3}
\sum_{m' \neq m} G(\calC_{m}, \calC_{m'})   \leq 
  \frac{1}{M'_{z}}  \sum_{m=0}^{M'_{z}-1} \sum_{m' \neq m} \exp \Big\{ n [ B(R_{y}, t, \hat{P}_{\bu_{m}\bu_{m'}})   + \epsilon] \Big\} .     
\end{align}

\subsection{Part 3}
Now, we move to a rather technical part of calculating the conditional expectations (given the fixed set of cloud-centers) of the averages of the left-hand-sides of (\ref{LHD1}), (\ref{LHD2}) and (\ref{LHD3}). We start with the strong user.
\begin{align}
\overline{N^{\mbox{\tiny IN}}(Q_{UXX'})}  
&\overset{\Delta}{=} \frac{1}{M_{y}M_{z}}  \sum_{m=0}^{M_{z}-1} \sum_{i=0}^{M_{y}-1} \mathbb{E} \big\{ N_{mi}^{\mbox{\tiny IN}}(Q_{UXX'}, \calC) \big\} \\
&= \frac{1}{M_{y}M_{z}}  \sum_{m=0}^{M_{z}-1} \sum_{i=0}^{M_{y}-1} (M_{y}-1) \cdot \mathrm{Pr} \big\{(\bX,\bX')  \in \mathcal{T}(Q_{XX'|U}|\bu_{m}) \big\} \\
&= \frac{1}{M_{y}M_{z}}  \sum_{m=0}^{M_{z}-1} \sum_{i=0}^{M_{y}-1}  (M_{y}-1)  \cdot \frac {|\mathcal{T}(Q_{XX'|U}|\bu_{m}) |}{ |\mathcal{T}(P_{X|U}|\bu_{m})|^2}  \\
&\doteq \frac{1}{M_{y}M_{z}}  \sum_{m=0}^{M_{z}-1} \sum_{i=0}^{M_{y}-1} (M_{y}-1) \cdot \frac { \exp\{nH_{Q}(X,X'|U)\} }{ e^{nH(X|U)} e^{nH(X'|U)} }  \\
&= \exp \Big\{n [R_{y}  - I_{Q}(X;X'|U)]  \Big\}.
\end{align}
Preparing to the next step, let us define the enumerator $N_{mi}^{m'}(Q_{UU'XX'}, \calC)$ to be the number of codewords $\bx_{m'j} \in \mathcal{C}_{m'}$, such that the joint empirical distribution of $\bx_{m'j}$ with $(\bu_{m}, \bu_{m'}, \bx_{mi})$ is $Q_{UU'XX'}$, where $(\bu_{m}, \bu_{m'}, \bx_{mi}) \in \calT(Q_{UU'X})$, that is
\begin{align}
N_{mi}^{m'}(Q_{UU'XX'}, \calC) =    
   \sum_{j=0}^{M_{y}-1} 
 \mathcal{I} \Big\{ (\bu_{m}, \bu_{m'}, \bx_{mi}, \bx_{m'j} ) \in \mathcal{T}(Q_{UU'XX'}) \Big\}.
\end{align}
Also note that $N_{mi}^{\mbox{\tiny OUT}}(Q_{UU'XX'}, \calC) = \sum_{m' \neq m} N_{mi}^{m'}(Q_{UU'XX'}, \calC)$.
We have
\begin{align} 
\mathbb{E} \left\{ N_{mi}^{m'}(Q_{UU'XX'}, \calC) \middle| \bu_{m}, \bu_{m'}, \bx_{mi}  \right\} 
&=  M_{y} \cdot \mathrm{Pr} \big\{ \bX' \in \mathcal{T}(Q_{X'|UU'X}| \bu_{m}, \bu_{m'}, \bx_{mi}) \big\} \\
&=  M_{y} \cdot \frac {|\mathcal{T}(Q_{X'|UU'X}| \bu_{m}, \bu_{m'}, \bx_{mi})|}{ |\mathcal{T}(P_{X|U}|\bu_{m'})|}  \\
&\doteq  M_{y} \cdot \frac { \exp\{nH_{Q}(X'|U,U',X)\} }{ e^{nH(X'|U')} }  \\
&= \exp \Big\{n [R_{y}  -  I_{Q}(UX;X'|U')]  \Big\},
\end{align}
such that we can evaluate the second expression of the strong user
\begin{align}
& \mathbb{E} \left\{  \sum_{m' \neq m} \sum_{j=0}^{M_{y}-1}  \exp \left\{- d_{s}(\bX_{mi}, \bX_{m'j} )  \right\} \middle| \{ \bu_{l} \}_{l=0}^{M'_{z}-1} \right\} \nonumber \\
&= \sum_{m' \neq m} \mathbb{E} \left\{ \sum_{j=0}^{M_{y}-1}  \exp \left\{- d_{s}(\bX_{mi}, \bX_{m'j} )  \right\} \middle| \{ \bu_{l} \}_{l=0}^{M'_{z}-1} \right\}  \\
&=  \sum_{m' \neq m} \mathbb{E} \left[ \mathbb{E} \left\{ \sum_{j=0}^{M_{y}-1}  \exp \left\{- d_{s}(\bx_{mi}, \bX_{m'j} )  \right\} \middle| \bX_{mi},  \{ \bu_{l} \}_{l=0}^{M'_{z}-1} \right\} \right] \\
\label{ConExp11}
&= \sum_{m' \neq m}  \sum_{Q_{X|UU'}} \sum_{\bx_{mi} \in \calT(Q_{X|UU'})} \frac{1}{|\calT(P_{X|U})|} \nonumber \\ &~~~~~\times  \mathbb{E} \left\{ \sum_{j=0}^{M_{y}-1}  \exp \left\{- d_{s}(\bx_{mi}, \bX_{m'j} )  \right\} \middle| \bx_{mi} \in \calT(Q_{X|UU'}),  \{ \bu_{l} \}_{l=0}^{M'_{z}-1} \right\} . 
\end{align}
For the conditional expectation in the last expression, we have the following
\begin{align}
&\mathbb{E} \left\{ \sum_{j=0}^{M_{y}-1}  \exp \Big\{- d_{s}(\bx_{mi}, \bX_{m'j} )  \Big\} \middle| \bx_{mi} \in \calT(Q_{X|UU'}), \{ \bu_{l} \}_{l=0}^{M'_{z}-1} \right\} \\
&= \mathbb{E} \left\{ \sum_{Q_{X'|UU'X}} N_{mi}^{m'}(Q_{UU'XX'}, \calC) \cdot  e^{- nD_{s}(Q_{XX'})} \middle| \bx_{mi} \in \calT(Q_{X|UU'}), \{ \bu_{l} \}_{l=0}^{M'_{z}-1} \right\}  \\
&=  \sum_{Q_{X'|UU'X}} 
\mathbb{E} \left\{ N_{mi}^{m'}(Q_{UU'XX'}, \calC) \middle|  \bu_{m}, \bu_{m'}, \bx_{mi}  \right\} \cdot  e^{- nD_{s}(Q_{XX'})}  \\
&\doteq  \sum_{Q_{X'|UU'X}} \exp \Big\{n [R_{y}  -  I_{Q}(UX;X'|U')]  \Big\} \cdot  e^{- nD_{s}(Q_{XX'})}  \\
&\doteq  \exp \left\{n \cdot \max_{Q_{X'|XUU'}}  [R_{y}  -  I_{Q}(UX;X'|U') - D_{s}(Q_{XX'}) ]  \right\} , 
\end{align}
and substituting it back into (\ref{ConExp11}) gives
\begin{align}
& \mathbb{E} \left\{  \sum_{m' \neq m} \sum_{j=0}^{M_{y}-1}  \exp \left\{- d_{s}(\bX_{mi}, \bX_{m'j} )  \right\} \middle| \{ \bu_{l} \}_{l=0}^{M'_{z}-1} \right\} \\
&\doteq  \sum_{m' \neq m}  \sum_{Q_{X|UU'}} \sum_{\bx_{mi} \in \calT(Q_{X|UU'})} \frac{1}{|\calT(P_{X|U})|} \nonumber \\ &~~~~~\times   \exp \left\{n \max_{Q_{X'|XUU'}}  [R_{y}  -  I_{Q}(UX;X'|U') - D_{s}(Q_{XX'}) ]  \right\} \\
&= \sum_{m' \neq m} \sum_{Q_{X|UU'}}  \frac{|\calT(Q_{X|UU'})|}{|\calT(P_{X|U})|} \cdot  \exp \left\{n \cdot \max_{Q_{X'|XUU'}}  [R_{y}  -  I_{Q}(UX;X'|U') - D_{s}(Q_{XX'}) ]  \right\} \\
&\doteq \sum_{m' \neq m} \max_{Q_{X|UU'}}  e^{ -n I_{Q}(X;U'|U) }
\exp \left\{n \cdot \max_{Q_{X'|XUU'}} \big[ R_{y}  -  I_{Q}(UX;X'|U') - D_{s}(Q_{XX'}) \big]  \right\} \\
&=  \sum_{m' \neq m} \exp \left\{ n K(R_{y}, s, \hat{P}_{\bu_{m}\bu_{m'}})  \right\},
\end{align}
where $K(R_{y}, s, \hat{P}_{\bu_{m}\bu_{m'}})$ has been defined in (\ref{K_definition}). Hence, 
\begin{align}
& \mathbb{E} \left\{  \frac{1}{M_{y}M'_{z}}  \sum_{m=0}^{M'_{z}-1} \sum_{i=0}^{M_{y}-1}  \sum_{m' \neq m} \sum_{j=0}^{M_{y}-1}  \exp \Big\{- d_{s}(\bX_{mi}, \bX_{m'j} )  \Big\} \middle| \{ \bu_{l} \}_{l=0}^{M'_{z}-1} \right\} \\
&= \frac{1}{M_{y}M'_{z}}  \sum_{m=0}^{M'_{z}-1} \sum_{i=0}^{M_{y}-1}  \mathbb{E} \left\{ \sum_{m' \neq m} \sum_{j=0}^{M_{y}-1}  \exp \Big\{- d_{s}(\bX_{mi}, \bX_{m'j} )  \Big\} \middle|   \{ \bu_{l} \}_{l=0}^{M'_{z}-1} \right\}  \\
&\doteq \frac{1}{M_{y}M'_{z}}  \sum_{m=0}^{M'_{z}-1} \sum_{i=0}^{M_{y}-1}  \sum_{m' \neq m} \exp \Big\{ n K(R_{y}, s, \hat{P}_{\bu_{m}\bu_{m'}})  \Big\}  \\
&= \frac{1}{M'_{z}}  \sum_{m=0}^{M'_{z}-1}  \sum_{m' \neq m} \exp \left\{ n K(R_{y}, s, \hat{P}_{\bu_{m}\bu_{m'}})  \right\}  .
\end{align}
Next, we assess the expectation of $G(\calC_{m}, \calC_{m'})$.
Let $N_{\bz, \bu_{m}}(Q_{UXZ}, \calC)$ denote the number of codewords $\bx_{mi} \in \mathcal{C}_{m}$, such that the joint empirical distribution of $\bx_{mi}$ with $(\bu_{m}, \bz)$ is $Q_{UXZ}$, that is
\begin{align}
N_{\bz, \bu_{m}}(Q_{UXZ}, \calC) =    
  \sum_{i=0}^{M_{y}-1}  
 \mathcal{I} \Big\{ (\bu_{m}, \bx_{mi}, \bz ) \in \mathcal{T}(Q_{UXZ}) \Big\}.
\end{align}
It follows from eq.\ (\ref{TEMdef}) that
\begin{align}
&\mathbb{E} \left\{ \left[ N_{\bz, \bu_{m}}(Q_{UXZ}, \calC)  \right]^{1-t} \right\} \nonumber \\
&\doteq         
              \left\{ 
               \begin{array}{l l}
              \exp \Big\{ n \big[ R_{y} -  I_{Q}(X;Z|U)  \big]  (1-t) \Big\}        & \quad \text{ $R_{y}   \geq  I_{Q}(X;Z|U)$  } \\
             \exp \Big\{ n \big[ R_{y} -  I_{Q}(X;Z|U)  \big] \Big\}                 & \quad \text{  $R_{y}  <  I_{Q}(X;Z|U)$  } 
               \end{array} \right.  \\
&= \exp \left\{ n E_{0}(R_{y},Q_{UXZ}, 1-t)  \right\}.
\end{align}
Hence, from the independence of codewords of different clouds, we get 
\begin{align}
& \mathbb{E} \left\{ \sum_{m' \neq m} G(\calC_{m}, \calC_{m'}) \middle| \{ \bu_{l} \}_{l=0}^{M'_{z}-1} \right\}  \nonumber \\
&= \mathbb{E} \left\{ \sum_{m' \neq m} \sum_{\bz}  
\left[ \left\{ \sum_{i=0}^{M_{y}-1} W_{2}(\bz| \bx_{m,i}) \right\}^{1-t}  \left\{\sum_{j=0}^{M_{y}-1} W_{2}(\bz| \bx_{m',j})  \right\}^{t} \right] \middle| \{ \bu_{l} \}_{l=0}^{M'_{z}-1}  \right\}  \\
&=  \mathbb{E} \left\{ \sum_{m' \neq m} \sum_{\bz}  
\left[ \left\{ \sum_{Q_{X|UZ}} N_{\bz, \bu_{m}}(Q_{UXZ}, \calC) e^{ nf(Q_{XZ})} \right\}^{1-t} \right. \right. \nonumber \\ 
&~~~~~~~~~~~~~~~~~~~~~~~~~~~~~~~~~~\left. \left. \times \left\{ \sum_{Q_{X'|U'Z}} N_{\bz, \bu_{m'}}(Q_{U'X'Z}, \calC) e^{ nf(Q_{X'Z})}  \right\}^{t} \right]  \middle| \{ \bu_{l} \}_{l=0}^{M'_{z}-1} \right\}  \\
&\overset{\mbox{\tiny PD}}{\doteq}  \mathbb{E} \left\{ \sum_{m' \neq m} \sum_{\bz}  
\left[  \sum_{Q_{X|UZ}} \left[ N_{\bz, \bu_{m}}(Q_{UXZ}, \calC)\right]^{1-t}  e^{ n(1-t)f(Q_{XZ})}  \right. \right. \nonumber \\ 
&~~~~~~~~~~~~~~~~~~~~~~~~~~~~~~~~~~~~~\left. \left. \times  \sum_{Q_{X'|U'Z}}  \left[N_{\bz, \bu_{m'}}(Q_{U'X'Z}, \calC)\right]^{t}   e^{ ntf(Q_{X'Z})}   \right] \middle| \{ \bu_{l} \}_{l=0}^{M'_{z}-1}  \right\} \\
&=  \sum_{m' \neq m}  \sum_{\bz}  
\left[  \sum_{Q_{X|UZ}} \mathbb{E}  \Big\{ \left[N_{\bz, \bu_{m}}(Q_{UXZ}, \calC) \right]^{1-t} \Big\}  e^{ n(1-t)f(Q_{XZ})}   \right. \nonumber \\ 
&~~~~~~~~~~~~~~~~~~~~~~~~~~~~~~~~~~~~~ \left. \times  \sum_{Q_{X'|U'Z}} \mathbb{E}  \Big\{ \left[N_{\bz, \bu_{m'}}(Q_{U'X'Z}, \calC)\right]^{t} \Big\} e^{ ntf(Q_{X'Z})}   \right]  \\
&\doteq  \sum_{m' \neq m} \sum_{\bz}  
\left[  \sum_{Q_{X|UZ}} \exp \Big\{ n E_{0}(R_{y},Q_{UXZ}, 1-t)  \Big\} \cdot  e^{ n(1-t)f(Q_{XZ})}   \right. \nonumber \\ 
&~~~~~~~~~~~~~~~~~~~~~~~~~~~~~~~~~~~~~ \left. \times  \sum_{Q_{X'|U'Z}} \exp \Big\{ n E_{0}(R_{y},Q_{U'X'Z}, t)  \Big\} \cdot  e^{ ntf(Q_{X'Z})}   \right]  \\
\label{up5}
&\doteq  \sum_{m' \neq m} \sum_{\bz}  
\left[  \max_{Q_{X|UZ}} \exp \Big\{ n E_{0}(R_{y},Q_{UXZ}, 1-t)  \Big\} \cdot  e^{ n(1-t)f(Q_{XZ})}   \right. \nonumber \\ 
&~~~~~~~~~~~~~~~~~~~~~~~~~~~~~~~~~~~~~ \left. \times  \max_{Q_{X'|U'Z}} \exp \Big\{ n E_{0}(R_{y},Q_{U'X'Z}, t)  \Big\} \cdot  e^{ ntf(Q_{X'Z})}   \right]  \\
&= \sum_{m' \neq m} \sum_{Q_{Z|UU'}} \sum_{\bz \in \calT(Q_{Z|UU'} |\bu_{m}, \bu_{m'} )}  
\Big[   \exp \big\{ n E_{1}(R_{y},Q_{UZ}, 1-t)  \big\} \nonumber \\
&~~~~~~~~~~~~~~~~~~~~~~~~~~~~~~~~~~~~~~~~~~~~
~~~~~~~~~~~~~~~~~~~~~~~ \times   \exp \big\{ n E_{1}(R_{y},Q_{U'Z}, t)  \big\}    \Big]  \\
&\doteq  \sum_{m' \neq m}  \sum_{Q_{Z|UU'}}  \exp \Big\{ n  \big[ H_{Q}(Z|U,U')  + E_{1}(R_{y},Q_{UZ}, 1-t) + E_{1}(R_{y},Q_{U'Z}, t)  \big] \Big\}   \\
&\doteq  \sum_{m' \neq m} \max_{Q_{Z|UU'}}  \exp \Big\{ n  \big[ H_{Q}(Z|U,U') + E_{1}(R_{y},Q_{UZ}, 1-t) + E_{1}(R_{y},Q_{U'Z}, t)  \big] \Big\}   \\
&= \sum_{m' \neq m} \exp \Big\{ n B(R_{y}, t, \hat{P}_{\bu_{m}\bu_{m'}})  \Big\},
\end{align}
which means that
\begin{align}
\mathbb{E} \left\{  \frac{1}{M'_{z}}  \sum_{m=0}^{M'_{z}-1}  \sum_{m' \neq m} G(\calC_{m}, \calC_{m'}) \middle| \{ \bu_{l} \}_{l=0}^{M'_{z}-1} \right\} 
&= \frac{1}{M'_{z}}  \sum_{m=0}^{M'_{z}-1} \sum_{m' \neq m} \exp \Big\{ n B(R_{y}, t, \hat{P}_{\bu_{m}\bu_{m'}})  \Big\}. 
\end{align}

\subsection{Part 4}
Define now the following three events. For the strong user, define
\begin{align}
\calA_{1} \overset{\Delta}{=} \bigcup_{Q_{UXX'}} \left\{\calC:~ \frac{1}{M_{y}M'_{z}}  \sum_{m=0}^{M'_{z}-1} \sum_{i=0}^{M_{y}-1}  N_{mi}^{\mbox{\tiny IN}}(Q_{UXX'}, \calC)  > e^ { n [R_{y} - I_{Q}(X;X'|U) + \epsilon/2] }  \right\},
\end{align}
and
\begin{align}
\calA_{2}  &\overset{\Delta}{=}  \left\{ \calC:~ \frac{1}{M_{y}M'_{z}}  \sum_{m=0}^{M'_{z}-1} \sum_{i=0}^{M_{y}-1}  \sum_{m' \neq m} \sum_{j=0}^{M_{y}-1}  \exp \left\{- d_{s}(\bx_{mi}, \bx_{m'j} )  \right\}  \right. \nonumber \\
& \left. ~~~~~~~~~~~~~~~~~~~~~~~~~~~~~~~~~~~~~~~~~~~~~  >  \frac{1}{M'_{z}}  \sum_{m=0}^{M'_{z}-1} \sum_{m' \neq m}  e^ { n [ K(R_{y}, s, \hat{P}_{\bu_{m}\bu_{m'}})   + \epsilon/2] }  \right\}.
\end{align}
For the weak user, define
\begin{align}
\calA_{3} \overset{\Delta}{=} \left\{ \calC:~ \frac{1}{M'_{z}}  \sum_{m=0}^{M'_{z}-1} \sum_{m' \neq m}  G(\calC_{m}, \calC_{m'})   > \frac{1}{M'_{z}}  \sum_{m=0}^{M'_{z}-1} \sum_{m' \neq m}  e^ { n [ B(R_{y}, t, \hat{P}_{\bu_{m}\bu_{m'}})   + \epsilon/2] }  \right\} .
\end{align}
Notice that the expression in the right hand side inside each of these events equals the conditional expectation (given the set of cloud centers) of the expression in the left hand side, multiplied by $e^{n \epsilon/2}$. Hence, by Markov's inequality,
\begin{align}
&\mathrm{Pr} \left\{ \calA_{1}  \cup   \calA_{2} \cup  \calA_{3} \middle|  \{ \bu_{l} \}_{l=0}^{M'_{z}-1} \right\} \\
&\overset{\mbox{\tiny UB}}{\leq}  \mathrm{Pr} \left\{\calA_{1}  \middle|  \{ \bu_{l} \}_{l=0}^{M'_{z}-1} \right\} 
+\mathrm{Pr} \left\{\calA_{2}  \middle|  \{ \bu_{l} \}_{l=0}^{M'_{z}-1} \right\}
+\mathrm{Pr} \left\{\calA_{3}  \middle|  \{ \bu_{l} \}_{l=0}^{M'_{z}-1} \right\}  \\
&\overset{\mbox{\tiny MI}}{\leq}   \sum_{Q_{UXX'}} e^{-n \epsilon/2} + e^{-n \epsilon/2}  +  e^{-n \epsilon/2}  \\
& \leq   3 \cdot  (n+1)^{|\calU| \cdot |\calX|^{2}} \cdot  e^{-n \epsilon/2} \to 0,
\end{align}
which means that there exists a code with
\begin{align}
\frac{1}{M'_{z}}  \sum_{m=0}^{M'_{z}-1} \sum_{m' \neq m}  G(\calC_{m}, \calC_{m'})   &\leq  \frac{1}{M'_{z}}  \sum_{m=0}^{M'_{z}-1} \sum_{m' \neq m} e^{ n [ B(R_{y}, t, \hat{P}_{\bu_{m}\bu_{m'}})   + \epsilon/2] } ,  \nonumber \\ 
\frac{1}{M_{y}M'_{z}}  \sum_{m=0}^{M'_{z}-1} \sum_{i=0}^{M_{y}-1}  \sum_{m' \neq m} \sum_{j=0}^{M_{y}-1}  e^{- d_{s}(\bx_{mi}, \bx_{m'j} )  }  &\leq  \frac{1}{M'_{z}}  \sum_{m=0}^{M'_{z}-1} \sum_{m' \neq m}  e^ { n [ K(R_{y}, s, \hat{P}_{\bu_{m}\bu_{m'}})   + \epsilon/2] }, \nonumber  \\ 
\frac{1}{M_{y}M'_{z}}  \sum_{m=0}^{M'_{z}-1} \sum_{i=0}^{M_{y}-1}  N_{mi}^{\mbox{\tiny IN}}(Q_{UXX'}, \calC)  &\leq  e^ { n [R_{y} - I_{Q}(X;X'|U) + \epsilon/2] }~~  \forall{Q_{UXX'}} . 
\end{align}
For a given such code and every given $Q_{UXX'}$, there must then exist at least $\big( 1 - 3e^{-n \epsilon/2} \big) \cdot M'_{z}$ values of $m$ and at least $(1 - 3e^{-n \epsilon/2}) \cdot M_{y}M'_{z}$ pairs $(m,i)$ such that 
\begin{align}
\sum_{m' \neq m}  G(\calC_{m}, \calC_{m'})   &\leq  \frac{1}{M'_{z}}  \sum_{m=0}^{M'_{z}-1} \sum_{m' \neq m} e^{ n [ B(R_{y}, t, \hat{P}_{\bu_{m}\bu_{m'}})   + \epsilon] } ,  \nonumber \\ 
\sum_{m' \neq m} \sum_{j=0}^{M_{y}-1}  \exp \left\{- d_{s}(\bx_{mi}, \bx_{m'j} )  \right\}  &\leq  \frac{1}{M'_{z}}  \sum_{m=0}^{M'_{z}-1} \sum_{m' \neq m}  e^ { n [ K(R_{y}, s, \hat{P}_{\bu_{m}\bu_{m'}})   + \epsilon] }, \nonumber  \\ 
 N_{mi}^{\mbox{\tiny IN}}(Q_{UXX'}, \calC)  &\leq  e^ { n [R_{y} - I_{Q}(X;X'|U) + \epsilon] },  
\end{align}
Upon eliminating the exceptional clouds and exceptional codewords from the code, for all $Q_{UXX'}$, we end up with at least $[1- 3(n+1)^{|\calU| \cdot |\calX|^{2}} \cdot  e^{-n \epsilon/2}] \cdot M'_{z}$ clouds and at least $[1- 3(n+1)^{|\calU| \cdot |\calX|^{2}} \cdot  e^{-n \epsilon/2}] \cdot M_{y}M'_{z}$ codewords for which
\begin{align}
\sum_{m' \neq m}  G(\calC_{m}, \calC_{m'})   &\leq  \frac{1}{M'_{z}}  \sum_{m=0}^{M'_{z}-1} \sum_{m' \neq m} e^{ n [ B(R_{y}, t, \hat{P}_{\bu_{m}\bu_{m'}})   + \epsilon] } ,  \nonumber \\ 
\sum_{m' \neq m} \sum_{j=0}^{M_{y}-1}  \exp \left\{- d_{s}(\bx_{mi}, \bx_{m'j} )  \right\}  &\leq  \frac{1}{M'_{z}}  \sum_{m=0}^{M'_{z}-1} \sum_{m' \neq m}  e^{ n [ K(R_{y}, s, \hat{P}_{\bu_{m}\bu_{m'}})   + \epsilon]  }, \nonumber  \\ 
 N_{mi}^{\mbox{\tiny IN}}(Q_{UXX'}, \calC)  &\leq  e^ { n [R_{y} - I_{Q}(X;X'|U) + \epsilon] }~~  \forall{Q_{UXX'}}.  
\end{align}
Let $\calC''$ denote the sub-code formed by these $[1- 3(n+1)^{|\calU| \cdot |\calX|^{2}} \cdot  e^{-n \epsilon/2}] \cdot M'_{z}$ remaining clouds and $[1- 3(n+1)^{|\calU| \cdot |\calX|^{2}} \cdot  e^{-n \epsilon/2}] \cdot M_{y}M'_{z}$ remaining codewords. Since   $N_{mi}^{\mbox{\tiny IN}}(Q_{UXX'}, \calC'') \leq N_{mi}^{\mbox{\tiny IN}}(Q_{UXX'}, \calC)$, then the sub-code certainly satisfies
\begin{align}
\label{upup1}
\sum_{m' \neq m}  G_{mm'}(\calC_{m}, \calC_{m'})   &\leq  \frac{1}{M'_{z}}  \sum_{m=0}^{M'_{z}-1} \sum_{m' \neq m} e^ { n [ B(R_{y}, t, \hat{P}_{\bu_{m}\bu_{m'}})   + \epsilon] }~~  \forall{m},   \\ 
\label{upup2}
\sum_{m' \neq m} \sum_{j=0}^{M_{y}-1}  \exp \left\{- d_{s}(\bx_{mi}, \bx_{m'j} )  \right\}  &\leq  \frac{1}{M'_{z}}  \sum_{m=0}^{M'_{z}-1} \sum_{m' \neq m}  e^ { n [ K(R_{y}, s, \hat{P}_{\bu_{m}\bu_{m'}})   + \epsilon]  }~~  \forall{(m,i)},  \\ 
 N_{mi}^{\mbox{\tiny IN}}(Q_{UXX'}, \calC'')  &\leq  e^ { n [R_{y} - I_{Q}(X;X'|U) + \epsilon] }~~  \forall{(m,i),Q_{UXX'}}.  
\end{align}
Thus, we have shown that there exists a code $\calC''$ consisting of $ [1- 3(n+1)^{|\calU| \cdot |\calX|^{2}} \cdot  e^{-n \epsilon/2}] \cdot e^{nR_{z}}$ clouds and $[1- 3(n+1)^{|\calU| \cdot |\calX|^{2}} \cdot  e^{-n \epsilon/2}] \cdot e^{n(R_{y}+R_{z})}$ codewords, for which all clouds satisfy both (\ref{weakResult}) and (\ref{upup1}) and all codewords satisfy (\ref{upup2}) and $N_{mi}^{\mbox{\tiny IN}}(Q_{UXX'}, \calC'') \leq N_{\mbox{\tiny IN}}^{*}(Q_{UXX'})$ for all joint types $Q_{UXX'}$.

\subsection{Part 5}
As a consequence of the previous parts, we have seen the existence of a code $\calC''$ for which
\begin{align}
 &\max_{m}  P_{\mbox{\tiny e}|m} (\calC'') \nonumber \\
&\leq  \frac{1}{M_{y}}  \sum_{m' \neq m}  G(\calC_{m}, \calC_{m'}) \\
&\leq  \frac{1}{M_{y}}  \frac{1}{M'_{z}}  \sum_{m=0}^{M'_{z}-1} \sum_{m' \neq m} \exp \Big\{ n [ B(R_{y}, t, \hat{P}_{\bu_{m}\bu_{m'}})   + \epsilon] \Big\} \\
&=  \frac{1}{M_{y}} \frac{1}{M'_{z}}  \sum_{m=0}^{M'_{z}-1}  \sum_{Q_{UU'} \in \calQ} N_{m}(Q_{UU'}, \calC') \cdot \exp \Big\{ n [ B(R_{y}, t, Q_{UU'}) + \epsilon] \Big\} \\
&\leq  \frac{1}{M_{y}}  \sum_{\substack{Q_{UU'} \in \calQ \\ I_{Q}(U;U')  \leq R_{z} + \epsilon}} \exp \Big\{ n [ R_{z} - I_{Q}(U;U') + \epsilon] \Big\} \cdot  \exp \Big\{ n [ B(R_{y}, t, Q_{UU'})   + \epsilon] \Big\}   \\ 
&=  \sum_{\substack{Q_{UU'} \in \calQ \\ I_{Q}(U;U')  \leq R_{z} + \epsilon}} \exp \Big\{ n [ R_{z} - I_{Q}(U;U') + B(R_{y}, t, Q_{UU'}) - R_{y}   + 2 \epsilon ] \Big\}  ,
\end{align}
as well as
\begin{align}
 &\max_{m,i}  P_{\mbox{\tiny e}|mi} (\calC'') \nonumber \\  
&\leq   \sum_{Q_{UXX'} \in \calS} N_{mi}^{\mbox{\tiny IN}}(Q_{UXX'}, \calC'') \cdot  e^{- nD_{s}(Q_{XX'})}   +  \sum_{m' \neq m} \sum_{j=0}^{M_{y}-1}  \exp \left\{- d_{s}(\bx_{mi}, \bx_{m'j} )  \right\}    \\
&\leq  \sum_{Q_{UXX'} \in \calS} N_{\mbox{\tiny IN}}^{*}(Q_{UXX'}) \cdot  e^{- nD_{s}(Q_{XX'})} + \frac{1}{M'_{z}}  \sum_{m=0}^{M'_{z}-1} \sum_{m' \neq m} \exp \Big\{ n [ K(R_{y}, s, \hat{P}_{\bu_{m}\bu_{m'}})   + \epsilon] \Big\}  \\
&=  \sum_{\substack{Q_{UXX'} \in \calS \\ I_{Q}(X;X'|U)  \leq R_{y} + \epsilon}} \exp \Big\{ n[R_{y}  - I_{Q}(X;X'|U)  - D_{s}(Q_{XX'}) + \epsilon ]  \Big\} \nonumber \\
&~~~~~+ \frac{1}{M'_{z}}  \sum_{m=0}^{M'_{z}-1} \sum_{Q_{UU'} \in \calQ} N_{m}(Q_{UU'}, \calC') \cdot \exp \Big\{ n [ K(R_{y}, s, Q_{UU'}) + \epsilon] \Big\} \\
&\leq  \sum_{\substack{Q_{UXX'} \in \calS \\ I_{Q}(X;X'|U)  \leq R_{y} + \epsilon}} \exp \Big\{ n[R_{y}  - I_{Q}(X;X'|U)  - D_{s}(Q_{XX'}) + \epsilon ]  \Big\} \nonumber \\
&~~~~~+ \sum_{\substack{Q_{UU'} \in \calQ \\ I_{Q}(U;U')  \leq R_{z} + \epsilon }} \exp \Big\{ n [ R_{z} - I_{Q}(U;U') + \epsilon] \Big\} \cdot  \exp \Big\{ n [ K(R_{y}, s, Q_{UU'})   + \epsilon] \Big\}
\end{align}
and due to the arbitrariness of $\epsilon > 0$, this means that there exists a sequence of codes with a rate pair $(R_{y}, R_{z})$ for which (\ref{THM2STRONG}) and (\ref{THM2WEAK}) hold. As for the final expression for the weak user, one can easily move from the expressions of $E_{0}$, $E_{1}$ and $B$ [defined in (\ref{E0DEF})--(\ref{BDEF})] to the expression of $\hat{D}_{t}$ [defined in (\ref{DhatDEF})] by using standard information--measures identities.  
This completes the proof of Theorem 2.

\section{Proof of Theorem 3}
Consider first the two expressions
\begin{align}
&\Phi_{m,i}(\bu_{m},\by) \overset{\Delta}{=}  \sum_{j \neq i} \exp \{n g( \hat{P}_{\bu_{m} \bx_{mj} \by } ) \} ,\\
&\Psi_{m}(\by) \overset{\Delta}{=}  \sum_{m' \neq m} \sum_{j=0}^{M_{y}-1} \exp \{n g( \hat{P}_{\bu_{m'} \bx_{m'j} \by } ) \},
\end{align}
where the sum in $\Phi_{m,i}(\bu_{m},\by)$ is over all wrong codewords in the true cloud, while the sum in $\Psi_{m}(\by)$ is over all codewords of the incorrect clouds.  
Let $\epsilon >0$ be arbitrarily small, and for every $\by \in \calY^{n}$ and $\bu_{m} \in \calU^{n}$, define the events
\begin{align}
\calS_{\epsilon}(m,i,\bu_{m},\by) &= \Big\{\calC:~ \Phi_{m,i}(\bu_{m},\by) \leq \exp \{n \phi(R_{y}-\epsilon, \hat{P}_{\bu_{m}\by}) \}   \Big\},  \\
\calK_{\epsilon}(m,\by) &= \Big\{\calC:~ \Psi_{m}(\by) \leq \exp \{n \psi(R_{y}-\epsilon, R_{z}-\epsilon, \hat{P}_{\by}) \}   \Big\}, 
\end{align}
where $\phi$ and $\psi$ are defined in (\ref{phiDEF}) and (\ref{psiDEF}), respectively. 
In Appendixes C and D, we show that the vast majority of constant composition codes $\{\calC\}$ are outside both $\calS_{\epsilon}(m,i,\bu_{m},\by)$ and $\calK_{\epsilon}(m,\by)$, simultaneously for all $(m,i)$, all $\bu_{m}$ and all $\by$. More precisely, it is shown that
\begin{align}
\mathrm{Pr} \{ \calS_{\epsilon}(m,i,\bu_{m},\by) \} &\leq  \exp \left\{ -e^{n \epsilon}  + n \epsilon + 1  \right\}, \\
\mathrm{Pr} \{ \calK_{\epsilon}(m,\by) \}  &\leq \exp \left\{ -e^{n \epsilon}  + n \epsilon + 1  \right\},
\end{align}
for every $(m,i)$, $\bu_{m}$ and $\by$, and hence,
\begin{align}
&\mathrm{Pr} \left\{  \bigcup_{m} \bigcup_{i} \bigcup_{\bu_{m} \in \calU^{n}} \bigcup_{\by \in \calY^{n}} \big\{ \calS_{\epsilon}(m,i,\bu_{m},\by) \cup \calK_{\epsilon}(m,\by) \big\}   \right\}  \\
&\overset{\Delta}{=} \mathrm{Pr} \left\{  \calB_{\epsilon}   \right\} \\
&\overset{\mbox{\tiny UB}}{\leq} 2 e^{n(R_{z} + R_{y})} |\calU|^{n} |\calY|^{n}  \exp \left\{ -e^{n \epsilon}  + n \epsilon + 1  \right\},
\end{align}
which still decays double-exponentially. Thus, for all codes in $\calG_{\epsilon} = \calB_{\epsilon}^{c}$, which is the vast majority of codes, we have both $\Phi_{m,i}(\bu_{m},\by) \geq \exp \{n \phi(R_{y}-\epsilon, \hat{P}_{\bu_{m}\by}) \}$ and $\Psi_{m}(\by) \geq \exp \{n \psi(R_{y}-\epsilon, R_{z}-\epsilon, \hat{P}_{\by}) \}$ simultaneously for all $m = 0,1, \ldots , M_{z} - 1$, $i = 0,1, \ldots , M_{y} - 1$, $\by \in \calY^{n}$ and $\bu_{m} \in \calU^{n}$.

\subsection{Analysis for the strong user} 
For a given code $\calC$, the probability of error given that message $(m,i)$ was transmitted is given by the following summation over all wrong codewords in the codebook
\begin{align}
   P_{\mbox{\tiny e}|mi} (\calC)
&=  \sum_{(m',j) \neq (m,i)} \sum_{\by} W_{1}(\by| \bx_{mi}) 
\cdot \mathrm{Pr} \left\{ \hat{m}=m', \hat{i}=j \middle| \by \right\}  \\
&=  \sum_{(m',j) \neq (m,i)} \sum_{\by} W_{1}(\by| \bx_{mi}) \nonumber \\
&~~~~~\times \frac{  \exp \{n g( \hat{P}_{\bu_{m'} \bx_{m'j} \by } ) \} }{ \exp \{n g( \hat{P}_{\bu_{m} \bx_{mi} \by } ) \} 
+ \sum_{(\tilde{m},k) \neq (m,i)} \exp \{n g( \hat{P}_{\bu_{\tilde{m}} \bx_{\tilde{m}k} \by } ) \} }.
\end{align}
Now, trivially,
\begin{align}
 \frac{  \exp \{n g( \hat{P}_{\bu_{m'} \bx_{m'j} \by } ) \} }{ \exp \{n g( \hat{P}_{\bu_{m} \bx_{mi} \by } ) \} + \sum_{(\tilde{m},k) \neq (m,i)} \exp \{n g( \hat{P}_{\bu_{\tilde{m}} \bx_{\tilde{m}k} \by } ) \} } \leq 1, 
\end{align}
and for a code in $\calG_{\epsilon}$, we also have
\begin{align}
&\frac{  \exp \{n g( \hat{P}_{\bu_{m'} \bx_{m'j} \by } ) \} }{ \exp \{n g( \hat{P}_{\bu_{m} \bx_{mi} \by } ) \} + \sum_{(\tilde{m},k) \neq (m,i)} \exp \{n g( \hat{P}_{\bu_{\tilde{m}} \bx_{\tilde{m}k} \by } ) \} } \nonumber \\
&= \frac{  \exp \{n g( \hat{P}_{\bu_{m'} \bx_{m'j} \by } ) \} }{ \exp \{n g( \hat{P}_{\bu_{m} \bx_{mi} \by } ) \} + \sum_{k \neq i} \exp \{n g( \hat{P}_{\bu_{m} \bx_{mk} \by } ) \} + \sum_{\tilde{m} \neq m} \sum_{k} \exp \{n g( \hat{P}_{\bu_{\tilde{m}} \bx_{\tilde{m}k} \by } ) \}  } \nonumber \\
&\leq \frac{  \exp \{n g( \hat{P}_{\bu_{m'} \bx_{m'j} \by } ) \} }{ \exp \{n g( \hat{P}_{\bu_{m} \bx_{mi} \by } ) \} + \exp \{n \phi(R_{y}-\epsilon, \hat{P}_{\bu_{m}\by}) \} +  \exp \{n \psi(R_{y}-\epsilon, R_{z}-\epsilon, \hat{P}_{\by}) \}}.  
\end{align}
Thus, for such a code
\begin{align}
 &\frac{  \exp \{n g( \hat{P}_{\bu_{m'} \bx_{m'j} \by } ) \} }{ \exp \{n g( \hat{P}_{\bu_{m} \bx_{mi} \by } ) \} + \sum_{(\tilde{m},k) \neq (m,i)} \exp \{n g( \hat{P}_{\bu_{\tilde{m}} \bx_{\tilde{m}k} \by } ) \} } \nonumber \\
&\leq \min \left\{1, \frac{  \exp \{n g( \hat{P}_{\bu_{m'} \bx_{m'j} \by } ) \} }{ \exp \{n g( \hat{P}_{\bu_{m} \bx_{mi} \by } ) \} + \exp \{n \phi(R_{y}-\epsilon, \hat{P}_{\bu_{m}\by}) \} +  \exp \{n \psi(R_{y}-\epsilon, R_{z}-\epsilon, \hat{P}_{\by}) \}} \right\}  \nonumber \\
&\doteq \exp \left\{-n \left[ \max \big\{ g( \hat{P}_{\bu_{m} \bx_{mi} \by }), \phi(R_{y}-\epsilon, \hat{P}_{\bu_{m}\by}), \psi(R_{y}-\epsilon, R_{z}-\epsilon, \hat{P}_{\by})  \big\} - g( \hat{P}_{\bu_{m'} \bx_{m'j} \by } ) \right]_{+}  \right\}  \nonumber \\
\label{up775}
&\overset{\Delta}{=} \exp \left\{-n \xi(R_{y}-\epsilon, R_{z}-\epsilon, \hat{P}_{\bu_{m} \bu_{m'} \bx_{mi} \bx_{m'j} \by})  \right\}. 
\end{align}
For any $\rho \geq 1$, we take the expectation over the randomness of the incorrect part of the codebook, $\calC^{mi} = \calC \setminus \{\bu_{m}, \bx_{mi}\}$, where $\bu_{m}$ and $\bx_{mi}$ are kept fixed for now. Let $P(\calC^{mi})$ denote the conditional probability of $\calC^{mi}$ given $\bu_{m}, \bx_{mi}$. Then, 
\begin{align}
& \mathbb{E} \left\{ \left[ P_{\mbox{\tiny e}|mi} (\calC) \right] ^{1 / \rho} \middle| \bu_{m}, \bx_{mi} \right\}  \nonumber \\  
&= \sum_{\calC^{mi}} P(\calC^{mi}) \left[ P_{\mbox{\tiny e}|mi} (\calC) \right] ^{1 / \rho}  \\
\label{up771}
&=  \sum_{\calC^{mi} \in \calG_{\epsilon}} P(\calC^{mi}) \left[ P_{\mbox{\tiny e}|mi} (\calC) \right] ^{1 / \rho} +  \sum_{\calC^{mi} \in \calB_{\epsilon}} P(\calC^{mi}) \left[ P_{\mbox{\tiny e}|mi} (\calC) \right] ^{1 / \rho} \\
\label{up772}
&\overset{\mbox{\tiny (a)}}{\leq}  \sum_{\calC^{mi} \in \calG_{\epsilon}} P(\calC^{mi}) \left[\sum_{(m',j) \neq (m,i)} \sum_{\by} W(\by| \bx_{mi})  \exp \left\{-n \xi(R_{y}-\epsilon, R_{z}-\epsilon, \hat{P}_{\bu_{m} \bu_{m'} \bx_{mi} \bx_{m'j} \by})  \right\} \right] ^{1 / \rho} \nonumber \\
&~~~~~+  \sum_{\calC^{mi} \in \calB_{\epsilon}} P(\calC^{mi}) \cdot 1^{1 / \rho} \\
&\overset{\cdot}{\leq}   \mathbb{E} \left( \left[ \sum_{(m',j) \neq (m,i)} \sum_{\by} W(\by| \bx_{mi})  \exp \left\{-n \xi(R_{y}-\epsilon, R_{z}-\epsilon, \hat{P}_{\bu_{m} \bu_{m'} \bx_{mi} \bx_{m'j} \by})  \right\} \right] ^{1 / \rho} \middle| \bu_{m}, \bx_{mi}  \right) \\
\label{up773}
&\leq \mathbb{E} \left( \left[\sum_{j \neq i} \sum_{\by} W(\by| \bx_{mi})  \exp \left\{-n \xi(R_{y}-\epsilon, R_{z}-\epsilon, \hat{P}_{\bu_{m} \bu_{m} \bx_{mi} \bx_{mj} \by})  \right\} \right] ^{1 / \rho} \middle| \bu_{m}, \bx_{mi}  \right) \nonumber \\
&+ \mathbb{E} \left( \left[ \sum_{m' \neq m} \sum_{j=0}^{M_{y}-1} \sum_{\by} W(\by| \bx_{mi})   \exp \left\{-n \xi(R_{y}-\epsilon, R_{z}-\epsilon, \hat{P}_{\bu_{m} \bu_{m'} \bx_{mi} \bx_{m'j} \by})  \right\} \right] ^{1 / \rho} \middle| \bu_{m}, \bx_{mi}  \right) \\
\label{up774}
&\overset{\mbox{\tiny (b)}}{\doteq}   \mathbb{E} \left( \left[ \sum_{j \neq i} \exp \Big\{-n \Upsilon (\hat{P}_{\bu_{m} \bx_{mi} \bx_{mj}}, R_{y}, R_{z})  \Big\} \right]^{1/ \rho} \middle| \bu_{m}, \bx_{mi}  \right) \nonumber \\
&~~~~~+  \mathbb{E} \left( \left[ \sum_{m' \neq m} \sum_{j=0}^{M_{y}-1}\exp \Big\{-n \Omega (\hat{P}_{\bu_{m} \bu_{m'} \bx_{mi} \bx_{m'j}}, R_{y}, R_{z})   \Big\} \right]^{1/ \rho} \middle| \bu_{m}, \bx_{mi}  \right) \\
&=  \mathbb{E} \left( \left[ \sum_{Q_{UXX'} \in \calS} N_{mi}^{\mbox{\tiny IN}}(Q_{UXX'}, \calC)  \exp \left\{-n \Upsilon (Q_{UXX'}, R_{y}, R_{z})  \right\} \right]^{1/ \rho} \middle| \bu_{m}, \bx_{mi}  \right) \nonumber \\
&~~~~~+  \mathbb{E} \left( \left[ \sum_{Q_{UU'XX'} \in \calP} N_{mi}^{\mbox{\tiny OUT}}(Q_{UU'XX'}, \calC)  \exp \left\{-n \Omega (Q_{UU'XX'}, R_{y}, R_{z})  \right\} \right]^{1/ \rho} \middle| \bu_{m}, \bx_{mi}  \right) \\
&\overset{\mbox{\tiny PD}}{\doteq}  \mathbb{E} \left( \sum_{Q_{UXX'} \in \calS} \left[ N_{mi}^{\mbox{\tiny IN}}(Q_{UXX'}, \calC) \right]^{1/ \rho}  \exp \left\{-n \Upsilon (Q_{UXX'}, R_{y}, R_{z})/ \rho  \right\}  \middle| \bu_{m}, \bx_{mi}  \right) \nonumber \\
&~~~~~+  \mathbb{E} \left( \sum_{Q_{UU'XX'} \in \calP} \left[ N_{mi}^{\mbox{\tiny OUT}}(Q_{UU'XX'}, \calC) \right]^{1/ \rho}  \exp \left\{-n \Omega (Q_{UU'XX'}, R_{y}, R_{z})/ \rho  \right\}  \middle| \bu_{m}, \bx_{mi}  \right) \\
\label{up10}
&= \sum_{Q_{UXX'} \in \calS} \mathbb{E} \left( \left[ N_{mi}^{\mbox{\tiny IN}}(Q_{UXX'}, \calC) \right]^{1/ \rho} \middle| \bu_{m}, \bx_{mi}  \right) \exp \left\{-n \Upsilon (Q_{UXX'}, R_{y}, R_{z})/ \rho  \right\}  \nonumber \\
&~~~~~+ \sum_{Q_{UU'XX'} \in \calP} \mathbb{E} \left( \left[ N_{mi}^{\mbox{\tiny OUT}}(Q_{UU'XX'}, \calC) \right]^{1/ \rho} \middle| \bu_{m}, \bx_{mi}  \right) \exp \left\{-n \Omega (Q_{UU'XX'}, R_{y}, R_{z})/ \rho  \right\} ,
\end{align}
where $N_{mi}^{\mbox{\tiny IN}}(Q_{UXX'}, \calC)$ and $N_{mi}^{\mbox{\tiny OUT}}(Q_{UU'XX'}, \calC)$ are defined in (\ref{ENUMstrong1}) and (\ref{ENUMstrong2}), respectively. 
The passage (a) is due to (\ref{up775}), and the passage (b) is due to the following derivation:
\begin{align}
& \sum_{\by} W(\by| \bx_{mi}) \exp \bigg\{-n 
\Big[ \max \big\{ g( \hat{P}_{\bu_{m} \bx_{mi} \by }), \phi(R_{y}-\epsilon, \hat{P}_{\bu_{m}\by}), \nonumber \\
&~~~~~~~~~~~~~~~~~~~~~~~~~~~~~~~~\psi(R_{y}-\epsilon, R_{z}-\epsilon, \hat{P}_{\by})  \big\} - g( \hat{P}_{\bu_{m'} \bx_{m'j} \by } ) \Big]_{+}  \bigg\} \nonumber \\
&= \sum_{Q_{Y|UU'XX'}} \sum_{\by \in \calT(Q_{Y|UU'XX'})} W(\by| \bx_{mi}) \exp \bigg\{-n 
\Big[ \max \big\{ g( \hat{P}_{\bu_{m} \bx_{mi} \by }), \phi(R_{y}-\epsilon, \hat{P}_{\bu_{m}\by}), \nonumber \\
&~~~~~~~~~~~~~~~~~~~~~~~~~~~~~~~~\psi(R_{y}-\epsilon, R_{z}-\epsilon, \hat{P}_{\by})  \big\} - g( \hat{P}_{\bu_{m'} \bx_{m'j} \by } ) \Big]_{+}  \bigg\}  \\
&\doteq \max_{Q_{Y|UU'XX'}} \exp \Bigg\{n \bigg( H_{Q}(Y|U,U',X,X') + \mathbb{E}_{Q} \log [W(Y|X)] \nonumber \\ 
&~~~~~ -\Big[ \max \big\{ g( Q_{UXY} ), \phi(R_{y}-\epsilon, Q_{UY}), \psi(R_{y}-\epsilon, R_{z}-\epsilon, Q_{Y})  \big\} - g( Q_{U'X'Y} ) \Big]_{+}  \bigg) \Bigg\} \\
&= \max_{Q_{Y|UU'XX'}} \exp \Bigg\{n \bigg( H_{Q}(Y|U,U',X,X') - H_{Q}(Y|U,X) - D(Q_{Y|UX} \| W_{Y|X} | Q_{UX})
\nonumber \\ 
&~~~~~ -\Big[ \max \big\{ g( Q_{UXY} ), \phi(R_{y}-\epsilon, Q_{UY}), \psi(R_{y}-\epsilon, R_{z}-\epsilon, Q_{Y})  \big\} - g( Q_{U'X'Y} ) \Big]_{+}  \bigg) \Bigg\} \\
&=  \exp \Bigg\{-n \min_{Q_{Y|UU'XX'}} \bigg( D(Q_{Y|UX} \| W_{Y|X} | Q_{UX}) + I_{Q}(U'X';Y|UX) 
\nonumber \\ 
&~~~~~ + \Big[ \max \big\{ g( Q_{UXY} ), \phi(R_{y}-\epsilon, Q_{UY}), \psi(R_{y}-\epsilon, R_{z}-\epsilon, Q_{Y})  \big\} - g( Q_{U'X'Y} ) \Big]_{+}  \bigg) \Bigg\} \\
&= \exp \Big\{-n \Omega (Q_{UU'XX'}, R_{y}-\epsilon, R_{z}-\epsilon)  \Big\},
\end{align}
and similarly
\begin{align}
&\sum_{\by} W(\by| \bx_{mi})  \exp \bigg\{-n 
\Big[ \max \big\{ g( \hat{P}_{\bu_{m} \bx_{mi} \by }), \phi(R_{y}-\epsilon, \hat{P}_{\bu_{m}\by}), \nonumber \\
&~~~~~~~~~~~~~~~~~~~~~~~~~~~~~~~~\psi(R_{y}-\epsilon, R_{z}-\epsilon, \hat{P}_{\by})  \big\} - g( \hat{P}_{\bu_{m} \bx_{mj} \by } ) \Big]_{+}  \bigg\} \nonumber \\
&\doteq \exp \Big\{-n \Upsilon (Q_{UXX'}, R_{y}-\epsilon, R_{z}-\epsilon)  \Big\}.
\end{align}
Now, it only remains to assess the conditional expectations of (\ref{up10}). For the left one, we have according to eq.\ (\ref{TEMdef})
\begin{align}
\label{STRONGconditional1}
&\mathbb{E} \left\{ \left[ N_{mi}^{\mbox{\tiny IN}}(Q_{UXX'}, \calC) \right]^{1/ \rho} \middle| \bu_{m}, \bx_{mi}  \right\}  \nonumber \\ 
&\doteq    
           \left\{   
               \begin{array}{l l}
  \exp\{n[R_{y} - I_{Q}(X;X'|U)]/ \rho\}   & \quad \text{  $R_{y} > I_{Q}(X;X'|U)$  }\\
  \exp\{n[R_{y} - I_{Q}(X;X'|U)]\}   & \quad \text{  $R_{y} \leq I_{Q}(X;X'|U)$  } 
               \end{array} \right.  .
\end{align}
In order to assess the second conditional expectation, $\mathbb{E} \left( \left[ N_{mi}^{\mbox{\tiny OUT}}(Q_{UU'XX'}, \calC) \right]^{1/ \rho} \middle| \bu_{m}, \bx_{mi}  \right)$, we make the following definitions. 
Let $N_{\bu_{m}, \bu_{m'}, \bx_{mi}}(Q)$ denote the number of codewords $\bx_{m'j} \in \mathcal{C}_{m'}$, such that $(\bu_{m}, \bu_{m'}, \bx_{mi}, \bx_{m'j}) \in \mathcal{T}(Q_{UU'XX'})$, that is
\begin{align}
N_{\bu_{m}, \bu_{m'}, \bx_{mi}}(Q) = \sum_{j=0}^{M_{y}-1} 
 \mathcal{I} \Big\{ (\bu_{m}, \bu_{m'}, \bx_{mi}, \bx_{m'j} ) \in \mathcal{T}(Q_{UU'XX'}) \Big\}.
\end{align}
Let $\bar{\calA}_{\bu_{m},\bx_{mi}}(Q_{UU'X})$ denote the set of cloud-centers $\bu_{m'}$, such that $(\bu_{m}, \bu_{m'}, \bx_{mi}) \in \mathcal{T}(Q_{UU'X})$.
In addition, let $\bar{N}_{\bu_{m}, \bx_{mi}}(Q_{UU'X})$ be the total amount of them, that is
\begin{align}
\bar{N}_{\bu_{m},\bx_{mi}}(Q_{UU'X}) = \sum_{m' \neq m}  \mathcal{I} \Big\{ (\bu_{m}, \bu_{m'}, \bx_{mi} ) \in \mathcal{T}(Q_{UU'X}) \Big\}.
\end{align}
We condition on the set $\bar{\calA}_{\bu_{m},\bx_{mi}}(Q_{UU'X})$ and assume that $\bar{N}_{\bu_{m},\bx_{mi}}(Q_{UU'X}) = e^{n \lambda}$. For a given $\rho > 1$, let $s \in [1, \rho]$. then, 
\begin{align}
&  \mathbb{E} \left\{ \left[ N_{mi}^{\mbox{\tiny OUT}}(Q_{UU'XX'}, \calC) \right]^{1/ \rho} \middle| \bar{\calA}_{\bu_{m},\bx_{mi}}(Q_{UU'X})  \right\}  \nonumber  \\
&=  \mathbb{E} \left\{ \left[ \sum_{\btildeu \in \bar{\calA}_{\bu_{m},\bx_{mi}}(Q_{UU'X}) } N_{\bu_{m},\btildeu,\bx_{mi}}(Q)\right]^{1 / \rho} \middle| \bar{\calA}_{\bu_{m},\bx_{mi}}(Q_{UU'X})  \right\}    \\
&=  \mathbb{E} \left\{ \left( \left[ \sum_{\btildeu \in \bar{\calA}_{\bu_{m},\bx_{mi}}(Q_{UU'X}) } N_{\bu_{m},\btildeu,\bx_{mi}}(Q) \right]^{1 / s} \right)^{s / \rho}  \middle| \bar{\calA}_{\bu_{m},\bx_{mi}}(Q_{UU'X})  \right\}    \\
&\overset{\mbox{\tiny PD}}{\leq}  \mathbb{E} \left\{ \left( \sum_{\btildeu \in \bar{\calA}_{\bu_{m},\bx_{mi}}(Q_{UU'X}) }  
\left[ N_{\bu_{m},\btildeu,\bx_{mi}}(Q) \right]^{1 / s} \right)^{s / \rho}  \middle| \bar{\calA}_{\bu_{m},\bx_{mi}}(Q_{UU'X})  \right\}   \\
&\leq \left( \mathbb{E} \left\{  \sum_{\btildeu \in \bar{\calA}_{\bu_{m},\bx_{mi}}(Q_{UU'X}) }  
\left[ N_{\bu_{m},\btildeu,\bx_{mi}}(Q) \right]^{1 / s} \middle| \bar{\calA}_{\bu_{m},\bx_{mi}}(Q_{UU'X})  \right\} \right)^{s / \rho}    \\
&= \left(  \sum_{\btildeu \in \bar{\calA}_{\bu_{m},\bx_{mi}}(Q_{UU'X}) }  
\mathbb{E} \left\{ \left[ N_{\bu_{m},\btildeu,\bx_{mi}}(Q) \right]^{1 / s} \right\}  \right)^{s / \rho} ,
\end{align}
where the the second inequality is due to Jensen's inequality (JI) and the concavity of the function $f(v) = v^{s/ \rho}$ when $0 < s/ \rho \leq 1$. 
Now, by eq.\ (\ref{TEMdef}), we have
\begin{align}
&\mathbb{E} \left\{ \left[ N_{\bu_{m},\btildeu,\bx_{mi}}(Q) \right]^{1 / s} \right\}  \nonumber \\ 
&\doteq    
           \left\{   
               \begin{array}{l l}
  \exp\{n[R_{y} - I_{Q}(UX;X'|U')]/s\}   & \quad \text{  $R_{y} > I_{Q}(UX;X'|U')$  }\\
  \exp\{n[R_{y} - I_{Q}(UX;X'|U')]\}   & \quad \text{  $R_{y} \leq I_{Q}(UX;X'|U')$  } 
               \end{array} \right.  ,  
\end{align}
and so,
\begin{align}
&\mathbb{E} \left\{ \left[ N_{mi}^{\mbox{\tiny OUT}}(Q_{UU'XX'}, \calC) \right]^{1/ \rho} \middle| \bar{\calA}_{\bu_{m},\bx_{mi}}(Q_{UU'X})  \right\} \nonumber \\ 
&\leq e^{n \lambda  s  / \rho} \cdot  \left( \mathbb{E} \left\{ \left[ N_{\bu_{m},\btildeu,\bx_{mi}}(Q) \right]^{1 / s} \right\} \right)^{s  / \rho}  \\
&\doteq e^{n \lambda  s  / \rho} \cdot
           \left\{   
               \begin{array}{l l}
  \exp\{n [R_{y} - I_{Q}(UX;X'|U')] / \rho\}   & \quad \text{  $R_{y} > I_{Q}(UX;X'|U')$  }\\
  \exp\{n [R_{y} - I_{Q}(UX;X'|U')] s / \rho \}   & \quad \text{  $R_{y} \leq I_{Q}(UX;X'|U')$  } 
               \end{array} \right. \\
&= 
           \left\{   
               \begin{array}{l l}
  \exp\{n [\lambda s + R_{y} - I_{Q}(UX;X'|U')]/ \rho \}  & \quad \text{  $R_{y} > I_{Q}(UX;X'|U')$  }\\
  \exp\{n  [\lambda + R_{y} - I_{Q}(UX;X'|U')]s / \rho \} & \quad \text{  $R_{y} \leq I_{Q}(UX;X'|U')$  } 
               \end{array} \right.  .
\end{align}
Optimizing over $s$, we get
\begin{align}
&\frac{1}{n} \log \mathbb{E} \left\{ \left[ N_{mi}^{\mbox{\tiny OUT}}(Q_{UU'XX'}, \calC) \right]^{1/ \rho} \middle| \bar{\calA}_{\bu_{m},\bx_{mi}}(Q_{UU'X})  \right\}  \nonumber \\
&\leq  \min_{1 \leq s \leq \rho}             \left\{   
               \begin{array}{l l}
   \left[\lambda s + R_{y} - I_{Q}(UX;X'|U')\right]/ \rho   & \quad \text{  $R_{y} > I_{Q}(UX;X'|U')$  }\\
   \left[\lambda + R_{y} - I_{Q}(UX;X'|U')\right]s / \rho    & \quad \text{  $R_{y} \leq I_{Q}(UX;X'|U'), \lambda + R_{y} > I_{Q}(UX;X'|U')$  }   \\
   \left[\lambda + R_{y} - I_{Q}(UX;X'|U')\right]s / \rho    & \quad \text{  $R_{y} \leq I_{Q}(UX;X'|U'), \lambda + R_{y} \leq I_{Q}(UX;X'|U')$  }
               \end{array} \right.  \\
&=      \left\{   
               \begin{array}{l l}
   \left[\lambda + R_{y} - I_{Q}(UX;X'|U')\right]/ \rho   & \quad \text{  $R_{y} > I_{Q}(UX;X'|U')$  }\\
   \left[\lambda + R_{y} - I_{Q}(UX;X'|U')\right]/ \rho    & \quad \text{  $R_{y} \leq I_{Q}(UX;X'|U'), \lambda + R_{y} > I_{Q}(UX;X'|U')$  }   \\
   \left[\lambda + R_{y} - I_{Q}(UX;X'|U')\right] \rho  / \rho    & \quad \text{  $R_{y} \leq I_{Q}(UX;X'|U'), \lambda + R_{y} \leq I_{Q}(UX;X'|U')$  }
               \end{array} \right.  \\
\label{upup29}
&=      \left\{   
               \begin{array}{l l}
    \left[\lambda + R_{y} - I_{Q}(UX;X'|U')\right] / \rho    & \quad \text{  $\lambda + R_{y} > I_{Q}(UX;X'|U')$  }   \\
   \left[\lambda + R_{y} - I_{Q}(UX;X'|U')\right]    & \quad \text{  $\lambda + R_{y} \leq I_{Q}(UX;X'|U')$  }
               \end{array} \right.  \\
&\overset{\Delta}{=} \bar{E}_{1}(R_{y}, Q_{UU'XX'}, \rho, \lambda).
\end{align}
Next, we have to take the expectation over the set $\bar{\calA}_{\bu_{m},\bx_{mi}}(Q_{UU'X})$. Since the expression of (\ref{upup29}) is independent of the specific cloud-centers, we only have to average it over the cardinality of $\bar{\calA}_{\bu_{m},\bx_{mi}}(Q_{UU'X})$, namely, over $\bar{N}_{\bu_{m}, \bx_{mi}}(Q_{UU'X})$.
\begin{align}
&\mathbb{E}_{\bar{N}} \left[  \exp \Big\{ n  \bar{E}_{1}(R_{y}, Q_{UU'XX'}, \rho, \lambda) \Big\} \right] \nonumber \\
\label{Origin43434}
&\doteq  \sum_{i=0}^{R_{z}/ \epsilon} \mathrm{Pr} \Big\{ e^{n i \epsilon}  \leq  \bar{N}_{\bu_{m}, \bx_{mi}}(Q_{UU'X}) \leq  e^{n (i+1) \epsilon}  \Big\} \cdot  \exp \Big\{ n  \bar{E}_{1}(R_{y}, Q_{UU'XX'}, \rho , i\epsilon)   \Big\} .
\end{align} 
Now, notice that the type-class enumerator $\bar{N}_{\bu_{m}, \bx_{mi}}(Q_{UU'X})$ is
a binomial random variable, pertaining to $e^{nR_{z}}$ trials and probability of success $e^{-n I_{Q}(UX;U')}$. On the one hand, we have \cite{MERHAV2014}
\begin{align}
\mathrm{Pr} \Big\{ \bar{N}_{\bu_{m}, \bx_{mi}}(Q_{UU'X})  \leq  e^{n t}  \Big\} \doteq \mathcal{I} \left\{ R_{z} \leq I_{Q}(UX;U') + t  \right\},
\end{align}
and on the other hand, we have \cite[pp. 167--169]{MERHAV09}
\begin{align}
\mathrm{Pr} \Big\{ \bar{N}_{\bu_{m}, \bx_{mi}}(Q_{UU'X})  \geq  e^{n t}  \Big\} \doteq e^{-n \bar{E}(Q_{UU'X}, t)} ,
\end{align}
where,
\begin{align}
\bar{E}(Q_{UU'X}, t)     =     
           \left\{ 
               \begin{array}{l l}
     \big[ I_{Q}(UX;U')  -  R_{z}  \big]_{+}   & \quad \text{  $\big[ R_{z}  -  I_{Q}(UX;U') \big]_{+} \geq t$  }\\
      \infty                                             & \quad \text{ $\big[ R_{z} - I_{Q}(UX;U') \big]_{+} < t$  } 
               \end{array} \right.  .
\end{align}
First, assume that $R_{z} > I_{Q}(UX;U')$. We conclude that the event $\big\{ e^{n t}  \leq  \bar{N}_{\bu_{m}, \bx_{mi}}(Q_{UU'X}) \leq  e^{n (t + \epsilon) }  \big\}$ occurs with very high probability if and only if $\big( R_{z}  -  I_{Q}(UX;U')  \big) \in [t, t + \epsilon)$, otherwise, its probability has a double exponential decay. Therefore, it turns out that the sum in (\ref{Origin43434}) is dominated by one summand, the one for which $i = \big( R_{z}  -  I_{Q}(UX;U') \big)/ \epsilon$.
Otherwise, if $R_{z} \leq I_{Q}(UX;U')$, then the sum is dominated by the term $i=0$, which has an exponent of $I_{Q}(UX;U')  -  R_{z}$. By using the fact that $\epsilon > 0$ is arbitrarily small, we get that the expectation in question behaves like
\begin{align}
&\mathbb{E} \left\{ \left[ N_{mi}^{\mbox{\tiny OUT}}(Q_{UU'XX'}, \calC) \right]^{1/ \rho} \middle| \bu_{m},\bx_{mi} \right\} \nonumber \\
&\leq \mathbb{E}_{\bar{N}} \left[  \exp \left\{ n  \bar{E}_{1}(R_{y}, Q_{UU'XX'}, \rho, \lambda) \right\} \right]  \\ 
&\doteq   \exp \left\{ n \cdot \left( \bar{E}_{1}(R_{y}, Q_{UU'XX'}, \rho, \bar{\lambda}) - \left[ I_{Q}(UX;U')  -  R_{z}  \right]_{+}  \right)  \right\} ,
\end{align} 
where $\bar{\lambda} \overset{\Delta}{=} [ R_{z} -  I_{Q}(UX;U') ]_{+}$. It can also be written as 
\begin{align}
&\frac{1}{n} \log \mathbb{E} \left\{ \left[ N_{mi}^{\mbox{\tiny OUT}}(Q_{UU'XX'}, \calC) \right]^{1/ \rho} \middle| \bu_{m},\bx_{mi} \right\} \nonumber \\
&\leq     \left\{   
               \begin{array}{l l}
    \left( [ R_{z} -  I_{Q}(UX;U') ]_{+} + R_{y} - I_{Q}(UX;X'|U') \right) / \rho  - \left[ I_{Q}(UX;U')  -  R_{z}  \right]_{+}  \\
  \quad \text{~~~~~~~~ If ~~~  $[ R_{z} -  I_{Q}(UX;U') ]_{+} + R_{y} > I_{Q}(UX;X'|U')$  }   \\
   \left( [ R_{z} -  I_{Q}(UX;U') ]_{+} + R_{y} - I_{Q}(UX;X'|U') - \left[ I_{Q}(UX;U')  -  R_{z}  \right]_{+} \right)    \\
  \quad \text{~~~~~~~~ If ~~~  $[ R_{z} -  I_{Q}(UX;U') ]_{+} + R_{y} \leq I_{Q}(UX;X'|U')$  }
               \end{array} \right.  \\
&=     \left\{   
               \begin{array}{l l}
    \left( [ R_{z} -  I_{Q}(UX;U') ]_{+} + R_{y} - I_{Q}(UX;X'|U') \right) / \rho  - \left[ I_{Q}(UX;U')  -  R_{z}  \right]_{+}  \\
  \quad \text{~~~~~~~~ If ~~~  $[ R_{z} -  I_{Q}(UX;U') ]_{+} + R_{y} > I_{Q}(UX;X'|U'), ~R_{z} >  I_{Q}(UX;U')$  }   \\
   \left( [ R_{z} -  I_{Q}(UX;U') ]_{+} + R_{y} - I_{Q}(UX;X'|U') \right) / \rho  - \left[ I_{Q}(UX;U')  -  R_{z}  \right]_{+}  \\
  \quad \text{~~~~~~~~ If ~~~  $[ R_{z} -  I_{Q}(UX;U') ]_{+} + R_{y} > I_{Q}(UX;X'|U'), ~R_{z} \leq  I_{Q}(UX;U')$  }   \\
   \left( [ R_{z} -  I_{Q}(UX;U') ]_{+} + R_{y} - I_{Q}(UX;X'|U') - \left[ I_{Q}(UX;U')  -  R_{z}  \right]_{+} \right)   \\
  \quad \text{~~~~~~~~ If ~~~  $[ R_{z} -  I_{Q}(UX;U') ]_{+} + R_{y} \leq I_{Q}(UX;X'|U')$  }
               \end{array} \right. \\ 
\label{STRONGconditional2}
&=     \left\{   
               \begin{array}{l l}
    \left(  R_{z} + R_{y} -  I_{Q}(UX;U'X')  \right) / \rho   \\
  \quad \text{~~~~~~~~ If ~~~  $ R_{z} + R_{y} > I_{Q}(UX;U'X'), ~R_{z} >  I_{Q}(UX;U')$  }   \\
    \left(  R_{y} - I_{Q}(UX;X'|U') \right) / \rho  - \left[ I_{Q}(UX;U')  -  R_{z}  \right]  \\
  \quad \text{~~~~~~~~ If ~~~  $ R_{y} > I_{Q}(UX;X'|U'), ~R_{z} \leq  I_{Q}(UX;U')$  }   \\
   \left( [ R_{z} -  I_{Q}(UX;U') ]_{+} + R_{y} - I_{Q}(UX;X'|U') - \left[ I_{Q}(UX;U')  -  R_{z}  \right]_{+} \right)    \\
  \quad \text{~~~~~~~~ If ~~~  $[ R_{z} -  I_{Q}(UX;U') ]_{+} + R_{y} \leq I_{Q}(UX;X'|U')$  }
               \end{array} \right.  .
\end{align}
Finally, both conditional expectations of (\ref{up10}) have been calculated, and it should be noticed that the expressions of (\ref{STRONGconditional1}) and (\ref{STRONGconditional2}) are, in fact, independent of $(\bu_{m}, \bx_{mi})$. Substituting them back into (\ref{up10}) provides an upper bound on $\mathbb{E} \left\{ \left[ P_{\mbox{\tiny e}|mi} (\calC) \right] ^{1 / \rho} \middle| \bu_{m}, \bx_{mi} \right\}$, which is independent of $(\bu_{m}, \bx_{mi})$.

\subsection{Analysis for the weak user}
For a given code $\calC$, the average probability of error given that one of the messages from cloud $m$ was transmitted is given by
\begin{align}
P_{\mbox{\tiny e}|m} (\calC) 
&= \frac {1}{M_{y}} \sum_{i=0}^{M_{y}-1} \sum_{m' \neq m} \sum_{\bz} W(\bz| \bx_{mi})
\cdot \mathrm{Pr} \left\{ \tilde{m}=m' | \bz \right\} \\
\label{up132}
&= \frac {1}{M_{y}} \sum_{i=0}^{M_{y}-1} \sum_{m' \neq m} \sum_{\bz} W(\bz| \bx_{mi}) \nonumber \\
&~~~~~\times \frac{\sum_{j=0}^{M_{y}-1} \exp \{n g( \hat{P}_{\bu_{m'} \bx_{m'j} \bz } ) \}} {\sum_{l=0}^{M_{y}-1} \exp \{n g( \hat{P}_{\bu_{m} \bx_{ml} \bz } ) \} + \sum_{\tilde{m} \neq m} \sum_{k=0}^{M_{y}-1} \exp \{n g( \hat{P}_{\bu_{\tilde{m}} \bx_{\tilde{m}k} \bz } ) \} }.
\end{align}
Notice that for every codebook within the set $\calG_{\epsilon}$, and a given index $j$ [appearing in the nominator of the last term of (\ref{up132})], we have
\begin{align}
&\sum_{l =0}^{M_{y}-1} \exp \{n g( \hat{P}_{\bu_{m} \bx_{ml} \bz } ) \} + \sum_{\tilde{m} \neq m} \sum_{k=0}^{M_{y}-1} \exp \{n g( \hat{P}_{\bu_{\tilde{m}} \bx_{\tilde{m}k} \bz } ) \} \\
&= \exp \{n g( \hat{P}_{\bu_{m} \bx_{mi} \bz } ) \} + \sum_{l \neq i} \exp \{n g( \hat{P}_{\bu_{m} \bx_{ml} \bz } ) \} \nonumber \\
&~~~~~+ \exp \{n g( \hat{P}_{\bu_{m'} \bx_{m'j} \bz } ) \} +  \sum_{\substack{\tilde{m} \neq m \\ (\tilde{m},k) \neq (m',j) }} \exp \{n g( \hat{P}_{\bu_{\tilde{m}} \bx_{\tilde{m}k} \bz } ) \} \\
&\geq \exp \{n g( \hat{P}_{\bu_{m} \bx_{mi} \bz } ) \} + \exp \{n \phi(R_{y}-\epsilon, \hat{P}_{\bu_{m}\bz})  \} \nonumber \\
&~~~~~+ \exp \{n g( \hat{P}_{\bu_{m'} \bx_{m'j} \bz } ) \} +  \exp \{n \psi(R_{y}-\epsilon, R_{z}-\epsilon, \hat{P}_{\bz}) \} \\
&\doteq \exp \left\{n \max \left[ g( \hat{P}_{\bu_{m} \bx_{mi} \bz } ) , \phi(R_{y}-\epsilon, \hat{P}_{\bu_{m}\bz}) ,  g( \hat{P}_{\bu_{m'} \bx_{m'j} \bz } ) , \psi(R_{y}-\epsilon, R_{z}-\epsilon, \hat{P}_{\bz})  \right]   \right\} .
\end{align}
For any $\rho \geq 1$, we take the expectation over the randomness of $\calC^{m} = \calC \setminus \{\bu_{m}\}$, while $\bu_{m}$ is kept fixed. 
\begin{align}
& \mathbb{E} \left\{ \left[ P_{\mbox{\tiny e}|m} (\calC) \right] ^{1 / \rho} \middle| \bu_{m} \right\}  \nonumber \\  
&=   \sum_{\calC^{m}} P(\calC^{m}) \left[ P_{\mbox{\tiny e}|m} (\calC) \right] ^{1 / \rho} \\
&=  \sum_{\calC^{m} \in \calB_{\epsilon}} P(\calC^{m}) \left[ P_{\mbox{\tiny e}|m} (\calC) \right] ^{1 / \rho} + \sum_{\calC^{m} \in \calG_{\epsilon}} P(\calC^{m}) \left[ P_{\mbox{\tiny e}|m} (\calC) \right] ^{1 / \rho}   \\
&=  \sum_{\calC^{m} \in \calB_{\epsilon}} P(\calC^{m}) \left[ P_{\mbox{\tiny e}|m} (\calC) \right] ^{1 / \rho} +  \sum_{\calC^{m} \in \calG_{\epsilon}} P(\calC^{m}) \left( \frac {1}{M_{y}} \sum_{i=0}^{M_{y}-1} \sum_{m' \neq m} \sum_{\bz} W(\bz| \bx_{mi})  \right. \nonumber \\
& \left. ~~~~~\times  \sum_{j=0}^{M_{y}-1} \frac{ \exp \{n g( \hat{P}_{\bu_{m'} \bx_{m'j} \bz } ) \}} {\sum_{l=0}^{M_{y}-1} \exp \{n g( \hat{P}_{\bu_{m} \bx_{ml} \bz } ) \} + \sum_{\tilde{m} \neq m} \sum_{k=0}^{M_{y}-1} \exp \{n g( \hat{P}_{\bu_{\tilde{m}} \bx_{\tilde{m}k} \bz } ) \} } \right) ^{1 / \rho}  \\
&\leq  \sum_{\calC^{m}  \in \calB_{\epsilon}} P(\calC^{m})
 \cdot 1^{1 / \rho} + \sum_{\calC^{m} \in \calG_{\epsilon}} P(\calC^{m}) \left( \frac {1}{M_{y}} \sum_{i=0}^{M_{y}-1} \sum_{m' \neq m} \sum_{\bz} W(\bz| \bx_{mi})  \right. \nonumber \\
& \left. \times  \sum_{j=0}^{M_{y}-1} \frac{ \exp \{n g( \hat{P}_{\bu_{m'} \bx_{m'j} \bz } ) \} } {  \exp \left\{n \max \left[ g( \hat{P}_{\bu_{m} \bx_{mi} \bz } ) , \phi(R_{y}-\epsilon, \hat{P}_{\bu_{m}\bz}) ,  g( \hat{P}_{\bu_{m'} \bx_{m'j} \bz } ) , \psi(R_{y}-\epsilon, R_{z}-\epsilon, \hat{P}_{\bz})  \right]   \right\} } \right) ^{1 / \rho} \\ 
&\overset{\cdot}{\leq}  \sum_{\calC^{m}} P(\calC^{m}) 
\left( \frac {1}{M_{y}} \sum_{i=0}^{M_{y}-1} \sum_{m' \neq m} \sum_{j=0}^{M_{y}-1} \sum_{\bz} W(\bz| \bx_{mi})  \right. \nonumber \\
& \left. ~~~~~ \times   \frac{ \exp \{n g( \hat{P}_{\bu_{m'} \bx_{m'j} \bz } ) \} } {  \exp \left\{n \max \left[ g( \hat{P}_{\bu_{m} \bx_{mi} \bz } ) , \phi(R_{y}-\epsilon, \hat{P}_{\bu_{m}\bz}) ,  g( \hat{P}_{\bu_{m'} \bx_{m'j} \bz } ) , \psi(R_{y}-\epsilon, R_{z}-\epsilon, \hat{P}_{\bz})  \right]   \right\} } \right) ^{1 / \rho} \\
\label{ConExp56}
&\doteq   \mathbb{E} \left[
\left( \frac {1}{M_{y}} \sum_{i=0}^{M_{y}-1} \sum_{m' \neq m} \sum_{j=0}^{M_{y}-1} \exp \big\{-n \Omega (\hat{P}_{\bu_{m} \bu_{m'} \bx_{mi} \bx_{m'j}}, R_{y}, R_{z})  \big\}  \right) ^{1 / \rho} \middle| \bu_{m}  \right]   .
\end{align}
In order to continue, we have to define the following enumerators.
Let $N_{\bu_{m}}(Q)$ denote the number of triplets $(\bx_{mi}, \bu_{m'}, \bx_{m'j})$, such that $(\bu_{m}, \bu_{m'}, \bx_{mi}, \bx_{m'j} ) \in \mathcal{T}(Q_{UU'XX'})$, that is
\begin{align}
N_{\bu_{m}}(Q) = \sum_{m' \neq m} \sum_{i=0}^{M_{y}-1}  \sum_{j=0}^{M_{y}-1} 
 \mathcal{I} \Big\{ (\bu_{m}, \bu_{m'}, \bx_{mi}, \bx_{m'j} ) \in \mathcal{T}(Q_{UU'XX'}) \Big\}.
\end{align}
Let $N_{\bu_{m}, \bu_{m'}}(Q)$ denote the number of pairs $(\bx_{mi}, \bx_{m'j})$, such that $(\bu_{m}, \bu_{m'}, \bx_{mi}, \bx_{m'j} ) \in \mathcal{T}(Q_{UU'XX'})$, that is
\begin{align}
N_{\bu_{m}, \bu_{m'}}(Q) = \sum_{i=0}^{M_{y}-1} \sum_{j=0}^{M_{y}-1} 
 \mathcal{I} \Big\{ (\bu_{m}, \bu_{m'}, \bx_{mi}, \bx_{m'j} ) \in \mathcal{T}(Q_{UU'XX'}) \Big\}.
\end{align}
Let $N_{\bu_{m}, \bu_{m'}, \bx_{mi}}(Q)$ denote the number of codewords $\bx_{m'j} \in \mathcal{C}_{m'}$, such that $(\bu_{m}, \bu_{m'}, \bx_{mi}, \bx_{m'j} ) \in \mathcal{T}(Q_{UU'XX'})$, that is
\begin{align}
N_{\bu_{m}, \bu_{m'}, \bx_{mi}}(Q) = \sum_{j=0}^{M_{y}-1} 
 \mathcal{I} \Big\{ (\bu_{m}, \bu_{m'}, \bx_{mi}, \bx_{m'j} ) \in \mathcal{T}(Q_{UU'XX'}) \Big\}.
\end{align}
Let $\hat{\calA}_{\bu_{m}}(Q_{UU'})$ denote the set of cloud-centers $\bu_{m'}$, such that the joint empirical distribution of $\bu_{m'}$ with $\bu_{m}$ is $Q_{UU'}$. In addition, let $\hat{N}_{\bu_{m}}(Q_{UU'})$ be the total amount of them, that is
\begin{align}
\hat{N}_{\bu_{m}}(Q_{UU'}) = \sum_{m' \neq m}  \mathcal{I} \Big\{ (\bu_{m}, \bu_{m'}) \in \mathcal{T}(Q_{UU'}) \Big\}.
\end{align} 
Let $\tilde{\calA}_{\bu_{m},\bu_{m'}}(Q_{UU'X})$ denote the set of codewords $\bx_{mi} \in \calC_{m}$, such that the joint empirical distribution of $\bx_{mi}$ with $(\bu_{m},\bu_{m'})$ is $Q_{UU'X}$. In addition, let $\tilde{N}_{\bu_{m}, \bu_{m'}}(Q_{UU'X})$ be the total amount of them, that is
\begin{align}
\tilde{N}_{\bu_{m},\bu_{m'}}(Q_{UU'X}) = \sum_{i = 0}^{M_{y}-1}  \mathcal{I} \Big\{ (\bu_{m}, \bu_{m'}, \bx_{mi} ) \in \mathcal{T}(Q_{UU'X}) \Big\}.
\end{align} 
Continuing with the conditional expectation of (\ref{ConExp56}), we have
\begin{align}
&\mathbb{E} \left\{ \left( \frac {1}{M_{y}} \sum_{i=0}^{M_{y}-1} \sum_{m' \neq m} \sum_{j=0}^{M_{y}-1} \exp \big\{-n \Omega (\hat{P}_{\bu_{m} \bu_{m'} \bx_{mi} \bx_{m'j}}, R_{y}, R_{z})  \big\} \right) ^{1 / \rho} \middle| \bu_{m} \right\} \\
&= \mathbb{E} \left\{  \left( e^{-n R_{y}} \sum_{Q_{UU'XX'} \in \calP} N_{\bu_{m}}(Q) e^ {-n \Omega (Q, R_{y}, R_{z}) } \right) ^{1 / \rho}  \middle| \bu_{m} \right\} \\
&= \mathbb{E} \left\{  \left(  \sum_{Q_{UU'XX'} \in \calP} N_{\bu_{m}}(Q) e^ {-n [\Omega (Q, R_{y}, R_{z}) + R_{y}] } \right) ^{1 / \rho} \middle| \bu_{m}  \right\} \\
&\doteq   \mathbb{E} \left\{  \sum_{Q_{UU'XX'} \in \calP} \left[ N_{\bu_{m}}(Q) \right] ^{1 / \rho}  e^ {-n [\Omega (Q, R_{y}, R_{z}) + R_{y}]  / \rho }  \middle| \bu_{m}  \right\} \\
\label{up20}  
&=   \sum_{Q_{UU'XX'} \in \calP} \mathbb{E} \left\{ \left[ N_{\bu_{m}}(Q) \right] ^{1 / \rho} \middle| \bu_{m} \right\}  e^ {-n [\Omega (Q, R_{y}, R_{z}) + R_{y}] /  \rho } .
\end{align}
In order to evaluate the moments $\mathbb{E} \left\{ \left[ N_{\bu_{m}}(Q) \right] ^{1 / \rho} \middle| \bu_{m} \right\}$, we condition on the set $\hat{\calA}_{\bu_{m}}(Q_{UU'})$ and assume that $\hat{N}_{\bu_{m}}(Q_{UU'}) = e^{n \mu}$. For a given $\rho > 1$, let $s \in [1, \rho]$. then, 
\begin{align}
&  \mathbb{E} \left\{ \left[N_{\bu_{m}}(Q)\right]^{1 / \rho} \middle| \hat{\calA}_{\bu_{m}}(Q_{UU'})  \right\}  \nonumber  \\
&=  \mathbb{E} \left\{ \left[ \sum_{\btildeu \in \hat{\calA}_{\bu_{m}}(Q_{UU'})} N_{\bu_{m},\btildeu}(Q)\right]^{1 / \rho} \middle| \hat{\calA}_{\bu_{m}}(Q_{UU'})  \right\}    \\
&=  \mathbb{E} \left\{ \left( \left[ \sum_{\btildeu \in \hat{\calA}_{\bu_{m}}(Q_{UU'})} N_{\bu_{m},\btildeu}(Q) \right]^{1 / s} \right)^{s / \rho}  \middle| \hat{\calA}_{\bu_{m}}(Q_{UU'})  \right\}    \\
&\overset{\mbox{\tiny PD}}{\leq}  \mathbb{E} \left\{ \left( \sum_{\btildeu \in \hat{\calA}_{\bu_{m}}(Q_{UU'})}  
\left[ N_{\bu_{m},\btildeu}(Q) \right]^{1 / s} \right)^{s / \rho}  \middle| \hat{\calA}_{\bu_{m}}(Q_{UU'})  \right\}    \\
&\overset{\mbox{\tiny JI}}{\leq} \left( \mathbb{E} \left\{  \sum_{\btildeu \in \hat{\calA}_{\bu_{m}}(Q_{UU'})}  
\left[ N_{\bu_{m},\btildeu}(Q) \right]^{1 / s} \middle| \hat{\calA}_{\bu_{m}}(Q_{UU'})  \right\} \right)^{s / \rho}    \\
\label{Stage23}
&= \left(  \sum_{\btildeu \in \hat{\calA}_{\bu_{m}}(Q_{UU'})}  
\mathbb{E} \left\{ \left[ N_{\bu_{m},\btildeu}(Q) \right]^{1 / s} \right\}  \right)^{s / \rho} .
\end{align}
In turn, in order to evaluate the moments $\mathbb{E} \left[N_{\bu_{m},\btildeu}(Q)\right]^{1 / s}$, we condition on the set $\tilde{\calA}_{\bu_{m},\btildeu}(Q_{UU'X})$ and assume that $\tilde{N}_{\bu_{m},\btildeu}(Q_{UU'X}) = e^{n \lambda}$. For a given $s \in [1, \rho]$, let $t \in [1, s]$. then, 
\begin{align}
&  \mathbb{E} \left\{ \left[N_{\bu_{m},\btildeu}(Q)\right]^{1 / s} \middle| \tilde{\calA}_{\bu_{m},\btildeu}(Q_{UU'X})  \right\}  \nonumber  \\
&=  \mathbb{E} \left\{ \left[ \sum_{\btildex \in \tilde{\calA}_{\bu_{m},\btildeu}(Q_{UU'X}) } N_{\bu_{m},\btildeu,\btildex}(Q)\right]^{1 / s} \middle| \tilde{\calA}_{\bu_{m},\btildeu}(Q_{UU'X})  \right\}    \\
&=  \mathbb{E} \left\{ \left( \left[ \sum_{\btildex \in \tilde{\calA}_{\bu_{m},\btildeu}(Q_{UU'X}) } N_{\bu_{m},\btildeu,\btildex}(Q) \right]^{1 / t} \right)^{t / s}  \middle| \tilde{\calA}_{\bu_{m},\btildeu}(Q_{UU'X})  \right\}    \\
&\overset{\mbox{\tiny PD}}{\leq}  \mathbb{E} \left\{ \left( \sum_{\btildex \in \tilde{\calA}_{\bu_{m},\btildeu}(Q_{UU'X}) }  
\left[ N_{\bu_{m},\btildeu,\btildex}(Q) \right]^{1 / t} \right)^{t / s}  \middle| \tilde{\calA}_{\bu_{m},\btildeu}(Q_{UU'X})  \right\}   \\
&\overset{\mbox{\tiny JI}}{\leq} \left( \mathbb{E} \left\{  \sum_{\btildex \in \tilde{\calA}_{\bu_{m},\btildeu}(Q_{UU'X}) }  
\left[ N_{\bu_{m},\btildeu,\btildex}(Q) \right]^{1 / t} \middle| \tilde{\calA}_{\bu_{m},\btildeu}(Q_{UU'X})  \right\} \right)^{t / s}    \\
&= \left(  \sum_{\btildex \in \tilde{\calA}_{\bu_{m},\btildeu}(Q_{UU'X}) }  
\mathbb{E} \left\{ \left[ N_{\bu_{m},\btildeu,\btildex}(Q) \right]^{1 / t} \right\}  \right)^{t / s} .
\end{align}
According to eq.\ (\ref{TEMdef}), 
\begin{align}
&\mathbb{E} \left\{ \left[ N_{\bu_{m}, \btildeu, \btildex}(Q) \right]^{1/t}  \right\}  \nonumber \\ 
&\doteq    
           \left\{   
               \begin{array}{l l}
  \exp\{n[R_{y} - I_{Q}(UX;X'|U')]/t\}   & \quad \text{  $R_{y} > I_{Q}(UX;X'|U')$  }\\
  \exp\{n[R_{y} - I_{Q}(UX;X'|U')]\}   & \quad \text{  $R_{y} \leq I_{Q}(UX;X'|U')$  } 
               \end{array} \right. ,  
\end{align}
and so,
\begin{align}
&\mathbb{E} \left\{ \left[N_{\bu_{m},\btildeu}(Q)\right]^{1 / s} \middle| \tilde{\calA}_{\bu_{m},\btildeu}(Q_{UU'X})  \right\} \nonumber \\ 
&\leq e^{n \lambda  t  / s} \cdot  \left( \mathbb{E} \left\{ \left[ N_{\bu_{m}, \btildeu, \btildex}(Q) \right]^{1/t}  \right\} \right)^{t  / s}  \\
&\doteq e^{n \lambda  t  / s} \cdot
           \left\{   
               \begin{array}{l l}
  \exp\{n [R_{y} - I_{Q}(UX;X'|U')] / s\}   & \quad \text{  $R_{y} > I_{Q}(UX;X'|U')$  }\\
  \exp\{n [R_{y} - I_{Q}(UX;X'|U')] t / s \}   & \quad \text{  $R_{y} \leq I_{Q}(UX;X'|U')$  } 
               \end{array} \right. \\
&= 
           \left\{   
               \begin{array}{l l}
  \exp\{n [\lambda t + R_{y} - I_{Q}(UX;X'|U')]/ s \}   & \quad \text{  $R_{y} > I_{Q}(UX;X'|U')$  }\\
  \exp\{n  [\lambda + R_{y} - I_{Q}(UX;X'|U')]t / s \}   & \quad \text{  $R_{y} \leq I_{Q}(UX;X'|U')$  } 
               \end{array} \right.  . 
\end{align}
After optimizing over $t$, we get
\begin{align}
&\frac{1}{n} \log \mathbb{E} \left\{ \left[N_{\bu_{m},\btildeu}(Q)\right]^{1 / s} \middle| \tilde{\calA}_{\bu_{m},\btildeu}(Q_{UU'X})  \right\} \nonumber \\
&\leq  \min_{1 \leq t \leq s}             \left\{   
               \begin{array}{l l}
  \left[\lambda t + R_{y} - I_{Q}(UX;X'|U')\right]/ s   & \quad \text{  $R_{y} > I_{Q}(UX;X'|U')$  }\\
  \left[\lambda + R_{y} - I_{Q}(UX;X'|U')\right]t / s    & \quad \text{  $R_{y} \leq I_{Q}(UX;X'|U'), \lambda + R_{y} > I_{Q}(UX;X'|U')$  }   \\
  \left[\lambda + R_{y} - I_{Q}(UX;X'|U')\right]t / s    & \quad \text{  $R_{y} \leq I_{Q}(UX;X'|U'), \lambda + R_{y} \leq I_{Q}(UX;X'|U')$  }
               \end{array} \right.  \\
&=      \left\{   
               \begin{array}{l l}
  \left[\lambda + R_{y} - I_{Q}(UX;X'|U')\right]/ s   & \quad \text{  $R_{y} > I_{Q}(UX;X'|U')$  }\\
  \left[\lambda + R_{y} - I_{Q}(UX;X'|U')\right]/ s    & \quad \text{  $R_{y} \leq I_{Q}(UX;X'|U'), \lambda + R_{y} > I_{Q}(UX;X'|U')$  }   \\
  \left[\lambda + R_{y} - I_{Q}(UX;X'|U')\right] s  / s    & \quad \text{  $R_{y} \leq I_{Q}(UX;X'|U'), \lambda + R_{y} \leq I_{Q}(UX;X'|U')$  }
               \end{array} \right.  \\
\label{upup19}
&=      \left\{   
               \begin{array}{l l}
  \left[\lambda + R_{y} - I_{Q}(UX;X'|U')\right] / s    & \quad \text{  $\lambda + R_{y} > I_{Q}(UX;X'|U')$  }   \\
  \left[\lambda + R_{y} - I_{Q}(UX;X'|U')\right]    & \quad \text{  $\lambda + R_{y} \leq I_{Q}(UX;X'|U')$  }
               \end{array} \right. \\
&\overset{\Delta}{=} \tilde{E}_{1}(R_{y}, Q_{UU'XX'}, s, \lambda).
\end{align}
Next, we take the expectation over the set $\tilde{\calA}_{\bu_{m},\btildeu}(Q_{UU'X})$. Since the expression of (\ref{upup19}) is independent of the specific codewords, we only have to average it over the cardinality of $\tilde{\calA}_{\bu_{m},\btildeu}(Q_{UU'X})$, namely, over $\tilde{N}_{\bu_{m}, \btildeu}(Q_{UU'X})$.
\begin{align}
&\mathbb{E}_{\tilde{N}} \left[  \exp \Big\{ n  \tilde{E}_{1}(R_{y}, Q_{UU'XX'}, s, \lambda) \Big\} \right] \nonumber \\
\label{Origin434}
&\doteq  \sum_{i=0}^{R_{y}/ \epsilon} \mathrm{Pr} \left\{ e^{n i \epsilon}  \leq  \tilde{N}_{\bu_{m}, \btildeu}(Q_{UU'X})  \leq  e^{n (i+1) \epsilon}  \right\} \cdot  \exp \Big\{ n  \tilde{E}_{1}(R_{y}, Q_{UU'XX'}, s, i\epsilon)   \Big\} .
\end{align} 
Notice that the type--class enumerator $\tilde{N}_{\bu_{m}, \btildeu}(Q_{UU'X})$ is a binomial random variable, pertaining to $e^{nR_{y}}$ trials and probability of success $e^{-n I_{Q}(X;U'|U)}$.
By making a similar large deviations analysis as we did for the sum (\ref{Origin43434}), we get that the expectation in question behaves like
\begin{align}
\mathbb{E} \left\{ \left[ N_{\bu_{m},\btildeu}(Q) \right]^{1 / s} \right\} 
&\leq \mathbb{E}_{\tilde{N}} \left[  \exp \left\{ n  \tilde{E}_{1}(R_{y}, Q_{UU'XX'}, s, \lambda) \right\} \right]  \\ 
&\doteq   \exp \left\{ n \cdot \left( \tilde{E}_{1}(R_{y}, Q_{UU'XX'}, s, \lambda^{*}) - \left[ I_{Q}(X;U'|U)  -  R_{y}  \right]_{+}  \right)  \right\} ,
\end{align} 
where $\lambda^{*} \overset{\Delta}{=} [ R_{y} -  I_{Q}(X;U'|U) ]_{+}$. It can also be written as 
\begin{align}
&\mathbb{E} \left\{ \left[ N_{\bu_{m},\btildeu}(Q) \right]^{1 / s} \right\} \nonumber \\
&\leq     \left\{   
               \begin{array}{l l}
  \exp\left\{n  \left( [ R_{y} -  I_{Q}(X;U'|U) ]_{+} + R_{y} - I_{Q}(UX;X'|U') \right) / s  -n \left[ I_{Q}(X;U'|U)  -  R_{y}  \right]_{+}\right\}  \\
  \quad \text{~~~~~~~~~~~ If ~~~  $[ R_{y} -  I_{Q}(X;U'|U) ]_{+} + R_{y} > I_{Q}(UX;X'|U')$  }   \\
  \exp\left\{n \left( [ R_{y} -  I_{Q}(X;U'|U) ]_{+} + R_{y} - I_{Q}(UX;X'|U') - \left[ I_{Q}(X;U'|U)  -  R_{y}  \right]_{+} \right)  \right\}  \\
  \quad \text{~~~~~~~~~~~ If ~~~  $[ R_{y} -  I_{Q}(X;U'|U) ]_{+} + R_{y} \leq I_{Q}(UX;X'|U')$  }
               \end{array} \right. 
\end{align}
We substitute it back into (\ref{Stage23}) and get that
\begin{align}
&  \mathbb{E} \left\{ \left[N_{\bu_{m}}(Q)\right]^{1 / \rho} \middle| \hat{\calA}_{\bu_{m}}(Q_{UU'})  \right\}  \nonumber  \\
&\leq  \left(  \sum_{\btildeu \in \hat{\calA}_{\bu_{m}}(Q_{UU'})}  
\mathbb{E} \left\{ \left[ N_{\bu_{m},\btildeu}(Q) \right]^{1 / s} \right\}  \right)^{s / \rho}  \\
&\leq   e^{n \mu s/ \rho} \cdot  \left\{   
               \begin{array}{l l}
  \exp\left\{n  \left( [ R_{y} -  I_{Q}(X;U'|U) ]_{+} + R_{y} - I_{Q}(UX;X'|U') \right) / \rho  -n \left[ I_{Q}(X;U'|U)  -  R_{y}  \right]_{+} s/ \rho \right\}  \\
  \quad \text{~~~~~~~~~~~ If ~~~  $[ R_{y} -  I_{Q}(X;U'|U) ]_{+} + R_{y} > I_{Q}(UX;X'|U')$  }   \\
  \exp\left\{n \left( [ R_{y} -  I_{Q}(X;U'|U) ]_{+} + R_{y} - I_{Q}(UX;X'|U') - \left[ I_{Q}(X;U'|U)  -  R_{y}  \right]_{+} \right) s/ \rho  \right\}  \\
  \quad \text{~~~~~~~~~~~ If ~~~  $[ R_{y} -  I_{Q}(X;U'|U) ]_{+} + R_{y} \leq I_{Q}(UX;X'|U')$  }
               \end{array} \right. \\
&=   \left\{   
               \begin{array}{l l}
  \exp\left\{n  \left( [ R_{y} -  I_{Q}(X;U'|U) ]_{+} + R_{y} - I_{Q}(UX;X'|U') 
  + s \left( \mu  - \left[ I_{Q}(X;U'|U)  -  R_{y}  \right]_{+} \right)   \right) / \rho  \right\}  \\
  \quad \text{~~~~~~~~~~~ If ~~~  $[ R_{y} -  I_{Q}(X;U'|U) ]_{+} + R_{y} > I_{Q}(UX;X'|U')$  }   \\
  \exp\left\{n \left(  2R_{y} -  I_{Q}(X;U'|U)  -  I_{Q}(UX;X'|U')  + \mu \right) s/ \rho  \right\}  \\
  \quad \text{~~~~~~~~~~~ If ~~~  $[ R_{y} -  I_{Q}(X;U'|U) ]_{+} + R_{y} \leq I_{Q}(UX;X'|U')$  }
               \end{array} \right. .
\end{align}
Optimizing over $s$, we get
\begin{align}
& \frac{1}{n} \log \mathbb{E} \left\{ \left[N_{\bu_{m}}(Q)\right]^{1 / \rho} \middle| \hat{\calA}_{\bu_{m}}(Q_{UU'})  \right\}  \nonumber  \\
&\leq   \min_{1 \leq s \leq \rho} \left\{   
               \begin{array}{l l}
 \left( [ R_{y} -  I_{Q}(X;U'|U) ]_{+} + R_{y} - I_{Q}(UX;X'|U') 
  + s \left( \mu  - \left[ I_{Q}(X;U'|U)  -  R_{y}  \right]_{+} \right)   \right) / \rho  \\
  \quad \text{If $[ R_{y} -  I_{Q}(X;U'|U) ]_{+} + R_{y} > I_{Q}(UX;X'|U'), ~\mu  - \left[ I_{Q}(X;U'|U)  -  R_{y}  \right]_{+} > 0$  }   \\
 \left( [ R_{y} -  I_{Q}(X;U'|U) ]_{+} + R_{y} - I_{Q}(UX;X'|U') 
  + s \left( \mu  - \left[ I_{Q}(X;U'|U)  -  R_{y}  \right]_{+} \right)   \right) / \rho   \\
  \quad \text{If $[ R_{y} -  I_{Q}(X;U'|U) ]_{+} + R_{y} > I_{Q}(UX;X'|U'), ~\mu  - \left[ I_{Q}(X;U'|U)  -  R_{y}  \right]_{+} \leq 0$  }   \\
 \left(  2R_{y} -  I_{Q}(X;U'|U)  -  I_{Q}(UX;X'|U')  + \mu \right) s/ \rho  \\
  \quad \text{If $[ R_{y} -  I_{Q}(X;U'|U) ]_{+} + R_{y} \leq I_{Q}(UX;X'|U'), ~2R_{y} -  I_{Q}(X;U'|U)  -  I_{Q}(UX;X'|U')  + \mu > 0$  } \\
\left(  2R_{y} -  I_{Q}(X;U'|U)  -  I_{Q}(UX;X'|U')  + \mu \right) s/ \rho \\
  \quad \text{If $[ R_{y} -  I_{Q}(X;U'|U) ]_{+} + R_{y} \leq I_{Q}(UX;X'|U'), ~2R_{y} -  I_{Q}(X;U'|U)  -  I_{Q}(UX;X'|U')  + \mu \leq 0$  }
               \end{array} \right. \\
&=   \left\{   
               \begin{array}{l l}
 \left( [ R_{y} -  I_{Q}(X;U'|U) ]_{+} + R_{y} - I_{Q}(UX;X'|U') 
  +  \left( \mu  - \left[ I_{Q}(X;U'|U)  -  R_{y}  \right]_{+} \right)   \right) / \rho  \\
  \quad \text{If $[ R_{y} -  I_{Q}(X;U'|U) ]_{+} + R_{y} > I_{Q}(UX;X'|U'), ~\mu  - \left[ I_{Q}(X;U'|U)  -  R_{y}  \right]_{+} > 0$  }   \\
 \left( [ R_{y} -  I_{Q}(X;U'|U) ]_{+} + R_{y} - I_{Q}(UX;X'|U') 
  + \rho \left( \mu  - \left[ I_{Q}(X;U'|U)  -  R_{y}  \right]_{+} \right)   \right) / \rho  \\
  \quad \text{If $[ R_{y} -  I_{Q}(X;U'|U) ]_{+} + R_{y} > I_{Q}(UX;X'|U'), ~\mu  - \left[ I_{Q}(X;U'|U)  -  R_{y}  \right]_{+} \leq 0$  }   \\
\left(  2R_{y} -  I_{Q}(X;U'|U)  -  I_{Q}(UX;X'|U')  + \mu \right) / \rho  \\
  \quad \text{If $[ R_{y} -  I_{Q}(X;U'|U) ]_{+} + R_{y} \leq I_{Q}(UX;X'|U'), ~2R_{y} -  I_{Q}(X;U'|U)  -  I_{Q}(UX;X'|U')  + \mu > 0$  } \\
\left(  2R_{y} -  I_{Q}(X;U'|U)  -  I_{Q}(UX;X'|U')  + \mu \right) \rho / \rho   \\
  \quad \text{If $[ R_{y} -  I_{Q}(X;U'|U) ]_{+} + R_{y} \leq I_{Q}(UX;X'|U'), ~2R_{y} -  I_{Q}(X;U'|U)  -  I_{Q}(UX;X'|U')  + \mu \leq 0$  }
               \end{array} \right. \\
&=   \left\{   
               \begin{array}{l l}
\left(  2R_{y} -  I_{Q}(X;U'|U)  - I_{Q}(UX;X'|U') + \mu  \right) / \rho  \\
  \quad \text{If $[ R_{y} -  I_{Q}(X;U'|U) ]_{+} + R_{y} > I_{Q}(UX;X'|U'), ~\mu  - \left[ I_{Q}(X;U'|U)  -  R_{y}  \right]_{+} > 0$  }   \\
\left( [ R_{y} -  I_{Q}(X;U'|U) ]_{+} + R_{y} - I_{Q}(UX;X'|U') 
  + \rho \left( \mu  - \left[ I_{Q}(X;U'|U)  -  R_{y}  \right]_{+} \right)   \right) / \rho   \\
  \quad \text{If $[ R_{y} -  I_{Q}(X;U'|U) ]_{+} + R_{y} > I_{Q}(UX;X'|U'), ~\mu  - \left[ I_{Q}(X;U'|U)  -  R_{y}  \right]_{+} \leq 0$  }   \\
\left(  2R_{y} -  I_{Q}(X;U'|U)  -  I_{Q}(UX;X'|U')  + \mu \right) / \rho  \\
  \quad \text{If $[ R_{y} -  I_{Q}(X;U'|U) ]_{+} + R_{y} \leq I_{Q}(UX;X'|U'), ~2R_{y} -  I_{Q}(X;U'|U)  -  I_{Q}(UX;X'|U')  + \mu > 0$  } \\
\left(  2R_{y} -  I_{Q}(X;U'|U)  -  I_{Q}(UX;X'|U')  + \mu \right)  \\
  \quad \text{If $[ R_{y} -  I_{Q}(X;U'|U) ]_{+} + R_{y} \leq I_{Q}(UX;X'|U'), ~2R_{y} -  I_{Q}(X;U'|U)  -  I_{Q}(UX;X'|U')  + \mu \leq 0$  }
               \end{array} \right. ,
\end{align}
where in the last passage, we have used the identity $a = [a]_{+} - [-a]_{+}$ on the objective of the first line. Let us look on the constraints of the first line. We can sum the inequalities $[ R_{y} -  I_{Q}(X;U'|U) ]_{+} + R_{y} > I_{Q}(UX;X'|U')$ and $\mu  - \left[ I_{Q}(X;U'|U)  -  R_{y}  \right]_{+} > 0$ to get $2R_{y} -  I_{Q}(X;U'|U)  -  I_{Q}(UX;X'|U')  + \mu > 0$, and hence, the first and third lines can be unified. Thus,
\begin{align}
& \frac{1}{n} \log \mathbb{E} \left\{ \left[N_{\bu_{m}}(Q)\right]^{1 / \rho} \middle| \hat{\calA}_{\bu_{m}}(Q_{UU'})  \right\}  \nonumber  \\
&\leq    \left\{   
               \begin{array}{l l}
\left(  2R_{y} -  I_{Q}(X;U'|U)  - I_{Q}(UX;X'|U') + \mu  \right) / \rho \\
  \quad \text{If $[ R_{y} -  I_{Q}(X;U'|U) ]_{+} + R_{y} > I_{Q}(UX;X'|U'), ~2R_{y} -  I_{Q}(X;U'|U)  -  I_{Q}(UX;X'|U')  + \mu > 0$  }   \\
\left( [ R_{y} -  I_{Q}(X;U'|U) ]_{+} + R_{y} - I_{Q}(UX;X'|U') 
  + \rho \left( \mu  - \left[ I_{Q}(X;U'|U)  -  R_{y}  \right]_{+} \right)   \right) / \rho  \\
  \quad \text{If $[ R_{y} -  I_{Q}(X;U'|U) ]_{+} + R_{y} > I_{Q}(UX;X'|U'), ~\mu  - \left[ I_{Q}(X;U'|U)  -  R_{y}  \right]_{+} \leq 0$  }   \\
\left(  2R_{y} -  I_{Q}(X;U'|U)  -  I_{Q}(UX;X'|U')  + \mu \right) / \rho  \\
  \quad \text{If $[ R_{y} -  I_{Q}(X;U'|U) ]_{+} + R_{y} \leq I_{Q}(UX;X'|U'), ~2R_{y} -  I_{Q}(X;U'|U)  -  I_{Q}(UX;X'|U')  + \mu > 0$  } \\
 \left(  2R_{y} -  I_{Q}(X;U'|U)  -  I_{Q}(UX;X'|U')  + \mu \right)  \\
  \quad \text{If $[ R_{y} -  I_{Q}(X;U'|U) ]_{+} + R_{y} \leq I_{Q}(UX;X'|U'), ~2R_{y} -  I_{Q}(X;U'|U)  -  I_{Q}(UX;X'|U')  + \mu \leq 0$  }
               \end{array} \right. \\
\label{upup21}
&=    \left\{   
               \begin{array}{l l}
 \left(  2R_{y} -  I_{Q}(X;U'|U)  - I_{Q}(UX;X'|U') 
  +   \mu  \right) / \rho   \\
  \quad \text{If $2R_{y} -  I_{Q}(X;U'|U)  -  I_{Q}(UX;X'|U')  + \mu > 0$  }   \\
\left( [ R_{y} -  I_{Q}(X;U'|U) ]_{+} + R_{y} - I_{Q}(UX;X'|U') 
  + \rho \left( \mu  - \left[ I_{Q}(X;U'|U)  -  R_{y}  \right]_{+} \right)   \right) / \rho  \\
  \quad \text{If $[ R_{y} -  I_{Q}(X;U'|U) ]_{+} + R_{y} > I_{Q}(UX;X'|U'), ~\mu  - \left[ I_{Q}(X;U'|U)  -  R_{y}  \right]_{+} \leq 0$  }   \\
 \left(  2R_{y} -  I_{Q}(X;U'|U)  -  I_{Q}(UX;X'|U')  + \mu \right)  \\
  \quad \text{If $[ R_{y} -  I_{Q}(X;U'|U) ]_{+} + R_{y} \leq I_{Q}(UX;X'|U'), ~2R_{y} -  I_{Q}(X;U'|U)  -  I_{Q}(UX;X'|U')  + \mu \leq 0$  }
               \end{array} \right. .
\end{align}
Next, we have to take the expectation over the set $\hat{\calA}_{\bu_{m}}(Q_{UU'})$. Since the expression of (\ref{upup21}) is independent of the specific cloud-centers, we only have to average it over the size of the set $\hat{\calA}_{\bu_{m}}(Q_{UU'})$, namely, over $\hat{N}_{\bu_{m}}(Q_{UU'})$.
Define,
\begin{align}
&\tilde{E}_{2}(R_{y}, Q_{UU'XX'}, \rho, \mu)   \nonumber \\
&\overset{\Delta}{=}      \left\{   
               \begin{array}{l l}
\left[  2R_{y} -  I_{Q}(X;U'|U)  - I_{Q}(UX;X'|U')  +  \mu  \right] / \rho    \\
  \quad \text{If $2R_{y} -  I_{Q}(X;U'|U)  -  I_{Q}(UX;X'|U')  + \mu > 0$  }   \\
 \left( [ R_{y} -  I_{Q}(X;U'|U) ]_{+} + R_{y} - I_{Q}(UX;X'|U')  \right) / \rho
  +  \mu  - \left[ I_{Q}(X;U'|U)  -  R_{y}  \right]_{+}   \\
  \quad \text{If $[ R_{y} -  I_{Q}(X;U'|U) ]_{+} + R_{y} > I_{Q}(UX;X'|U'), ~\mu  - \left[ I_{Q}(X;U'|U)  -  R_{y}  \right]_{+} \leq 0$  }   \\
  \left[  2R_{y} -  I_{Q}(X;U'|U)  -  I_{Q}(UX;X'|U')  + \mu \right]    \\
  \quad \text{If $[ R_{y} -  I_{Q}(X;U'|U) ]_{+} + R_{y} \leq I_{Q}(UX;X'|U'), ~2R_{y} -  I_{Q}(X;U'|U)  -  I_{Q}(UX;X'|U')  + \mu \leq 0$  }
               \end{array} \right. .
\end{align}
We assess the expectation as
\begin{align}
&\mathbb{E}_{\hat{N}} \left[  \exp \Big\{ n  \tilde{E}_{2}(R_{y}, Q_{UU'XX'}, \rho, \mu) \Big\} \right] \nonumber \\
&\doteq  \sum_{i=0}^{R_{z}/ \epsilon} \mathrm{Pr} \Big\{ e^{n i \epsilon}  \leq  \hat{N}_{\bu_{m}}(Q_{UU'})  \leq  e^{n (i+1) \epsilon}  \Big\} \cdot  \exp \Big\{ n  \tilde{E}_{2}(R_{y}, Q_{UU'XX'}, \rho, i\epsilon)   \Big\} .
\end{align} 
The type--class enumerator $\hat{N}_{\bu_{m}}(Q_{UU'})$ is a binomial random variable, pertaining to $e^{nR_{z}}$ trials and success rate of $e^{-n I_{Q}(U;U')}$. 
Repeating once again the large deviations analysis as we did for the sum (\ref{Origin43434}), we get that the expectation in question behaves like
\begin{align}
\mathbb{E} \left\{ \left[ N_{\bu_{m}}(Q) \right]^{1 / \rho} \middle| \bu_{m} \right\} 
&\leq \mathbb{E}_{\hat{N}} \left[  \exp \left\{ n  \tilde{E}_{2}(R_{y}, Q_{UU'XX'}, \rho, \mu) \right\} \right]  \\ 
&\doteq   \exp \left\{ n \cdot \left( \tilde{E}_{2}(R_{y}, Q_{UU'XX'}, \rho, \mu^{*}) - \left[ I_{Q}(U;U')  -  R_{z}  \right]_{+}  \right)  \right\} ,
\end{align}
where $\mu^{*} \overset{\Delta}{=} \left[ R_{z} -  I_{Q}(U;U') \right]_{+}$. It can also be written as
\begin{align}
&  \mathbb{E} \left\{ \left[ N_{\bu_{m}}(Q) \right]^{1 / \rho} \middle| \bu_{m} \right\} \nonumber \\ 
&\leq  \left\{   
               \begin{array}{l l}
  e^{n  \left(  2R_{y} -  I_{Q}(X;U'|U)  - I_{Q}(UX;X'|U') 
  +   \mu^{*}  \right) / \rho  - n\left[ I_{Q}(U;U')  -  R_{z}  \right]_{+}  }  \\
  \quad \text{If $2R_{y} -  I_{Q}(X;U'|U)  -  I_{Q}(UX;X'|U')  + \mu^{*} > 0$  }   \\
  e^{n  \left( [ R_{y} -  I_{Q}(X;U'|U) ]_{+} + R_{y} - I_{Q}(UX;X'|U') 
  + \rho \left( \mu^{*}  - \left[ I_{Q}(X;U'|U)  -  R_{y}  \right]_{+} \right)   \right) / \rho  - n\left[ I_{Q}(U;U')  -  R_{z}  \right]_{+}  }  \\
  \quad \text{If $[ R_{y} -  I_{Q}(X;U'|U) ]_{+} + R_{y} > I_{Q}(UX;X'|U'), ~\mu^{*}  - \left[ I_{Q}(X;U'|U)  -  R_{y}  \right]_{+} \leq 0$  }   \\
  e^{n \left(  2R_{y} -  I_{Q}(X;U'|U)  -  I_{Q}(UX;X'|U')  + \mu^{*} \right) - n\left[ I_{Q}(U;U')  -  R_{z}  \right]_{+}  }  \\
  \quad \text{If $[ R_{y} -  I_{Q}(X;U'|U) ]_{+} + R_{y} \leq I_{Q}(UX;X'|U'), ~2R_{y} -  I_{Q}(X;U'|U)  -  I_{Q}(UX;X'|U')  + \mu^{*} \leq 0$  }
               \end{array} \right.  \\
&=  \left\{   
               \begin{array}{l l}
  e^{n  \left(  2R_{y} -  I_{Q}(X;U'|U)  - I_{Q}(UX;X'|U') 
  +   \left[ R_{z} -  I_{Q}(U;U') \right]_{+}  \right) / \rho  - n\left[ I_{Q}(U;U')  -  R_{z}  \right]_{+}  }  \\
  \quad \text{If $2R_{y} -  I_{Q}(X;U'|U)  -  I_{Q}(UX;X'|U')  + \left[ R_{z} -  I_{Q}(U;U') \right]_{+} > 0, ~ R_{z}>I_{Q}(U;U')$  }   \\
  e^{n  \left(  2R_{y} -  I_{Q}(X;U'|U)  - I_{Q}(UX;X'|U') 
  +   \left[ R_{z} -  I_{Q}(U;U') \right]_{+}  \right) / \rho  - n\left[ I_{Q}(U;U')  -  R_{z}  \right]_{+}  }  \\
  \quad \text{If $2R_{y} -  I_{Q}(X;U'|U)  -  I_{Q}(UX;X'|U')  + \left[ R_{z} -  I_{Q}(U;U') \right]_{+} > 0, ~ R_{z} \leq I_{Q}(U;U')$  }   \\
  e^{n  \left( [ R_{y} -  I_{Q}(X;U'|U) ]_{+} + R_{y} - I_{Q}(UX;X'|U') 
  + \rho \left( \mu^{*}  - \left[ I_{Q}(X;U'|U)  -  R_{y}  \right]_{+}  -  \left[ I_{Q}(U;U')  -  R_{z}  \right]_{+}\right)   \right) / \rho    }  \\
  \quad \text{If $[ R_{y} -  I_{Q}(X;U'|U) ]_{+} + R_{y} > I_{Q}(UX;X'|U'), ~\mu^{*}  - \left[ I_{Q}(X;U'|U)  -  R_{y}  \right]_{+} \leq 0$  }   \\
  e^{n \left(  2R_{y} -  I_{Q}(X;U'|U)  -  I_{Q}(UX;X'|U')  + \mu^{*} - \left[ I_{Q}(U;U')  -  R_{z}  \right]_{+}\right)   }  \\
  \quad \text{If $[ R_{y} -  I_{Q}(X;U'|U) ]_{+} + R_{y} \leq I_{Q}(UX;X'|U'), ~2R_{y} -  I_{Q}(X;U'|U)  -  I_{Q}(UX;X'|U')  + \mu^{*} \leq 0$  }
               \end{array} \right. \\
&=  \left\{   
               \begin{array}{l l}
  e^{n  \left(  2R_{y} +   R_{z}  -   I_{Q}(X;U'|U)  - I_{Q}(UX;X'|U') 
   -  I_{Q}(U;U')   \right) / \rho    }  \\
  \quad \text{If $2R_{y} + R_{z}  -  I_{Q}(X;U'|U)  -  I_{Q}(UX;X'|U')   -  I_{Q}(U;U')  > 0, ~ R_{z}>I_{Q}(U;U')$  }   \\
  e^{n  \left(  2R_{y} -  I_{Q}(X;U'|U)  - I_{Q}(UX;X'|U') 
    \right) / \rho  - n\left[ I_{Q}(U;U')  -  R_{z}  \right]  }  \\
  \quad \text{If $2R_{y} -  I_{Q}(X;U'|U)  -  I_{Q}(UX;X'|U')  > 0, ~ R_{z} \leq I_{Q}(U;U')$  }   \\
  e^{n  \left( [ R_{y} -  I_{Q}(X;U'|U) ]_{+} + R_{y} - I_{Q}(UX;X'|U') 
  + \rho \left( \mu^{*}  - \left[ I_{Q}(X;U'|U)  -  R_{y}  \right]_{+}  -  \left[ I_{Q}(U;U')  -  R_{z}  \right]_{+}\right)   \right) / \rho    }  \\
  \quad \text{If $[ R_{y} -  I_{Q}(X;U'|U) ]_{+} + R_{y} > I_{Q}(UX;X'|U'), ~\mu^{*}  - \left[ I_{Q}(X;U'|U)  -  R_{y}  \right]_{+} \leq 0$  }   \\
  e^{n \left(  2R_{y} -  I_{Q}(X;U'|U)  -  I_{Q}(UX;X'|U')  + \mu^{*} - \left[ I_{Q}(U;U')  -  R_{z}  \right]_{+}\right)   }  \\
  \quad \text{If $[ R_{y} -  I_{Q}(X;U'|U) ]_{+} + R_{y} \leq I_{Q}(UX;X'|U'), ~2R_{y} -  I_{Q}(X;U'|U)  -  I_{Q}(UX;X'|U')  + \mu^{*} \leq 0$  }
               \end{array} \right. \\
&=  \left\{   
               \begin{array}{l l}
  e^{n  \left(  2R_{y} +   R_{z}  -   I_{Q}(UX;U'X')   \right) / \rho    }  \\
  \quad \text{If $2R_{y} + R_{z}  -  I_{Q}(UX;U'X')  > 0, ~ R_{z}>I_{Q}(U;U')$  }   \\
  e^{n  \left(  2R_{y} -  I_{Q}(X;U'|U)  - I_{Q}(UX;X'|U') 
    \right) / \rho  - n\left[ I_{Q}(U;U')  -  R_{z}  \right]  }  \\
  \quad \text{If $2R_{y} -  I_{Q}(X;U'|U)  -  I_{Q}(UX;X'|U')  > 0, ~ R_{z} \leq I_{Q}(U;U')$  }   \\
  e^{n  \left( [ R_{y} -  I_{Q}(X;U'|U) ]_{+} + R_{y} - I_{Q}(UX;X'|U') 
  + \rho \left( \mu^{*}  - \left[ I_{Q}(X;U'|U)  -  R_{y}  \right]_{+}  -  \left[ I_{Q}(U;U')  -  R_{z}  \right]_{+}\right)   \right) / \rho    }  \\
  \quad \text{If $[ R_{y} -  I_{Q}(X;U'|U) ]_{+} + R_{y} > I_{Q}(UX;X'|U'), ~\mu^{*}  - \left[ I_{Q}(X;U'|U)  -  R_{y}  \right]_{+} \leq 0$  }   \\
  e^{n \left(  2R_{y} -  I_{Q}(X;U'|U)  -  I_{Q}(UX;X'|U')  + \mu^{*} - \left[ I_{Q}(U;U')  -  R_{z}  \right]_{+}\right)   }  \\
  \quad \text{If $[ R_{y} -  I_{Q}(X;U'|U) ]_{+} + R_{y} \leq I_{Q}(UX;X'|U'), ~2R_{y} -  I_{Q}(X;U'|U)  -  I_{Q}(UX;X'|U')  + \mu^{*} \leq 0$  }
               \end{array} \right. .
\end{align}
Notice that this upper bound on $\mathbb{E} \left\{ \left[ N_{\bu_{m}}(Q) \right]^{1 / \rho} \middle| \bu_{m} \right\}$ is, in fact, independent of $\bu_{m}$. Thus, by substituting it back into (\ref{up20}), we get an upper bound on $\mathbb{E} \left\{ \left[ P_{\mbox{\tiny e}|m} (\calC) \right] ^{1 / \rho} \middle| \bu_{m} \right\}$, which is independent of $\bu_{m}$.

\subsection{Wrapping up} 
Since the bounds of (\ref{up10}) and (\ref{up20}) are independent of $\bu_{m}$ and $\bx_{mi}$, they also hold for the unconditional expectations, $\mathbb{E}  \left[ P_{\mbox{\tiny e}|m} (\calC) \right] ^{1 / \rho} $ and $\mathbb{E}  \left[ P_{\mbox{\tiny e}|mi} (\calC) \right] ^{1 / \rho} $, i.e.,
\begin{align}
\mathbb{E} \left\{ \left[ P_{\mbox{\tiny e}|m} (\calC) \right] ^{1 / \rho}  \right\}  
\leq \sum_{Q_{UU'XX'} \in \calP} \mathbb{E} \left\{ \left[ N_{\bu_{m}}(Q) \right] ^{1 / \rho} \middle| \bu_{m}  \right\}  e^ {-n [\Omega (Q, R_{y}, R_{z}) + R_{y}] /  \rho } 
\overset{\Delta}{=} \mathbf{\Delta}_{\mbox{\tiny w}} \nonumber ,
\end{align}
and
\begin{align}
&\mathbb{E} \left\{ \left[  P_{\mbox{\tiny e}|mi} (\calC) \right] ^{1 / \rho} \right\} \nonumber \\
&\leq  \sum_{Q_{UXX'} \in \calS} \mathbb{E} \left( \left[ N_{mi}^{\mbox{\tiny IN}}(Q_{UXX'}, \calC) \right]^{1/ \rho} \middle| \bu_{m}, \bx_{mi}  \right) \exp \left\{-n \Upsilon (Q_{UXX'}, R_{y}, R_{z})/ \rho  \right\}  \nonumber \\
&~~~~~+ \sum_{Q_{UU'XX'} \in \calP} \mathbb{E} \left( \left[ N_{mi}^{\mbox{\tiny OUT}}(Q_{UU'XX'}, \calC) \right]^{1/ \rho} \middle| \bu_{m}, \bx_{mi}  \right) \exp \left\{-n \Omega (Q_{UU'XX'}, R_{y}, R_{z})/ \rho  \right\} \overset{\Delta}{=} \mathbf{\Delta}_{\mbox{\tiny s}} \nonumber.
\end{align} 
According to MI, we get that
\begin{align}
&\mathrm{Pr} \left[ \left\{ \frac{1}{M_{z}M_{y}}\sum_{m=0}^{M_{z}-1} \sum_{i=0}^{M_{y}-1} \left[ P_{\mbox{\tiny e}|mi} (\calC) \right] ^{1 / \rho}  > 4 \cdot \mathbf{\Delta}_{\mbox{\tiny s}} \right\} \bigcup
\left\{ \frac{1}{M_{z}} \sum_{m=0}^{M_{z}-1} \left[ P_{\mbox{\tiny e}|m} (\calC) \right] ^{1 / \rho}  > 4 \cdot \mathbf{\Delta}_{\mbox{\tiny w}} \right\} \right] \nonumber \\
&\overset{\mbox{\tiny UB}}{\leq} \mathrm{Pr}  \left\{ \frac{1}{M_{z}M_{y}}\sum_{m=0}^{M_{z}-1} \sum_{i=0}^{M_{y}-1} \left[ P_{\mbox{\tiny e}|mi} (\calC) \right] ^{1 / \rho}  > 4 \cdot \mathbf{\Delta}_{\mbox{\tiny s}} \right\} 
+ \mathrm{Pr} \left\{ \frac{1}{M_{z}} \sum_{m=0}^{M_{z}-1} \left[ P_{\mbox{\tiny e}|m} (\calC) \right] ^{1 / \rho}  > 4 \cdot \mathbf{\Delta}_{\mbox{\tiny w}} \right\} \\
&\overset{\mbox{\tiny MI}}{\leq} \frac{1}{4} + \frac{1}{4}, 
\end{align}
which means that there exists a code with both
\begin{align}
&\frac{1}{M_{z}M_{y}} \sum_{m=0}^{M_{z}-1}  \sum_{i=0}^{M_{y}-1} \left[ P_{\mbox{\tiny e}|mi} (\calC) \right] ^{1 / \rho}  \leq  4 \cdot \mathbf{\Delta}_{\mbox{\tiny s}} ,  \nonumber \\ 
&~~~~~~\frac{1}{M_{z}} \sum_{m=0}^{M_{z}-1} \left[ P_{\mbox{\tiny e}|m} (\calC) \right] ^{1 / \rho}  \leq 4 \cdot \mathbf{\Delta}_{\mbox{\tiny w}}.  
\end{align}
Now, for a given code $\calC$, index the message pairs $\{(m,i)\}$ according to decreasing order of $\{P_{\mbox{\tiny e}|mi} (\calC)\}$, and furthermore, index the clouds $\{m\}$ according to decreasing order of $\{P_{\mbox{\tiny e}|m} (\calC)\}$. Expurgate half of the pairs that have the highest scores. Then, among the remaining codewords one can find at least $M_{z}/2$ different clouds, such that each one of them still contains at least $M_{y}/2$ codewords. Afterwards, expurgate from the original codebook quarter of the clouds that have the highest scores. Intersecting those two remaining sets, at least $M_{z}/4$ values of $m$ are in common. Denote this good eighth by $\calC'$. Then, for this sub--code
\begin{align}
\left[ P_{\mbox{\tiny e}|mi} (\calC') \right] ^{1 / \rho}  &\leq  16 \cdot \mathbf{\Delta}_{\mbox{\tiny s}} ~~ \forall{(m,i)},  \nonumber \\ 
\left[ P_{\mbox{\tiny e}|m} (\calC') \right] ^{1 / \rho}  &\leq 16 \cdot \mathbf{\Delta}_{\mbox{\tiny w}}~~ \forall{m}.  
\end{align}
Thus, for the strong user,
\begin{align}
&  \max_{m,i} P_{\mbox{\tiny e}|mi} (\calC')   \nonumber \\
&\leq  \left[ \sum_{Q_{UXX'} \in \calS} \mathbb{E} \left( \left[ N_{mi}^{\mbox{\tiny IN}}(Q_{UXX'}, \calC) \right]^{1/ \rho} \middle| \bu_{m}, \bx_{mi}  \right) \exp \left\{-n \Upsilon (Q_{UXX'}, R_{y}, R_{z})/ \rho  \right\} \right.  \nonumber \\
&\left. ~~ + \sum_{Q_{UU'XX'} \in \calP} \mathbb{E} \left( \left[ N_{mi}^{\mbox{\tiny OUT}}(Q_{UU'XX'}, \calC) \right]^{1/ \rho} \middle| \bu_{m}, \bx_{mi}  \right) \exp \left\{-n \Omega (Q_{UU'XX'}, R_{y}, R_{z})/ \rho  \right\} \right]^{\rho} \\
&\doteq \max \left[ \max_{Q_{UXX'} \in \calS} \left\{ \mathbb{E} \left( \left[ N_{mi}^{\mbox{\tiny IN}}(Q_{UXX'}, \calC) \right]^{1/ \rho} \middle| \bu_{m}, \bx_{mi}  \right) \right\}^{\rho} \exp \left\{-n \Upsilon (Q_{UXX'}, R_{y}, R_{z}) \right\} ,   \right.  \nonumber \\
&\left.  \max_{Q_{UU'XX'} \in \calP} \left\{ \mathbb{E} \left( \left[ N_{mi}^{\mbox{\tiny OUT}}(Q_{UU'XX'}, \calC) \right]^{1/ \rho} \middle| \bu_{m}, \bx_{mi}  \right) \right\}^{\rho} \exp \left\{-n \Omega (Q_{UU'XX'}, R_{y}, R_{z}) \right\} \right] .
\end{align}  
Since it holds for every $\rho \geq 1$, the maximal probability of error can be bounded as
\begin{align}
&  \max_{m,i} P_{\mbox{\tiny e}|mi} (\calC')   \nonumber \\
&\leq  \inf_{\rho > 1}  \max \left[ \max_{Q_{UXX'} \in \calS} \left\{ \mathbb{E} \left( \left[ N_{mi}^{\mbox{\tiny IN}}(Q_{UXX'}, \calC) \right]^{1/ \rho} \middle| \bu_{m}, \bx_{mi}  \right) \right\}^{\rho} \exp \left\{-n \Upsilon (Q_{UXX'}, R_{y}, R_{z}) \right\} ,   \right.  \nonumber \\
&\left.  \max_{Q_{UU'XX'} \in \calP} \left\{ \mathbb{E} \left( \left[ N_{mi}^{\mbox{\tiny OUT}}(Q_{UU'XX'}, \calC) \right]^{1/ \rho} \middle| \bu_{m}, \bx_{mi}  \right) \right\}^{\rho} \exp \left\{-n \Omega (Q_{UU'XX'}, R_{y}, R_{z}) \right\} \right] \\
&\leq  \lim_{\rho \to \infty}  \max \left[ \max_{Q_{UXX'} \in \calS} \left\{ \mathbb{E} \left( \left[ N_{mi}^{\mbox{\tiny IN}}(Q_{UXX'}, \calC) \right]^{1/ \rho} \middle| \bu_{m}, \bx_{mi}  \right) \right\}^{\rho} \exp \left\{-n \Upsilon (Q_{UXX'}, R_{y}, R_{z}) \right\} ,   \right.  \nonumber \\
&\left.  \max_{Q_{UU'XX'} \in \calP} \left\{ \mathbb{E} \left( \left[ N_{mi}^{\mbox{\tiny OUT}}(Q_{UU'XX'}, \calC) \right]^{1/ \rho} \middle| \bu_{m}, \bx_{mi}  \right) \right\}^{\rho} \exp \left\{-n \Omega (Q_{UU'XX'}, R_{y}, R_{z}) \right\} \right] \\
\label{subs}
&=  \max \left[ \max_{Q_{UXX'} \in \calS} \lim_{\rho \to \infty} \left\{ \mathbb{E} \left( \left[ N_{mi}^{\mbox{\tiny IN}}(Q_{UXX'}, \calC) \right]^{1/ \rho} \middle| \bu_{m}, \bx_{mi}  \right) \right\}^{\rho} \exp \left\{-n \Upsilon (Q_{UXX'}, R_{y}, R_{z}) \right\} ,   \right.  \nonumber \\
&\left.  \max_{Q_{UU'XX'} \in \calP} \lim_{\rho \to \infty} \left\{ \mathbb{E} \left( \left[ N_{mi}^{\mbox{\tiny OUT}}(Q_{UU'XX'}, \calC) \right]^{1/ \rho} \middle| \bu_{m}, \bx_{mi}  \right) \right\}^{\rho} \exp \left\{-n \Omega (Q_{UU'XX'}, R_{y}, R_{z}) \right\} \right] .
\end{align}
It is easy to see that 
\begin{align}
&\lim_{\rho \to \infty} \left\{ \mathbb{E} \left( \left[ N_{mi}^{\mbox{\tiny IN}}(Q_{UXX'}, \calC) \right]^{1/ \rho} \middle| \bu_{m}, \bx_{mi}  \right) \right\}^{\rho} \\
&= \lim_{\rho \to \infty}   
           \left\{   
               \begin{array}{l l}
  \exp\{n[R_{y} - I_{Q}(X;X'|U)]\}   & \quad \text{  $R_{y} > I_{Q}(X;X'|U)$  }\\
  \exp\{n[R_{y} - I_{Q}(X;X'|U)] \cdot \rho \}   & \quad \text{  $R_{y} \leq I_{Q}(X;X'|U)$  } 
               \end{array} \right.  \\
\label{subs1}
&=      \left\{   
               \begin{array}{l l}
  \exp\{n[R_{y} - I_{Q}(X;X'|U)]\}   & \quad \text{  $R_{y} > I_{Q}(X;X'|U)$  }\\
  0                                                   & \quad \text{  $R_{y} \leq I_{Q}(X;X'|U)$  } 
               \end{array} \right. ,
\end{align}
and similarly,
\begin{align}
&\lim_{\rho \to \infty} \left\{ \mathbb{E} \left( \left[ N_{mi}^{\mbox{\tiny OUT}}(Q_{UU'XX'}, \calC) \right]^{1/ \rho} \middle| \bu_{m}, \bx_{mi}  \right) \right\}^{\rho} \\
&\leq \lim_{\rho \to \infty}  \left\{   
               \begin{array}{l l}
  \exp\left\{n  \left(  R_{z} + R_{y} -  I_{Q}(UX;U'X')  \right)   \right\}  \\
  \quad \text{~~~~~~~~ If ~~~  $ R_{z} + R_{y} > I_{Q}(UX;U'X'), ~R_{z} >  I_{Q}(UX;U')$  }   \\
  \exp\left\{n  \left(  R_{y} - I_{Q}(UX;X'|U') \right) -n \left[ I_{Q}(UX;U')  -  R_{z}  \right] \cdot \rho \right\}  \\
  \quad \text{~~~~~~~~ If ~~~  $ R_{y} > I_{Q}(UX;X'|U'), ~R_{z} \leq  I_{Q}(UX;U')$  }   \\
  \exp\left\{n \left( [ R_{z} -  I_{Q}(UX;U') ]_{+} + R_{y} - I_{Q}(UX;X'|U') - \left[ I_{Q}(UX;U')  -  R_{z}  \right]_{+} \right) \cdot \rho  \right\}  \\
  \quad \text{~~~~~~~~ If ~~~  $[ R_{z} -  I_{Q}(UX;U') ]_{+} + R_{y} \leq I_{Q}(UX;X'|U')$  }
               \end{array} \right. \\ 
\label{subs2}
&=  \left\{   
               \begin{array}{l l}
  \exp\left\{n  \left(  R_{z} + R_{y} -  I_{Q}(UX;U'X')  \right)   \right\}  & \quad \text{$ R_{z} + R_{y} > I_{Q}(UX;U'X'), ~R_{z} >  I_{Q}(UX;U')$  }\\
0      & \quad \text{elsewhere}  
               \end{array} \right. .
\end{align}
Substituting (\ref{subs1}) and (\ref{subs2}) back into (\ref{subs}) gives the desired result (\ref{THM3STRONG1})--(\ref{THM3STRONG2}). For the weak user, we have that
\begin{align}
\label{WEAKbound}
\max_{m} P_{\mbox{\tiny e}|m} (\calC')  \leq \left( \sum_{Q_{UU'XX'} \in \calP} \mathbb{E} \left\{ \left[ N_{\bu_{m}}(Q) \right] ^{1 / \rho}  \middle| \bu_{m} \right\}  e^ {-n [\Omega (Q, R_{y}, R_{z}) + R_{y}] /  \rho } \right)^{\rho}.
\end{align}
Since it holds for every $\rho \geq 1$, the maximal probability of error can be bounded as
\begin{align}
\max_{m} P_{\mbox{\tiny e}|m} (\calC')   
&\leq  \inf_{\rho > 1}  \left( \sum_{Q_{UU'XX'} \in \calP} \mathbb{E} \left\{ \left[ N_{\bu_{m}}(Q) \right] ^{1 / \rho} \middle| \bu_{m} \right\}  e^ {-n [\Omega (Q, R_{y}, R_{z}) + R_{y}] /  \rho } \right)^{\rho}  \\
&\doteq  \inf_{\rho > 1}  \left( \max_{Q_{UU'XX'} \in \calP} \mathbb{E} \left\{ \left[ N_{\bu_{m}}(Q) \right] ^{1 / \rho} \middle| \bu_{m} \right\}  e^ {-n [\Omega (Q, R_{y}, R_{z}) + R_{y}] /  \rho } \right)^{\rho}  \\
&=  \inf_{\rho > 1}  \max_{Q_{UU'XX'} \in \calP} \left( \mathbb{E} \left\{ \left[ N_{\bu_{m}}(Q) \right] ^{1 / \rho} \middle| \bu_{m} \right\} \right)^{\rho}  e^ {-n [\Omega (Q, R_{y}, R_{z}) + R_{y}]  }   \\
&\leq  \lim_{\rho \to \infty}  \max_{Q_{UU'XX'} \in \calP}  \left( \mathbb{E} \left\{ \left[ N_{\bu_{m}}(Q) \right] ^{1 / \rho} \middle| \bu_{m} \right\} \right)^{\rho}  e^ {-n [\Omega (Q, R_{y}, R_{z}) + R_{y}]  }   \\
&=    \max_{Q_{UU'XX'} \in \calP} \lim_{\rho \to \infty} \left( \mathbb{E} \left\{ \left[ N_{\bu_{m}}(Q) \right] ^{1 / \rho} \middle| \bu_{m} \right\} \right)^{\rho}  e^ {-n [\Omega (Q, R_{y}, R_{z}) + R_{y}]  }  .
\end{align}
Finally,
\begin{align}
& \lim_{\rho \to \infty} \left( \mathbb{E} \left\{ \left[ N_{\bu_{m}}(Q) \right]^{1 / \rho} \right\} \middle| \bu_{m}  \right)^{\rho} \nonumber \\ 
&\leq  \lim_{\rho \to \infty} \left\{   
               \begin{array}{l l}
  e^{n  \left(  2R_{y} +   R_{z}  -   I_{Q}(UX;U'X')   \right)     }  \\
  \quad \text{If $2R_{y} + R_{z}  -  I_{Q}(UX;U'X')  > 0, ~ R_{z}>I_{Q}(U;U')$  }   \\
  e^{n  \left(  2R_{y} -  I_{Q}(X;U'|U)  - I_{Q}(UX;X'|U') 
    \right)   - n\left[ I_{Q}(U;U')  -  R_{z}  \right] \cdot \rho  }  \\
  \quad \text{If $2R_{y} -  I_{Q}(X;U'|U)  -  I_{Q}(UX;X'|U')  > 0, ~ R_{z} \leq I_{Q}(U;U')$  }   \\
  e^{n  \left( [ R_{y} -  I_{Q}(X;U'|U) ]_{+} + R_{y} - I_{Q}(UX;X'|U') 
  + \rho \left( \mu^{*}  - \left[ I_{Q}(X;U'|U)  -  R_{y}  \right]_{+}  -  \left[ I_{Q}(U;U')  -  R_{z}  \right]_{+}\right)   \right)     }  \\
  \quad \text{If $[ R_{y} -  I_{Q}(X;U'|U) ]_{+} + R_{y} > I_{Q}(UX;X'|U'), ~\mu^{*}  - \left[ I_{Q}(X;U'|U)  -  R_{y}  \right]_{+} \leq 0$  }   \\
  e^{n \left(  2R_{y} -  I_{Q}(X;U'|U)  -  I_{Q}(UX;X'|U')  + \mu^{*} - \left[ I_{Q}(U;U')  -  R_{z}  \right]_{+}\right) \cdot \rho  }  \\
  \quad \text{If $[ R_{y} -  I_{Q}(X;U'|U) ]_{+} + R_{y} \leq I_{Q}(UX;X'|U'), ~2R_{y} -  I_{Q}(X;U'|U)  -  I_{Q}(UX;X'|U')  + \mu^{*} \leq 0$  }
               \end{array} \right.  \\
&=   \left\{   
               \begin{array}{l l}
 e^{ n  \left(  2R_{y} +   R_{z}  -   I_{Q}(UX;U'X')   \right)  }  &
  \quad \text{$   I_{Q}(UX;U'X')  < 2R_{y} + R_{z}, ~ I_{Q}(U;U')<R_{z}$  }   \\
  0  &
  \quad \text{elsewhere}   \\
               \end{array} \right. ,
\end{align} 
and so,
\begin{align}
&\max_{m} P_{\mbox{\tiny e}|m} (\calC') \nonumber \\
&\leq   \max_{Q_{UU'XX'} \in \calP:~I_{Q}(UX;U'X')  < 2R_{y} + R_{z}, I_{Q}(U;U')<R_{z}}  
 e^ {-n [I_{Q}(UX;U'X') + \Omega (Q, R_{y}, R_{z}) - R_{y} - R_{z}]  }   \\
&=  \exp \left\{-n \min_{\substack{Q_{UU'XX'} \in \calP \\ I_{Q}(UX;U'X')  < 2R_{y} + R_{z}, \\ I_{Q}(U;U')<R_{z} }}  [I_{Q}(UX;U'X') + \Omega (Q, R_{y}, R_{z}) - R_{y} - R_{z}]  \right\}  ,
\end{align}
which is (\ref{THM3WEAK}). The proof of Theorem 3 is now complete.

\section*{Appendix A}
\renewcommand{\theequation}{A.\arabic{equation}}
    \setcounter{equation}{0}
\subsection*{Comparison Between $E_{\mbox{\tiny wu}}^{\mbox{\tiny GLD}}(R_{y},R_{z})$ and $E_{\mbox{\tiny wu}}^{\mbox{\tiny ML1}}(R_{y},R_{z})$}
In order to compare between the expurgated exponents, let us first compare between $D_{t}(Q_{XX'})$ [defined in (\ref{Ddefinition})] and $\Omega (Q_{UU'XX'}, R_{y}, R_{z})$ [defined in (\ref{OMEGAdefinition})]. Recall that   
\begin{align}
&\exp \Big\{-n \Omega (Q_{UU'XX'}, R_{y}, R_{z})  \Big\} \\
&\doteq \sum_{\bz} W(\bz | \bx) \exp \bigg\{-n 
\Big[ \max \big\{ g( \hat{P}_{\bu \bx \bz }), \phi(R_{y}, \hat{P}_{\bu \bz}), \psi(R_{y}, R_{z}, \hat{P}_{\bz})  \big\} - g( \hat{P}_{\bu' \bx' \bz } ) \Big]_{+}  \bigg\} \nonumber \\
&\doteq \sum_{\bz} W(\bz | \bx)  \cdot  \min \left\{1, \frac{ \exp \{n g( \hat{P}_{\bu' \bx' \bz } ) \} } {  \exp \{n g( \hat{P}_{\bu \bx \bz } )  \} + \exp \{n \phi(R_{y}, \hat{P}_{\bu \bz})  \} + \exp \{n \psi(R_{y}, R_{z}, \hat{P}_{\bz}) \} } \right\},
\end{align}
and,
\begin{align}
\exp \{-n D_{t}(Q_{XX'}) \} \doteq \sum_{\bz} W(\bz | \bx) \cdot   \left[ \frac {W(\bz| \bx' )}{W(\bz| \bx)} \right]^{t}, ~~~~~ t \in [0,1],  
\end{align}
which can also be written using the same notation as
\begin{align}
\exp \{-n D_{t}(Q_{XX'}) \} \doteq \sum_{\bz} W(\bz | \bx)  \cdot   \left[ \frac { e^ {n g( \hat{P}_{\bu' \bx' \bz } ) } }{ e^ {n g( \hat{P}_{\bu \bx \bz } ) } } \right]^{t}, ~~~~~ t \in [0,1].  
\end{align}
It is easy to argue that for $t \in [0,1]$:
\begin{align}
\label{CompareGLDtoML}
&\min \left\{1, \frac{ \exp \{n g( \hat{P}_{\bu' \bx' \bz } ) \} } {  \exp \{n g( \hat{P}_{\bu \bx \bz } )  \} + \exp \{n \phi(R_{y}, \hat{P}_{\bu \bz})  \} + \exp \{n \psi(R_{y}, R_{z}, \hat{P}_{\bz}) \} } \right\} \nonumber \\ 
&\leq  \left[ \frac { e^ {n g( \hat{P}_{\bu' \bx' \bz } ) } }{ e^ {n g( \hat{P}_{\bu \bx \bz } ) } } \right]^{t} .
\end{align}
To see why this is true, let us distinguish between the cases $g( \hat{P}_{\bu' \bx' \bz } ) \leq g( \hat{P}_{\bu \bx \bz } )$ and $g( \hat{P}_{\bu' \bx' \bz } ) > g( \hat{P}_{\bu \bx \bz } )$. In the former case,
\begin{align}
&\min \left\{1, \frac{ \exp \{n g( \hat{P}_{\bu' \bx' \bz } ) \} } {  \exp \{n g( \hat{P}_{\bu \bx \bz } )  \} + \exp \{n \phi(R_{y}, \hat{P}_{\bu \bz})  \} + \exp \{n \psi(R_{y}, R_{z}, \hat{P}_{\bz}) \} } \right\} \nonumber \\ 
&\leq  \frac { \exp \{n g( \hat{P}_{\bu' \bx' \bz } ) \} }{ \exp\{n g( \hat{P}_{\bu \bx \bz } ) \} }  \\
&\leq  \left[ \frac { \exp \{n g( \hat{P}_{\bu' \bx' \bz } ) \} }{ \exp\{n g( \hat{P}_{\bu \bx \bz } ) \} } \right]^{t} .
\end{align}
In the latter case, the right-hand side of (\ref{CompareGLDtoML}) exceeds unity, whereas the left-hand side is always less then unity. 
The conclusion from this observation is that at least for the choice $g(Q) = \sum_{u,x,z} Q(u,x,z) \log W(z|x)$, the inequality $\Omega (Q_{UU'XX'}, R_{y}, R_{z}) \geq D_{t}(Q_{XX'})$ holds for any $t \in [0,1]$. Now, we have the following
\begin{align}
&E_{\mbox{\tiny wu}}^{\mbox{\tiny ML1}}  (R_{y},R_{z}) \nonumber \\
&= \max_{0 \leq t \leq 1}
\min_{\substack{Q_{UU'XX'} \in \calP \\ I_{Q}(UX;U'X')  < 2R_{y} + R_{z} }}
[ I_{Q}(UX;U'X') + D_{t}(Q_{XX'}) ] - R_{y}- R_{z} \\
&\leq \min_{\substack{Q_{UU'XX'} \in \calP \\ I_{Q}(UX;U'X')  < 2R_{y} + R_{z} }}
[ I_{Q}(UX;U'X') + \Omega (Q_{UU'XX'}, R_{y}, R_{z}) ] - R_{y}- R_{z} \\
&\leq \min_{\substack{Q_{UU'XX'} \in \calP \\ I_{Q}(UX;U'X')  < 2R_{y} + R_{z} \\ I_{Q}(U;U')  < R_{z} }} [ I_{Q}(UX;U'X') + \Omega (Q_{UU'XX'}, R_{y}, R_{z}) ] - R_{y}- R_{z} \\
&= E_{\mbox{\tiny wu}}^{\mbox{\tiny GLD}}  (R_{y},R_{z}),
\end{align}
hence, the GLD--based expurgated bound, $E_{\mbox{\tiny wu}}^{\mbox{\tiny GLD}}  (R_{y},R_{z})$, is at least as tight as the ML--based expurgated bound of Theorem 1.

\section*{Appendix B}
\renewcommand{\theequation}{B.\arabic{equation}}
    \setcounter{equation}{0}
\subsection*{Comparison Between $E_{\mbox{\tiny wu}}^{\mbox{\tiny GLD}}(R_{y},R_{z})$ and $E_{\mbox{\tiny w}}(R_{y},R_{z})$}
The expurgated exponent of the weak user under the GLD is given by
\begin{align}
\label{ExGLD}
E_{\mbox{\tiny wu}}^{\mbox{\tiny GLD}}  (R_{y},R_{z})
&= \min_{Q_{UU'XX'} \in \tilde{\calP}} [I_{Q}(UX;U'X') + \Omega (Q, R_{y}, R_{z}) ] - R_{y} - R_{z} ,
\end{align}
where,
\begin{align}
\tilde{\calP} \overset{\Delta}{=} \{Q_{UU'XX'} :Q \in \calP,~I_{Q}(UX;U'X')  < 2R_{y} + R_{z},~ I_{Q}(U;U')<R_{z}  \}. 
\end{align}
Let us choose $g(Q) = I_{Q}(UX;Z)$. We then get
\begin{align}
\phi(R_{y}, Q_{UZ})
&= \max_{\{Q_{X|UZ}:~ I_{Q}(X;Z|U) \leq R_{y} \}} [g(Q)  - I_{Q}(X;Z|U)] + R_{y} 
=  I_{Q}(U;Z) + R_{y} ,
\end{align}
and
\begin{align}
\psi(R_{y}, R_{z}, Q_{Z})
&= \max_{\{Q_{UX|Z}:~ I_{Q}(U;Z) \leq R_{z} , ~ I_{Q}(UX;Z)  \leq R_{z} + R_{y} \}} [ g(Q)  - I_{Q}(UX;Z) ] + R_{z} + R_{y} \\
&=   R_{z} + R_{y} ,
\end{align}
and furthermore
\begin{align}
\Omega (Q_{UU'XX'}, R_{y}, R_{z})
&= \min_{Q_{Z|UU'XX'}} \bigg( D(Q_{Z|UX} \| W_{Z|X} | Q_{UX}) + I_{Q}(U'X';Z|UX) 
\nonumber \\ & ~+ \Big[ \max \big\{ g( Q_{UXZ} ), \phi(R_{y}, Q_{UZ}), \psi(R_{y}, R_{z}, Q_{Z})  \big\} - g( Q_{U'X'Z} ) \Big]_{+}  \bigg) \\
&= \min_{Q_{Z|UU'XX'}} \bigg( D(Q_{Z|UX} \| W_{Z|X} | Q_{UX}) + I_{Q}(U'X';Z|UX) 
\nonumber \\ & ~+ \Big[ \max \big\{ I_{Q}(UX;Z), I_{Q}(U;Z) + R_{y}, R_{z} + R_{y}  \big\} - I_{Q}(U'X';Z)  \Big]_{+}  \bigg).   
\end{align}
Define
\begin{align}
\tilde{\calP}_{z} &\overset{\Delta}{=} \{Q_{UU'XX'Z} :Q \in \calP,~I_{Q}(UX;U'X')  < 2R_{y} + R_{z}, I_{Q}(U;U')<R_{z}  \}, \\ 
\hat{\calP}_{z} &\overset{\Delta}{=} \{Q_{U'X'|UXZ} :Q \in \calP,~I_{Q}(UX;U'X')  < 2R_{y} + R_{z}, ~I_{Q}(U;U')<R_{z} \}.
\end{align}
Substituting into (\ref{ExGLD}) gives
\begin{align}
&E_{\mbox{\tiny wu}}^{\mbox{\tiny GLD}}  (R_{y},R_{z}) \nonumber \\
&\geq  \min_{Q_{UU'XX'} \in \tilde{\calP}} \Bigg\{ \min_{Q_{Z|UU'XX'}} \bigg( D(Q_{Z|UX} \| W_{Z|X} | Q_{UX}) + I_{Q}(U'X';Z|UX) 
\nonumber \\ &~~~~~+ \Big[ \max \big\{ I_{Q}(UX;Z), I_{Q}(U;Z) + R_{y}, R_{z} + R_{y}  \big\} - I_{Q}(U'X';Z)  \Big]_{+}  \bigg) + I_{Q}(UX;U'X')  \Bigg\} \nonumber \\
&~~~~~~~~~~~~~~~~~~~~~~~~~~~~~~~~
~~~~~~~~~~~~~~~~~~~~~~~~~~~~~~~~~~
~~~~~~~~~~~~~~~~~~~~~~- R_{y} - R_{z} \\
&=  \min_{Q_{UU'XX'Z} \in \tilde{\calP}_{z}} \Bigg\{ D(Q_{Z|UX} \| W_{Z|X} | Q_{UX}) + I_{Q}(U'X';UXZ)  \nonumber \\ &~~~~~+ \Big[ \max \big\{ I_{Q}(UX;Z), I_{Q}(U;Z) + R_{y}, R_{z} + R_{y}  \big\} - I_{Q}(U'X';Z)  \Big]_{+}   \Bigg\} - R_{y} - R_{z} \\
&=  \min_{Q_{UU'XX'Z} \in \tilde{\calP}_{z}} \Bigg\{ D(Q_{Z|UX} \| W_{Z|X} | Q_{UX}) + I_{Q}(U'X';UX|Z) + I_{Q}(U'X';Z)  \nonumber \\ &~~~~~+ \Big[ \max \big\{ I_{Q}(UX;Z), I_{Q}(U;Z) + R_{y}, R_{z} + R_{y}  \big\} - I_{Q}(U'X';Z)  \Big]_{+}   \Bigg\} - R_{y} - R_{z} \\
&=  \min_{Q_{UU'XX'Z} \in \tilde{\calP}_{z}} \Bigg\{ D(Q_{Z|UX} \| W_{Z|X} | Q_{UX}) + I_{Q}(U'X';UX|Z)   \nonumber \\ &~~~~~+  \max \big\{ I_{Q}(UX;Z), I_{Q}(U'X';Z), I_{Q}(U;Z) + R_{y}, R_{z} + R_{y}  \big\}   \Bigg\} - R_{y} - R_{z} \\
&=  \min_{Q_{Z|UX} } \min_{Q_{U'X'|UXZ} \in \hat{\calP}_{z}} \Bigg\{ D(Q_{Z|UX} \| W_{Z|X} | Q_{UX}) + I_{Q}(U'X';UX|Z)   \nonumber \\ &~~~~~+  \max \big\{ I_{Q}(UX;Z), I_{Q}(U'X';Z), I_{Q}(U;Z) + R_{y}, R_{z} + R_{y}  \big\}   \Bigg\} - R_{y} - R_{z} \\
\label{infimum1}
&=  \min_{Q_{Z|UX} }  \Bigg\{ D(Q_{Z|UX} \| W_{Z|X} | Q_{UX}) + \min_{Q_{U'X'|UXZ} \in \hat{\calP}_{z}} \bigg( I_{Q}(U'X';UX|Z)   \nonumber \\ &~~~~~+  \max \big\{ I_{Q}(UX;Z), I_{Q}(U'X';Z), I_{Q}(U;Z) + R_{y}, R_{z} + R_{y}  \big\} \bigg)   \Bigg\} - R_{y} - R_{z} \\
&=  \min_{Q_{Z|UX} }  \Bigg\{ D(Q_{Z|UX} \| W_{Z|X} | Q_{UX})    \nonumber \\ 
&~~~~~+  \max \big\{ I_{Q}(U;Z) + I_{Q}(X;Z|U), I_{Q}(U;Z) + R_{y}, R_{z} + R_{y}  \big\}    \Bigg\} - R_{y} - R_{z} \\
&=  \min_{Q_{Z|UX} }  \Big\{ D(Q_{Z|UX} \| W_{Z|X} | Q_{UX})    \nonumber \\ 
&~~~~~+  \max \big\{ I_{Q}(U;Z) + I_{Q}(X;Z|U) - R_{y} - R_{z}, I_{Q}(U;Z) - R_{z}, 0  \big\}    \Big\}  \\
&=  \min_{Q_{Z|UX} }  \Big\{ D(Q_{Z|UX} \| W_{Z|X} | Q_{UX})    \nonumber \\ 
&~~~~~+  \left[ \max \big\{ I_{Q}(U;Z) + I_{Q}(X;Z|U) - R_{y} - R_{z}, I_{Q}(U;Z) - R_{z} \big\} \right]_{+}    \Big\}  \\
&=  \min_{Q_{Z|UX} }  \Big\{ D(Q_{Z|UX} \| W_{Z|X} | Q_{UX})    
+  \left[ I_{Q}(U;Z) + \max \big\{  I_{Q}(X;Z|U) - R_{y} , 0  \big\} - R_{z} \right]_{+}    \Big\}  \\
&=  \min_{Q_{Z|UX} }  \Big\{ D(Q_{Z|UX} \| W_{Z|X} | Q_{UX})    
+  \left[ I_{Q}(U;Z) + \left[ I_{Q}(X;Z|U) - R_{y} \right]_{+} - R_{z} \right]_{+}    \Big\}  \\
&= E_{\mbox{\tiny w}}(R_{y},R_{z}),
\end{align}
where the first inequality is because the metric $g(Q) = I_{Q}(UX;Z)$ may be sub--optimal, and where the inner minimum of (\ref{infimum1}) is attained for a random variable $(U',X')$ which is independent of $(U,X,Z)$, for which both $I_{Q}(U'X';UX|Z) = 0$ and $I_{Q}(U'X';Z) = 0$, and the constraints $I_{Q}(U;U') \leq R_{z}$ and $I_{Q}(UX;U'X') \leq 2R_{y}+R_{z}$ are obviously satisfied since $I_{Q}(U;U') = 0$ and $I_{Q}(UX;U'X') = 0$.

\section*{Appendix C}
\renewcommand{\theequation}{C.\arabic{equation}}
    \setcounter{equation}{0}
\subsection*{Proving that $\mathrm{Pr} \{\calS_{\epsilon}(m,i,\bu_{m},\by) \}$ decays double-exponentially fast}
Let $N_{\bu_{m},\by}(Q)$ denote the number of codewords $\bx_{mj} \in \mathcal{C}_{m}$, such that the joint empirical distribution of  $\bx_{mj}$ with $\bu_{m}$ and $\by$ is $Q_{UXY}$, that is
\begin{align}
N_{\bu_{m},\by}(Q) = \sum_{j \neq i}  
 \mathcal{I} \Big\{ (\by, \bu_{m}, \bx_{mj} ) \in \mathcal{T}(Q_{UXY}) \Big\}.
\end{align}
First, note that
\begin{align}
\Phi_{m,i}(\bu_{m},\by) = \sum_{j \neq i} \exp \{n g( \hat{P}_{\bu_{m} \bx_{mj} \by } ) \} = \sum_{Q} N_{\bu_{m},\by}(Q) e^{ng(Q)} .
\end{align}
Thus, taking the randomness of $\{\bX_{mj}\}_{j=0}^{M_{y}-1}$ into account,
\begin{align}
&\mathrm{Pr} \Big\{ \Phi_{m,i}(\bu_{m},\by) \leq \exp \{n \phi(R_{y}-\epsilon, \hat{P}_{\bu_{m}\by}) \}   \Big\} \\
&=  \mathrm{Pr} \Bigg\{ \sum_{Q} N_{\bu_{m},\by}(Q) e^{ng(Q)} \leq \exp \{n \phi(R_{y}-\epsilon, \hat{P}_{\bu_{m}\by}) \}   \Bigg\} \\
&\leq  \mathrm{Pr} \bigg\{ \max_{Q} N_{\bu_{m},\by}(Q) e^{ng(Q)} \leq \exp \{n \phi(R_{y}-\epsilon, \hat{P}_{\bu_{m}\by}) \}   \bigg\} \\
&=  \mathrm{Pr} \bigcap_{Q} \Big\{ N_{\bu_{m},\by}(Q) e^{ng(Q)} \leq \exp \{n \phi(R_{y}-\epsilon, \hat{P}_{\bu_{m}\by}) \}   \Big\} \\
&=  \mathrm{Pr} \bigcap_{Q} \Big\{ N_{\bu_{m},\by}(Q)  \leq \exp \{n [ \phi(R_{y}-\epsilon, \hat{P}_{\bu_{m}\by}) - g(Q) ] \}   \Big\} .
\end{align}
Now, $N_{\bu_{m},\by}(Q)$ is a binomial random variable with $e^{nR_{y}}$ trials and success rate which is of the exponential order of $e^{-nI_{Q}(X;Y|U)}$. We prove that by the very definition of the function $\phi(R_{y}-\epsilon, \hat{P}_{\bu_{m}\by})$, there must exist some conditional distribution $Q_{X|UY}^{*}$ such that for $Q^{*} = \hat{P}_{\bu_{m}\by} \times Q_{X|UY}^{*}$, the two inequalities $I_{Q^{*}}(X;Y|U) \leq R_{y}-\epsilon$ and $R_{y}-\epsilon - I_{Q^{*}}(X;Y|U) \geq \phi(R_{y}-\epsilon, \hat{P}_{\bu_{m}\by}) - g(Q^{*})$ hold. To show that, we assume conversely, i.e., that for every conditional distribution $Q_{X|UY}$, which defines $Q = \hat{P}_{\bu_{m}\by} \times Q_{X|UY}$, either $I_{Q}(X;Y|U) > R_{y}-\epsilon$ or $R_{y}- I_{Q}(X;Y|U)-\epsilon < \phi(R_{y}-\epsilon, \hat{P}_{\bu_{m}\by}) - g(Q)$, which means that for every distribution $Q$
\begin{align}
R_{y} - \epsilon &< \max \{I_{Q}(X;Y|U),  I_{Q}(X;Y|U) + \phi(R_{y}-\epsilon, \hat{P}_{\bu_{m}\by}) - g(Q) \}  \\
&= I_{Q}(X;Y|U) + [\phi(R_{y}-\epsilon, \hat{P}_{\bu_{m}\by}) - g(Q)]_{+}.
\end{align}
Writing it slightly differently, for every $Q_{X|UY}$ there exists some real number $t \in [0,1]$ such that
\begin{align}
R_{y} - \epsilon &< I_{Q}(X;Y|U) + t[\phi(R_{y}-\epsilon, \hat{P}_{\bu_{m}\by}) - g(Q)],
\end{align}
or equivalently, 
\begin{align}
\phi(R_{y}-\epsilon, \hat{P}_{\bu_{m}\by}) &>
\max_{Q_{X|UY}} \min_{t \in [0,1]}  g(Q) + \frac{R_{y} - I_{Q}(X;Y|U) - \epsilon}{t} \\
&= \max_{Q_{X|UY}} \left\{ 
               \begin{array}{l l}
                 g(Q) + R_{y} - I_{Q}(X;Y|U) - \epsilon                 & \quad \text{  $I_{Q}(X;Y|U) \leq R_{y} - \epsilon$  }\\
                 -\infty                & \quad \text{ $I_{Q}(X;Y|U) > R_{y} - \epsilon$  } 
               \end{array} \right. \\
&= \max_{\{Q_{X|UY}:~ I_{Q}(X;Y|U) \leq R_{y} - \epsilon\}} [g(Q)  - I_{Q}(X;Y|U)] + R_{y} - \epsilon \\
&\equiv \phi(R_{y}-\epsilon, \hat{P}_{\bu_{m}\by}),
\end{align}
which is a contradiction. Let the conditional distribution $Q_{X|UY}^{*}$ be as defined above. Then,
\begin{align}
&\mathrm{Pr} \bigcap_{Q} \Big\{ N_{\bu_{m},\by}(Q)  \leq \exp \{n [ \phi(R_{y}-\epsilon, \hat{P}_{\bu_{m}\by}) - g(Q) ] \}   \Big\} \\
\label{up7}
&\leq \mathrm{Pr} \Big\{ N_{\bu_{m},\by}(Q^{*})  \leq \exp \{n [ \phi(R_{y}-\epsilon, \hat{P}_{\bu_{m}\by}) - g(Q^{*}) ] \}   \Big\}.
\end{align}
Now, we know that both of the inequalities $I_{Q^{*}}(X;Y|U) \leq R_{y}-\epsilon$ and $R_{y} - I_{Q^{*}}(X;Y|U) - \epsilon  \geq \phi(R_{y}-\epsilon, \hat{P}_{\bu_{m}\by}) - g(Q^{*})$ hold. By the Chernoff bound, the probability of (\ref{up7}) is upper bounded by
\begin{align}
\exp \Big\{ -e^{nR_{y}} D( e^{-an} \| e^{-bn} ) \Big\},
\end{align}
where $a = R_{y} + g(Q^{*}) - \phi(R_{y}-\epsilon, \hat{P}_{\bu_{m}\by})$ and $b = I_{Q^{*}}(X;Y|U)$, and where $D( \alpha \| \beta )$, for $\alpha,\beta \in [0,1]$, is the binary divergence function, that is 
\begin{align}
D( \alpha \| \beta ) = \alpha \log \frac{\alpha}{\beta} + (1-\alpha) \log \frac{1-\alpha}{1-\beta}.  
\end{align}
Since $a - b \geq \epsilon$, the binary divergence is lower bounded as follows (\cite[Section 6.3]{MERHAV09}):
\begin{align}
D( e^{-an} \| e^{-bn} ) &\geq e^{-bn} \left\{ 1- e^{-(a-b)n} [1 + n(a-b)]  \right\} \\
&\geq e^{-n I_{Q^{*}}(X;Y|U)} [ 1- e^{-n \epsilon} (1 + n \epsilon)  ] ,
\end{align}
where in the second inequality, we invoked the decreasing monotonicity of the function $f(t) = (1+t) e^{-t}$ for $t \geq 0$. Finally, we get that
\begin{align}
& \mathrm{Pr} \Big\{ N_{\bu_{m},\by}(Q^{*})  \leq \exp \{n [ \phi(R_{y}-\epsilon, \hat{P}_{\bu_{m}\by}) - g(Q^{*}) ] \}   \Big\} \\
&\leq \exp \Big\{ -e^{nR_{y}} \cdot  e^{-n I_{Q^{*}}(X;Y|U)} [ 1- e^{-n \epsilon} (1 + n \epsilon)  ]  \Big\}  \\
&\leq \exp \big\{ -e^{n \epsilon}  [ 1- e^{-n \epsilon} (1 + n \epsilon)  ]  \big\} \\
&= \exp \big\{ -e^{n \epsilon}  + n \epsilon + 1  \big\} .
\end{align}

\section*{Appendix D}
\renewcommand{\theequation}{D.\arabic{equation}}
    \setcounter{equation}{0}
\subsection*{Proving that $\mathrm{Pr} \{\calK_{\epsilon}(m,\by) \}$ decays double-exponentially fast}
Let $\tilde{N}_{\by}(Q_{UY})$ denote the number of cloud-centers $\bu_{m'}$, $m' \neq m$, such that the joint empirical distribution of  $\bu_{m'}$ with $\by$ is $Q_{UY}$, that is
\begin{align}
\tilde{N}_{\by}(Q_{UY}) = \sum_{m' \neq m}  \mathcal{I} \Big\{ (\by, \bu_{m'} ) \in \mathcal{T}(Q_{UY}) \Big\}.
\end{align} 
In addition, let $N_{\by}(Q)$ denote the number of codewords $\bx_{m'j} \in \mathcal{C}_{m'}$, for any $m' \neq m$ for which $(\by, \bu_{m'}) \in \mathcal{T}(Q_{UY})$, such that the joint empirical distribution of  $\bx_{m'j}$ with $\bu_{m'}$  and  $\by$ is $Q_{UXY}$, that is
\begin{align}
N_{\by}(Q) = \sum_{m' \neq m} \sum_{j=0}^{M_{y}-1} 
 \mathcal{I} \Big\{ (\by, \bu_{m'}, \bx_{m'j} ) \in \mathcal{T}(Q_{UXY}) \Big\}.
\end{align}
As before, we start by observing that
\begin{align}
\Psi_{m}(\by) =  \sum_{m' \neq m} \sum_{j=0}^{M_{y}-1} \exp \{n g( \hat{P}_{\bu_{m'} \bx_{m'j} \by } ) \}  = \sum_{Q_{U|Y}} \sum_{Q_{X|UY}}  N_{\by}(Q) e^{ng(Q)} .
\end{align}
Thus, considering the randomness of the cloud--centers $\{\bU_{m'}\}$ and the codewords $\{\bX_{m'j}\}$,
\begin{align}
&\mathrm{Pr} \Big\{ \Psi_{m}(\by) \leq \exp \{n \psi(R_{y}-\epsilon, R_{z}-\epsilon, \hat{P}_{\by}) \}   \Big\} \\
&=  \mathrm{Pr} \Bigg\{ \sum_{Q_{U|Y}} \sum_{Q_{X|UY}}  N_{\by}(Q)  e^{ng(Q)} \leq \exp \{n \psi(R_{y}-\epsilon, R_{z}-\epsilon, \hat{P}_{\by}) \}   \Bigg\} \\
&\leq  \mathrm{Pr} \bigg\{ \max_{Q_{U|Y}} \max_{Q_{X|UY}}  N_{\by}(Q) e^{ng(Q)} \leq \exp \{n \psi(R_{y}-\epsilon, R_{z}-\epsilon, \hat{P}_{\by}) \}   \bigg\} \\
&=  \mathrm{Pr} \bigcap_{Q_{U|Y}} \bigcap_{Q_{X|UY}}  \Big\{ N_{\by}(Q) e^{ng(Q)} \leq \exp \{n \psi(R_{y}-\epsilon, R_{z}-\epsilon, \hat{P}_{\by}) \}   \Big\} \\
&=  \mathrm{Pr} \bigcap_{Q_{U|Y}} \bigcap_{Q_{X|UY}} \Big\{ N_{\by}(Q)  \leq \exp \{n [ \psi(R_{y}-\epsilon, R_{z}-\epsilon, \hat{P}_{\by}) - g(Q) ] \}   \Big\} .
\end{align}

We now show that by the very definition of the function $\psi(R_{y}-\epsilon, R_{z}-\epsilon, \hat{P}_{\by})$, there must exist some conditional distributions $Q_{U|Y}^{*}$ and $Q_{X|UY}^{*}$ such that for $Q^{*} = \hat{P}_{\by} \times Q_{U|Y}^{*} \times Q_{X|UY}^{*}$, the three inequalities $I_{Q^{*}}(U;Y) \leq R_{z}-\epsilon$, $I_{Q^{*}}(U;Y) + I_{Q^{*}}(X;Y|U) \leq R_{z} + R_{y}-2\epsilon$ and $R_{z} + R_{y} - I_{Q^{*}}(U;Y) - I_{Q^{*}}(X;Y|U) - 2\epsilon \geq \psi(R_{y}-\epsilon, R_{z}-\epsilon, \hat{P}_{\by}) - g(Q^{*})$ hold. Assume conversely, that for every $Q_{U|Y}$ and every $Q_{X|UY}$, which defines $Q = \hat{P}_{\by} \times Q_{U|Y} \times Q_{X|UY}$, 
either $I_{Q}(U;Y) > R_{z}-\epsilon$ 
or $I_{Q}(U;Y) + I_{Q}(X;Y|U) > R_{z} + R_{y}-2\epsilon$ 
or $R_{z} + R_{y} - I_{Q}(U;Y) - I_{Q}(X;Y|U) - 2\epsilon < \psi(R_{y}-\epsilon, R_{z}-\epsilon, \hat{P}_{\by}) - g(Q)$, which means that for every $Q$
\begin{align}
&R_{z} - \epsilon   \\
 &< \max \Big\{ I_{Q}(U;Y), I_{Q}(U;Y) + I_{Q}(X;Y|U) - R_{y}+\epsilon, \nonumber \\  &~~~~~~~~~~~~ I_{Q}(U;Y) + I_{Q}(X;Y|U) - R_{y}+\epsilon + \psi(R_{y}-\epsilon, R_{z}-\epsilon, \hat{P}_{\by}) - g(Q) \Big\}  \\
&= I_{Q}(U;Y)  +  \max \Big\{ 0 , I_{Q}(X;Y|U) - R_{y}+\epsilon, \nonumber \\ &~~~~~~~~~~~~~~~~~~~~~~~~~~~~ I_{Q}(X;Y|U) - R_{y}+\epsilon + \psi(R_{y}-\epsilon, R_{z}-\epsilon, \hat{P}_{\by}) - g(Q) \Big\}  \\
&= I_{Q}(U;Y) + \Big[ \max \Big\{  I_{Q}(X;Y|U) - R_{y}+\epsilon, \nonumber \\ &~~~~~~~~~~~~~~~~~~~~~~~~~~~~~  I_{Q}(X;Y|U) - R_{y}+\epsilon + \psi(R_{y}-\epsilon, R_{z}-\epsilon, \hat{P}_{\by}) - g(Q) \Big\} \Big]_{+}  \\
&= I_{Q}(U;Y) + \Big[ I_{Q}(X;Y|U) - R_{y}+\epsilon + \max \Big\{ 0 , \psi(R_{y}-\epsilon, R_{z}-\epsilon, \hat{P}_{\by}) - g(Q) \Big\} \Big]_{+}  \\
&= I_{Q}(U;Y) + \Big[ I_{Q}(X;Y|U) - R_{y}+\epsilon + \big[ \psi(R_{y}-\epsilon, R_{z}-\epsilon, \hat{P}_{\by}) - g(Q) \big]_{+} \Big]_{+}
\end{align}
or, in other words, that for every $Q_{U|Y}$ and $Q_{X|UY}$ there exists $(t,s) \in [0,1]^{2}$ such that
\begin{align}
R_{z} - \epsilon &< I_{Q}(U;Y) + t\Big[ I_{Q}(X;Y|U) - R_{y}+\epsilon + s\big[ \psi(R_{y}-\epsilon, R_{z}-\epsilon, \hat{P}_{\by}) - g(Q) \big] \Big],
\end{align}
or equivalently, 
\begin{align}
& \psi(R_{y}-\epsilon, R_{z}-\epsilon, \hat{P}_{\by}) \\ 
&> \max_{Q_{U|Y}} \max_{Q_{X|UY}} \min_{(t,s) \in [0,1]^{2}}  g(Q) + \frac{1}{s} \Bigg( \frac{R_{z} - I_{Q}(U;Y) - \epsilon}{t} + R_{y} - I_{Q}(X;Y|U) - \epsilon \Bigg)  \\
&= \max_{\{Q_{U|Y}:~ I_{Q}(U;Y) \leq R_{z} - \epsilon\}} \max_{Q_{X|UY}} \min_{s \in [0,1]}  g(Q) + \frac{1}{s} \Big( R_{z} + R_{y} - I_{Q}(UX;Y)  - 2\epsilon \Big)  \\
&= \max_{\{Q_{U|Y}:~ I_{Q}(U;Y) \leq R_{z} - \epsilon\}} \max_{\{Q_{X|UY}:~ I_{Q}(UX;Y)  \leq R_{z} + R_{y} - 2\epsilon\}} [ g(Q)  - I_{Q}(UX;Y) ] + R_{z} + R_{y} - 2\epsilon   \\
&= \max_{\{Q_{UX|Y}:~ I_{Q}(U;Y) \leq R_{z} - \epsilon, ~ I_{Q}(UX;Y)  \leq R_{z} + R_{y} - 2\epsilon\}} [ g(Q)  - I_{Q}(UX;Y) ] + R_{z} + R_{y} - 2\epsilon   \\
&\equiv \psi(R_{y}-\epsilon, R_{z}-\epsilon, \hat{P}_{\by}),
\end{align}
which is a contradiction. Let then $Q_{U|Y}^{*}$ and $Q_{X|UY}^{*}$ be two conditional distributions as defined above. Then,
\begin{align}
&\mathrm{Pr} \bigcap_{Q_{U|Y}} \bigcap_{Q_{X|UY}} \Big\{ N_{\by}(Q)  \leq \exp \{n [ \psi(R_{y}-\epsilon, R_{z}-\epsilon, \hat{P}_{\by}) - g(Q) ] \}   \Big\} \\
\label{up8}
&\leq \mathrm{Pr} \Big\{ N_{\by}(Q^{*})  \leq \exp \{n [ \psi(R_{y}-\epsilon, R_{z}-\epsilon, \hat{P}_{\by}) - g(Q^{*}) ] \}   \Big\}.
\end{align}
Now, remember that $I_{Q^{*}}(U;Y) \leq R_{z}-\epsilon$, $I_{Q^{*}}(UX;Y) \leq R_{z} + R_{y}-2\epsilon$ and $R_{z} + R_{y} - I_{Q^{*}}(UX;Y)  - 2\epsilon \geq \psi(R_{y}-\epsilon, R_{z}-\epsilon, \hat{P}_{\by}) - g(Q^{*})$. 
Next, given that $\tilde{N}_{\by}(Q_{UY}^{*}) = e^{n \lambda}$, $N_{\by}(Q^{*})$ is a binomial random variable with $e^{n(R_{y} + \lambda)}$ trials and success rate of the exponential order of $e^{-nI_{Q^{*}}(X;Y|U)}$. \\
By the Chernoff bound, the probability of (\ref{up8}) is upper bounded by
\begin{align}
&\mathrm{Pr} \Big\{ N_{\by}(Q^{*})  \leq \exp \{n [ \psi(R_{y}-\epsilon, R_{z}-\epsilon, \hat{P}_{\by}) - g(Q^{*}) ] \} \Big| \tilde{N}_{\by}(Q_{UY}^{*}) = e^{n \lambda}  \Big\} \\
\label{up9}
&\leq             \left\{ 
               \begin{array}{l l}
                \exp \Big\{ -e^{n(R_{y} + \lambda)} D( e^{-an} \| e^{-bn} ) \Big\}                                                                                                                                                                                 & \quad \text{  $\lambda \in \calG$  }\\
                1              & \quad \text{ $\lambda \in \calG ^{c}$  } 
               \end{array} \right. 
\end{align}
where $a = R_{y} + \lambda + g(Q^{*}) - \psi(R_{y}-\epsilon, R_{z}-\epsilon, \hat{P}_{\by})$, $b = I_{Q^{*}}(X;Y|U)$ and  
\begin{align}  
\calG = \Big\{\lambda ~\Big|~ I_{Q^{*}}(X;Y|U) \leq R_{y} - \epsilon + \lambda + g(Q^{*}) - \psi(R_{y}-\epsilon, R_{z}-\epsilon, \hat{P}_{\by})   \Big\}.
\end{align}
Noting that
\begin{align}
a - b &= R_{y} + \lambda  - I_{Q^{*}}(X;Y|U) + g(Q^{*}) - \psi(R_{y}-\epsilon, R_{z}-\epsilon, \hat{P}_{\by})  \\
&= R_{z} + R_{y}   - I_{Q^{*}}(UX;Y)   + g(Q^{*}) - \psi(R_{y}-\epsilon, R_{z}-\epsilon, \hat{P}_{\by})  + \lambda + I_{Q^{*}}(U;Y) - R_{z}  \\
&\geq 2 \epsilon  + \lambda + I_{Q^{*}}(U;Y) - R_{z}  \\
&\overset{\Delta}{=} \Theta(\lambda, R_{z})  ,
\end{align}
the binary divergence can be lower bounded as:
\begin{align}
D( e^{-an} \| e^{-bn} ) &\geq e^{-bn} \{ 1- e^{-(a-b)n} [1 + n(a-b)]  \} \\
&\geq e^{-n I_{Q^{*}}(X;Y|U)} \Big[ 1- e^{-n \Theta(\lambda, R_{z})} (1 + n \Theta(\lambda, R_{z}))  \Big] ,
\end{align}
where in the second inequality, we have used the decreasing monotonicity of the function $f(t) = (1+t) e^{-t}$ for $t \geq 0$. Thus,
\begin{align}
&\exp \Big\{ -e^{n(R_{y} + \lambda)} D( e^{-an} \| e^{-bn} ) \Big\}  \\
&\leq \exp \bigg\{ -e^{n(R_{y} + \lambda)} \cdot  e^{-n I_{Q^{*}}(X;Y|U)} \Big[ 1- e^{-n \Theta(\lambda, R_{z})} (1 + n \Theta(\lambda, R_{z}))  \Big]  \bigg\}  \\
&= \exp \bigg\{ -e^{n(R_{z} + R_{y} - I_{Q^{*}}(UX;Y) + I_{Q^{*}}(U;Y) - R_{z} + \lambda)} \cdot  \Big[ 1- e^{-n \Theta(\lambda, R_{z})} (1 + n \Theta(\lambda, R_{z}))  \Big]  \bigg\}  \\
&\leq \exp \bigg\{ -e^{n(2 \epsilon + I_{Q^{*}}(U;Y) - R_{z} + \lambda)} \cdot  \Big[ 1- e^{-n \Theta(\lambda, R_{z})} (1 + n \Theta(\lambda, R_{z}))  \Big]  \bigg\}  \\
&= \exp \bigg\{ -e^{n\Theta(\lambda, R_{z})} \cdot  \Big[ 1- e^{-n \Theta(\lambda, R_{z})} (1 + n \Theta(\lambda, R_{z}))  \Big]  \bigg\}  \\
&= \exp \Big\{ -e^{n \Theta(\lambda, R_{z})}  + n \Theta(\lambda, R_{z}) + 1  \Big\} .
\end{align}
Let us continue from (\ref{up9}) and upper bound as follows
\begin{align}
&\mathrm{Pr} \Big\{ N_{\by}(Q^{*})  \leq \exp \{n [ \psi(R_{y}-\epsilon, R_{z}-\epsilon, \hat{P}_{\by}) - g(Q^{*}) ] \} \Big| \tilde{N}_{\by}(Q_{UY}^{*}) = e^{n \lambda}  \Big\} \\
&\leq           
\exp \Big\{ -e^{n(R_{y} + \lambda)} D( e^{-an} \| e^{-bn} ) \Big\} 
\cdot \calI \big\{ \calG \big\} + \calI \big\{ \calG^{c} \big\}  \\
&\leq           
\exp \Big\{ -e^{n \Theta(\lambda, R_{z})}  + n \Theta(\lambda, R_{z}) + 1  \Big\}
+ \calI \big\{ \calG^{c} \big\}.
\end{align}
Then, the next step will be to evaluate the expectation over the enumerator $\tilde{N}_{\by}(Q_{UY}^{*})$.
\begin{align}
 &\mathbb{E} \bigg[ \exp \Big\{ -e^{n \Theta(\lambda, R_{z})}  + n \Theta(\lambda, R_{z}) + 1  \Big\} + \calI \big\{ \calG^{c} \big\}  \bigg] \\
\label{Origin4321}
&\doteq  \sum_{i=0}^{R_{z}/ \delta} \mathrm{Pr} \Big\{ e^{n i \delta}  \leq  \tilde{N}_{\by}(Q_{UY}^{*})  \leq  e^{n (i+1) \delta}  \Big\} \nonumber \\ 
&~~~~~~~~~~~~~~~~~~~~~\times  \bigg[ \exp \Big\{ -e^{n \Theta(i \delta, R_{z})}  + n \Theta(i \delta, R_{z}) + 1  \Big\} + \calI \big\{ \calG^{c} \big\}  \bigg] .
\end{align} 
Moving forward, notice that the type class enumerator $\tilde{N}_{\by}(Q_{UY}^{*})$ is a sum of $e^{nR_{z}}$ independent binary random variables, each one with an expectation of $e^{-n I_{Q^{*}}(U;Y)}$. On the one hand, we have
\begin{align}
\mathrm{Pr} \Big\{ \tilde{N}_{\by}(Q_{UY}^{*})  \leq  e^{n t}  \Big\} \doteq \mathcal{I} \Big\{ R_{z} \leq I_{Q^{*}}(U;Y) + t  \Big\},
\end{align}
and on the other hand, we have
\begin{align}
\mathrm{Pr} \Big\{ \tilde{N}_{\by}(Q_{UY}^{*})  \geq  e^{n t}  \Big\} \doteq e^{-n \hat{E}(Q_{UY}^{*}, t)}
\end{align}
where,
\begin{align}
\hat{E}(Q_{UY}^{*}, t)     =     
           \left\{ 
               \begin{array}{l l}
     \big[ I_{Q^{*}}(U;Y)  -  R_{z}  \big]_{+}   & \quad \text{  $\big[ R_{z}  -   I_{Q^{*}}(U;Y)   \big]_{+} \geq t$  }\\
      \infty                                             & \quad \text{ $\big[ R_{z}  -   I_{Q^{*}}(U;Y)   \big]_{+} < t.$  } 
               \end{array} \right.  
\end{align}
Since $R_{z} - \epsilon \geq I_{Q^{*}}(U;Y)$, we conclude that the event $\big\{ e^{n t}  \leq  \tilde{N}_{\by}(Q_{UY}^{*})  \leq  e^{n (t + \delta) }  \big\}$ occurs with very high probability if and only if $\big( R_{z} - \epsilon  -  I_{Q^{*}}(U;Y) \big) \in [t, t + \delta)$, otherwise, its probability has a double exponential decay. Therefore, it turns out that the sum in (\ref{Origin4321}) is dominated by one summand only, the one for which $i = \big( R_{z} - \epsilon  - I_{Q^{*}}(U;Y)  \big)/ \delta$. For this value of $i$, we have that
\begin{align}
\Theta(i \delta, R_{z}) &= 2 \epsilon  + i \delta + I_{Q^{*}}(U;Y) - R_{z} =  \epsilon, 
\end{align}
and, in addition,
\begin{align}
\calI \big\{ \calG^{c} \big\} &= \calI \big\{ I_{Q^{*}}(X;Y|U) > R_{y} - \epsilon + i \delta + g(Q^{*}) - \psi(R_{y}-\epsilon, R_{z}-\epsilon, \hat{P}_{\by}) \big\} \\
&= \calI \big\{ I_{Q^{*}}(UX;Y) > R_{y}  + R_{z} - 2\epsilon  + g(Q^{*}) - \psi(R_{y}-\epsilon, R_{z}-\epsilon, \hat{P}_{\by}) \big\} \\
&=0,
\end{align}
thanks to the fact that $R_{z} + R_{y} - I_{Q^{*}}(UX;Y)  - 2\epsilon \geq \psi(R_{y}-\epsilon, R_{z}-\epsilon, \hat{P}_{\by}) - g(Q^{*})$.
By using the fact that $\delta > 0$ is arbitrarily small, we get that the expectation in question behaves like
\begin{align}
\mathbb{E} \bigg[ \exp \Big\{ -e^{n \Theta(\lambda, R_{z})}  + n \Theta(\lambda, R_{z}) + 1  \Big\} + \calI \big\{ \calG^{c} \big\}  \bigg] \doteq  \exp \big\{ -e^{n \epsilon}  + n \epsilon + 1  \big\}.
\end{align}

\end{document}